\documentclass[fleqn,usenatbib]{mnras}

\usepackage{newtxtext,newtxmath}
\usepackage{rotating}

\usepackage[T1]{fontenc}

\DeclareRobustCommand{\VAN}[3]{#2}
\let\VANthebibliography\thebibliography
\def\thebibliography{\DeclareRobustCommand{\VAN}[3]{##3}\VANthebibliography}

\usepackage{braket}
\usepackage{float}
\usepackage[x11names,table]{xcolor}
\usepackage{booktabs}
\usepackage{times}
\usepackage[usenames,dvipsnames]{pstricks}
\usepackage{color}
\usepackage{subcaption}
\usepackage{comment}
\usepackage{ulem}
\usepackage[section]{placeins}
\usepackage{balance}
\usepackage{graphicx}	
\usepackage{amsmath}	
\usepackage{dblfloatfix} 
\usepackage{caption}
\usepackage{adjustbox}      

\definecolor{blazeorange}{rgb}{1.0, 0.6, 0.2}
\definecolor{seagreen}{rgb}{0.18, 0.55, 0.34}
\definecolor{rufous}{rgb}{0.66, 0.11, 0.03}
\definecolor{royalfuchsia}{rgb}{0.79, 0.17, 0.57}
\definecolor{scarlet}{rgb}{1.0, 0.13, 0.0}
\definecolor{royalpurple}{rgb}{0.47, 0.32, 0.66}
\definecolor{darkblue}{rgb}{0, 0, 0.66}
\definecolor{violet}{rgb}{0.5,0.,0.5}

\title[GRBs from Population III Stars]{On the Efficiency of Producing Gamma-Ray Bursts from Isolated Population III Stars}

\author[Morales-Rivera et al.]{
Gibran Morales-Rivera,$^{1}$\thanks{E-mail: gibranmr@gmail.com}
Ramandeep Gill,$^{1,3}$\thanks{E-mail: r.gill@irya.unam.mx}
S. Jane Arthur,$^{1}$
P. Beniamini,$^{2,3,4}$
J. Granot$^{2,3,4}$
\\
$^{1}$Instituto de Radioastronom\'ia y Astrof\'isica, Universidad Nacional Aut\'onoma de M\'exico, Antigua Carretera a P\'atzcuaro $\#$ 8701,  Ex-Hda. San Jos\'e de la \\ Huerta, Morelia, Michoac\'an, C.P. 58089, M\'exico\\
$^{2}$Department of Natural Sciences, The Open University of Israel, P.O Box 808, Ra'anana 4353701, Israel \\
$^{3}$Astrophysics Research Center of the Open university (ARCO), The Open University of Israel, P.O Box 808, Ra'anana 43537, Israel \\
$^{4}$Department of Physics, The George Washington University, 725 21st Street NW., Washington, DC 20052, USA
}

\date{Accepted XXX. Received YYY; in original form ZZZ}

\pubyear{\the\year{}}

\begin{document}
\label{firstpage}
\pagerange{\pageref{firstpage}--\pageref{lastpage}}
\maketitle

\begin{abstract}
The rate of long-duration gamma-ray bursts (GRBs) from isolated Pop III 
stars is not well known, as it depends on our poor understanding of their initial 
mass function (IMF), rotation rates, stellar evolution, and mass loss. Some massive 
($M_{\rm ZAMS}\gtrsim20M_\odot$) Pop III stars are expected to suffer core-collapse and launch 
a relativistic jet that would power a GRB. In the collapsar scenario, a key 
requirement is that the pre-supernova star imparts sufficient angular momentum 
to the remnant black hole to form an accretion disc and launch a relativistic jet, 
which demands rapid initial rotation of the progenitor 
star and suppression of line-driven mass loss during its chemically homogeneous 
evolution. Here we explore a grid of stellar evolution models of Pop III stars with 
masses $20\leq M_{\rm ZAMS}/M_\odot \leq 100$, which are initially rotating with surface angular 
velocities $0.6\leq \Omega_0/\Omega_{\rm crit}\leq 0.9$, where centrifugally-driven mass 
loss ensues for $\Omega>\Omega_{\rm crit}$. Realistic accretion and jet propagation 
models are used to derive the initial black hole masses and spins, and jet breakout times 
for these stars. The GRB production efficiency is obtained over a phase space comprising 
progenitor initial mass, rotation, and wind efficiency. For modest wind efficiency of $\eta_{\rm wind}=0.45-0.35$, 
the Pop III GRB production efficiency is $\eta_{\rm GRB}\sim10^{-5}-3\times10^{-4}\,M_\odot^{-1}$, 
respectively, for a top-heavy IMF. This yields an observable all-sky equivalent rate 
of $\sim2-40\,{\rm yr}^{-1}$ by \textit{Swift}, with 75\% of the GRBs located 
at $z\lesssim8$. If the actual observed rate is much lower, then this would imply $\eta_{\rm wind}>0.45$, which 
leads to significant loss of mass and angular momentum that renders isolated Pop III stars 
incapable of producing GRBs and favors a binary scenario instead.
\end{abstract}




\section{Introduction}
The first stars in the Universe, often called Population III (Pop III) stars, emerged from  
metal-free primordial gas (i.e. comprising only H and He) and ended the so-called dark-ages 
\citep[see, e.g.,][for reviews]{Bromm-Larson-04,Klessen-Glover-23}. The copious amounts of 
UV photons produced by these stars re-ionized the Universe \citep[e.g.][]{Gnedin-Ostriker-97,Tumlinson-Shull-00}, 
and the supernovae that ended the lives of massive Pop III stars enriched the intergalactic 
medium with heavy elements \citep[e.g.][]{Ostriker-Gnedin-96,Heger-Woosley-02,Yoshida+04} that affected the next 
generation of stars. In the $\Lambda$CDM model of cosmology, Pop III stars are predicted to 
form at redshift $z\gtrsim30$ and dominate the star-formation rate at $z\sim15-20$, after which 
epoch slightly metal-enriched Pop II stars begin to emerge \citep[e.g.][]{Hartwig+22}. Significant 
uncertainty exists in predicting the formation rate and scenarios of the first stars \citep{Barkana-Loeb-01} 
as exploration is only limited to advanced numerical simulations \citep[e.g.][]{Bromm+99,Bromm+02,Abel+02}, 
that in many cases are low resolution or lack relevant physical processes, with no near-term possibility of direct observations 
\citep{Schauer+20}. Alternative strategies involving surveys of local metal poor stellar populations 
to constrain the effect of the first stars on the nucleosynthetic makeup of Pop II stars \citep{Tumlinson-06} 
will have to wait future large, high-resolution spectroscopic surveys \citep{Jeon+21}. 
Therefore, at present, it is very challenging to probe the initial mass function and star 
formation rate at high redshifts.

Most simulations find an approximately top-heavy initial mass function of Pop III stars \citep{Wollenberg+20} 
in comparison to the present day one \citep{Chabrier-03}. Stars at the higher end 
of the mass distribution ($M\gtrsim8M_\odot$) are expected to explode as core-collapse supernovae, 
pair-instability supernovae \citep[PISNe;][]{Heger-Woosley-02}, or collapse directly to a black hole with no associated supernova \citep{Heger+03,Yoon+12}.
Out of these, some are expected to launch powerful relativistic jets if the collapsed core retains sufficient 
angular momentum to form an accretion disc, which would also power a more energetic supernova -- a hypernova with kinetic energy 
$E_{\rm kin}\gtrsim10^{52}$\,erg \citep{Iwamoto+98,Khokhlov+99,Nomoto+03}. If the jet manages to break out of 
the star then internal dissipation can produce a bright GRB \citep[see, e.g.,][for a review]{Kumar-Zhang-15}. 
The radio afterglows of hypernovae can be detected up to $z\sim20$ \citep{Ioka-Meszaros-05}, the 
prompt GRB to $z\sim100$ \citep{Lamb-Reichart-00} and their afterglows to $z\sim30$ in infrared, X-rays 
\citep{Ciardi-Loeb-00,Gou+04} and radio \citep{Ioka-03,Inoue-04}. The current record holders 
include GRB\,090423 at $z=8.2$ \citep{Tanvir+09} and GRB\,090429B at $z=9.4$ \citep{Cuchiara+11}. 
Therefore, by obtaining deeper and more sensitive observations, it is possible to probe the initial 
mass function of Pop III stars using GRBs and their afterglows \citep[see, e.g.,][for a review]{Toma+16}.

This prospect has attracted proposals of several new missions with greater sensitivity towards 
detecting high-redshift transients. The current front-runner in probing the distant Universe is 
the \textit{James Webb Space Telescope} (JWST) that will not detect the Pop III stars themselves 
but will be sensitive to intrinsically bright transients, such as GRBs and PISNe \citep{Whalen+13}, 
instead. Another upcoming observatory that has the potential for detecting supernovae from Pop III stars  
for $z\gtrsim6$ is the \textit{Roman Space Telescope} \citep{Spergel+15,Whalen+13}. A few other notable  
missions that have been proposed to probe the highest redshift include the High-$z$ Gamma-ray 
bursts for Unraveling the Dark Ages Mission \citep[HiZ-GUNDAM;][]{Yonetoku+14,Yoshida+16}, the Gamow Explorer 
\citep{White+21}, and the Transient High-Energy Sky and Early Universe Surveyor \citep[THESEUS;][]{Amati+18}. 

In anticipation of these missions, several works have calculated the intrinsic and observable 
rate of GRBs at high redshifts 
\citep{Bromm-Loeb-06, Naoz-Bromberg-07,Campisi+11,DeSouza+11,Kinugawa+19,Ghirlanda-Salvaterra-22,Fryer+22}. 
Such a calculation requires knowledge of the star-formation rate (SFR) at high redshifts, 
the efficiency of producing GRBs per unit stellar mass \citep[e.g.][]{,Lloyd-Ronning+20}, and the luminosity 
function. While the luminosity function can be constructed from existing observations of GRBs with redshifts 
measured using their afterglows \citep[e.g.][]{Lloyd-Ronning+02,Wanderman-Piran-10}, the SFR of Pop III stars and GRB production efficiency 
($\eta_{\rm GRB}$ $=$ number of GRBs per formed stellar mass) 
are not so well constrained. In particular, any inference of $\eta_{\rm GRB}$ is further 
complicated by our poor understanding of stellar evolution of metal-free stars which is 
significantly impacted by rotation and mass loss (see below for a discussion of mass loss 
in Pop III stars). The not so well understood physics of core-collapse, properties of the remnant, 
and jet launching mechanisms further add to the challenges in ascertaining $\eta_{\rm GRB}$. 

Therefore, many works fix this parameter by comparing the theoretically expected rate of GRBs 
for a given instrument with the observed one. For example, \citet{Bromm-Loeb-06} constrain 
$\eta_{\rm GRB}\simeq2\times10^{-9}\,M_\odot^{-1}$ for GRBs produced by Pop I/II progenitors, 
and assume that the same applies to Pop III stars, by comparing with the rate of \textit{Swift} 
GRBs. Although consistent with upper limits in later works, it is several orders of 
magnitude lower than expected. For example, the local observed all-sky rate of GRBs with luminosity 
$L_{\gamma,\rm iso} \geq 10^{50}\,{\rm erg\,s}^{-1}$ is $\simeq 1.3\,{\rm Gpc}^{-3}\,{\rm yr}^{-1}$ 
\citep{Wanderman-Piran-10}. Correcting this for beaming, with $\eta_{\rm beam}\simeq1/500$ 
\citep{Frail+01}, since not all GRBs are beamed towards us, 
we obtain the intrinsic rate of $\sim 6.5\times10^{-7}\,{\rm Mpc}^{-3}\,{\rm yr}^{-1}$. When 
taking the density of Milky way (MW) like galaxies in the local Universe to be $\sim10^{-2}\,{\rm Mpc}^{-3}$, this 
rate is equivalent to $\sim 6.5\times10^{-5}\,{\rm yr}^{-1}$ per MW-galaxy. The SFR of MW is 
$0.68-1.5\,M_\odot\,{\rm yr}^{-1}$ \citep{Robitaille-Whitnew-10}, which serves as a good proxy of the 
local SFR, then the local GRB production efficiency is $\eta_{\rm GRB, local}\sim 4.3\times10^{-5} - 10^{-4}\,M_\odot^{-1}$. 
In later works, \citet{Naoz-Bromberg-07} obtain an upper limit on the efficiency, with 
$\eta_{\rm GRB}<3.2\times10^{-4}\,M_\odot^{-1}$, by comparing with the observed GRB rate of 
$15\,{\rm yr}^{-1}$, out of the $90\,{\rm yr}^{-1}$, detected by \textit{Swift} with no optical 
counterpart and $T_{90}>50$\,s. The assumption of optically `dark' GRBs, and therefore no redshift 
information, is valid for very high-z GRBs as their optical afterglow photons observed at $z=0$, but emitted 
with energies $E_z = (1+z)E_{\rm obs}$ and thus with wavelengths smaller than the Lyman limit, 
would be absorbed en route by photoionizing H/He atoms \citep{Lamb-Reichart-00}. \citet{Campisi+11} calculate 
$\eta_{\rm GRB}=f_{\rm GRB}\times\eta_{\rm BH}\lesssim7\times10^{-5}M_\odot^{-1}$, where $\eta_{\rm BH}$ is 
the efficiency per unit solar mass for producing a BH and $f_{\rm GRB}$ is the fraction that 
produces a GRB. They arrive at this limit by assuming a Salpeter initial mass function for Pop III 
stars in the mass range of $100<M/M_\odot<140$ and $260<M/M_\odot<500$, which yields 
$\eta_{\rm BH}\sim3.2\times10^{-3}M_\odot^{-1}$, and that $f_{\rm GRB}\sim2.2\times10^{-2}$ as 
inferred from \textit{Swift} observations. Stellar evolution models show that very massive Pop III stars with 
$M>260M_\odot$ end their lives as red-supergiants that have extended stellar envelopes which 
makes it difficult for the jet to penetrate out over the engine lifetime \citep{Yoon+12, Yoon+15}.

In this work, we calculate the efficiency of producing a GRB from a distribution of {isolated (i.e. single and not in a binary)} Pop III 
stars based on stellar evolution numerical calculations done using the public code \texttt{MESA}. {We leave the investigation of Pop III stars in a binary to a future work.}  
After a brief review in \S\ref{sec:pop-3-evolution} of the stellar evolution expected, and as 
shown by earlier works using different stellar evolution codes, for massive and rapidly rotating 
Pop III stars, we describe our simulation setup in \S\ref{sec:setup}. The remaining \S\ref{sec:simulations} 
describes the results from our model grid spanning an initial mass range of $20\leq M_{\rm ZAMS}/M_\odot \leq 100$ 
as well as initial rotation of $0.6\leq \hat\Omega_0\equiv\Omega_0/\Omega_{\rm crit}\leq0.9$ and shows 
the radial profiles of density, angular velocity and specific angular momentum. We evolve most of 
the stars in our grid to advanced nuclear burning stages approaching imminent core-collapse, as 
shown in the HR diagram, which has not been achieved in many earlier works on Pop III stars. These 
stellar profiles are then used to analytically calculate the properties of core-collapse and 
accretion in \S\ref{sec:core-collapse-accretion}, where we give the mass and spin of the newly 
formed BH when the accretion disc forms. Based on the rate of accretion, we calculate the jet power 
in \S\ref{sec:jet-power} and its subsequent evolution through the stellar interior after launch  
using an analytic formalism in \S\ref{sec:criteria_GRB}. Using our jet breakout criteria, here we 
also present the efficiency of GRB production as a function of initial progenitor mass, rotation, and stellar 
wind efficiency. We use this phase space producing successful GRBs to calculate the rate of GRBs 
in \S\ref{sec:GRB-rate} as a function of redshift. In \S\ref{sec:discussion} we conclude that if 
the all-sky detection rate of Pop III GRBs by any instrument with \textit{Swift}/BAT sensitivity 
is significantly less than $\sim2\,{\rm yr}^{-1}$, then the stellar wind efficiency is constrained 
from below, which severely limits the prospects of producing GRBs from isolated Pop III stars.

\begin{figure*}
    \includegraphics[width=0.48\textwidth]{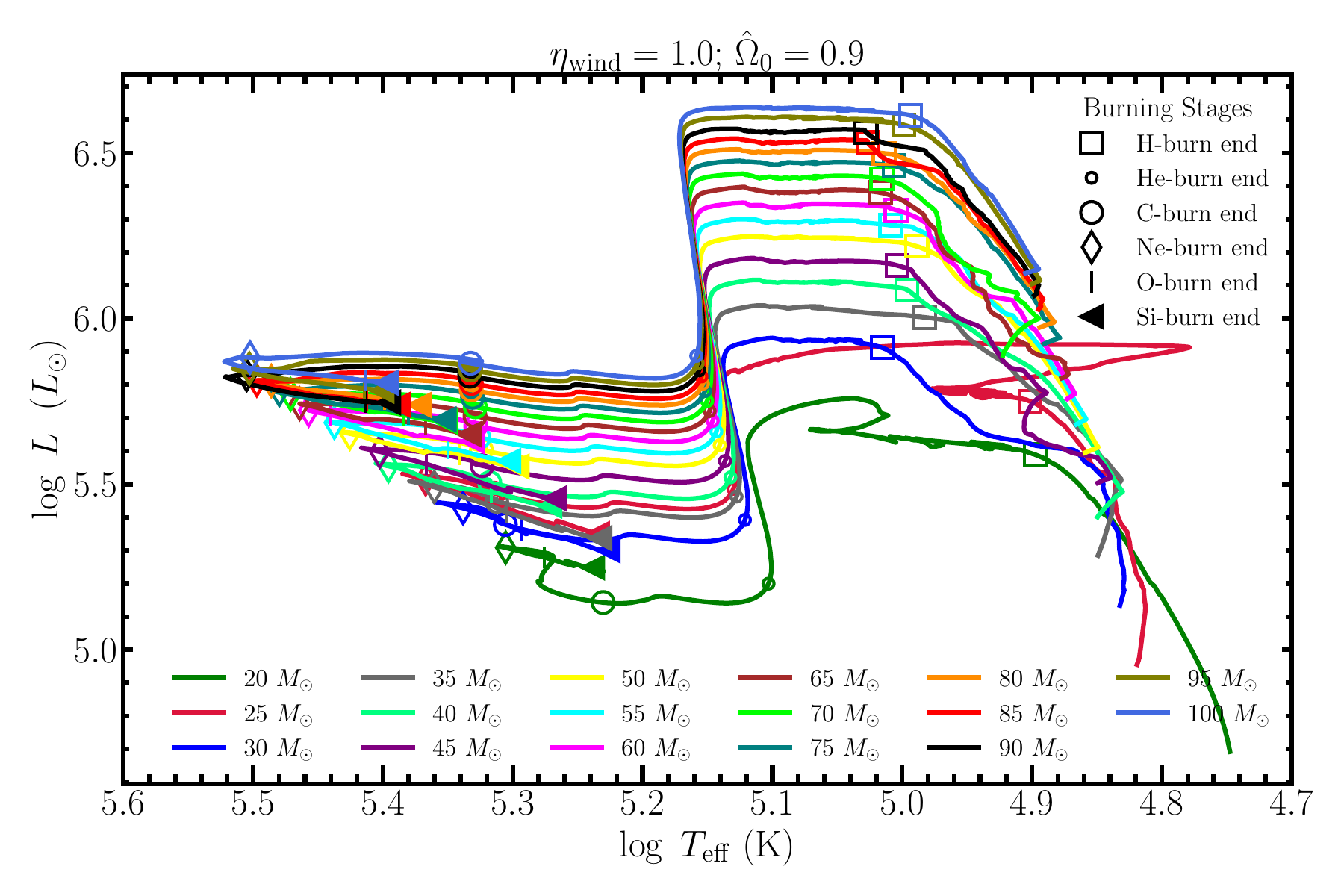}
    \includegraphics[width=0.48\textwidth]{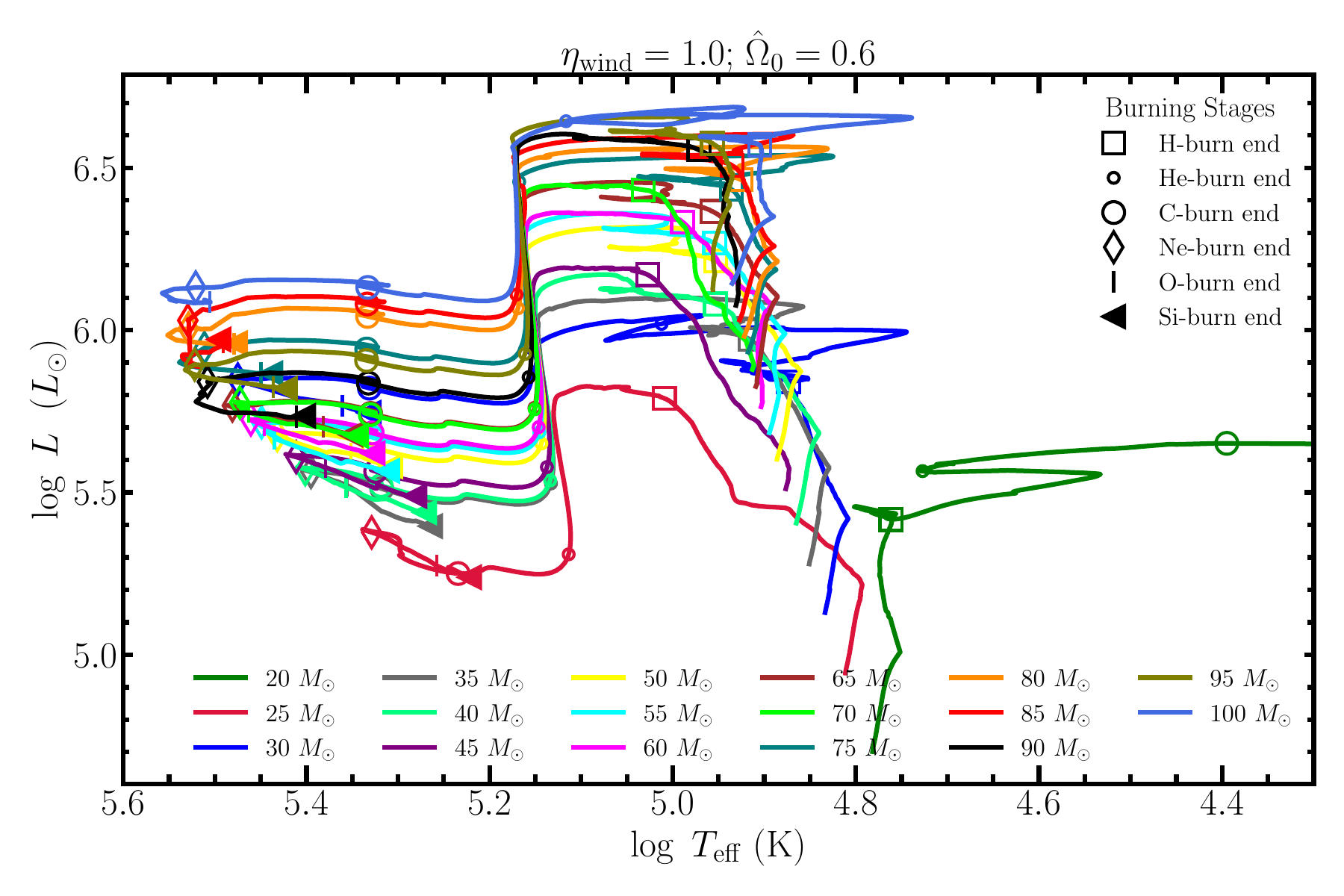}
    \includegraphics[width=0.48\textwidth]{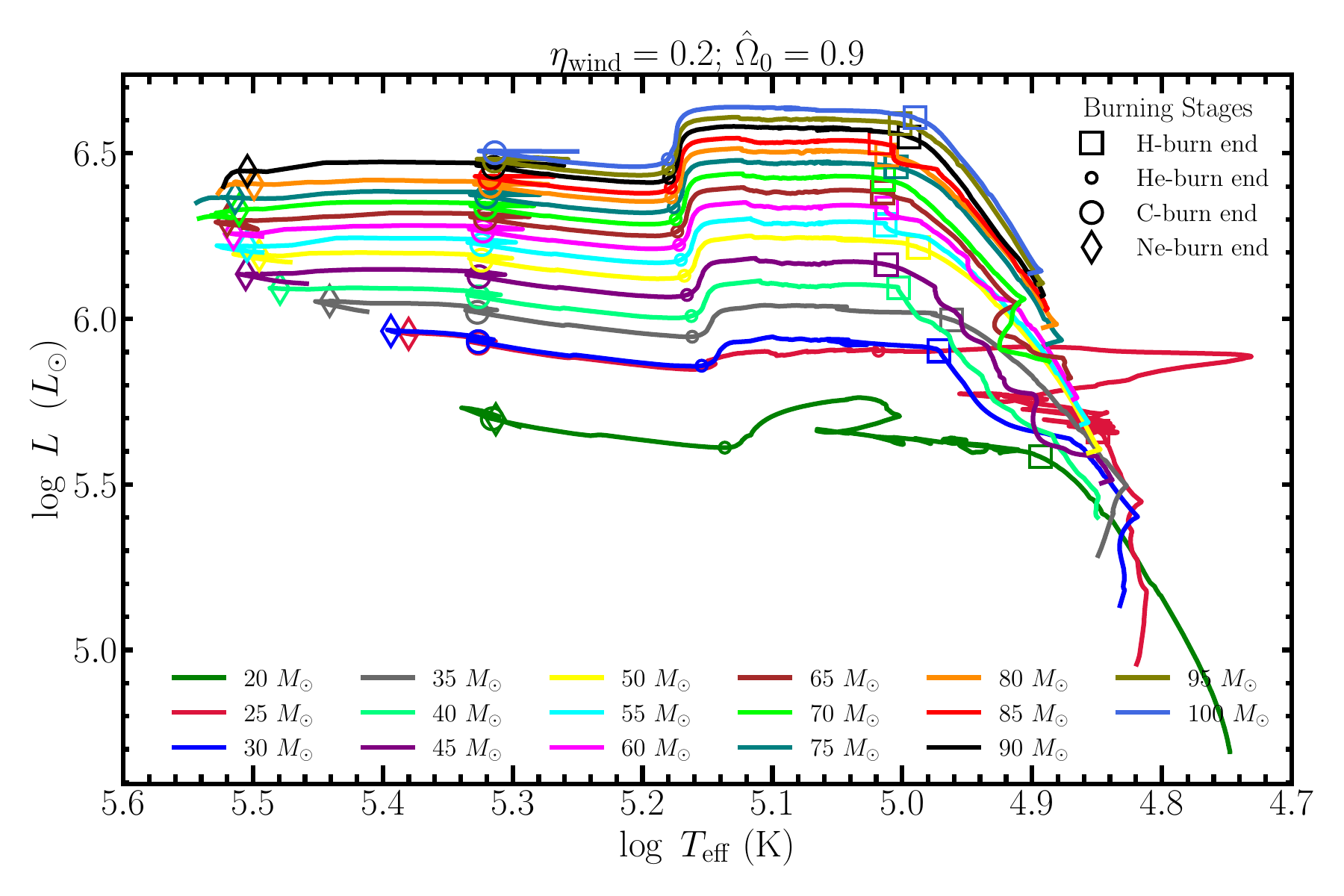}
    \includegraphics[width=0.48\textwidth]{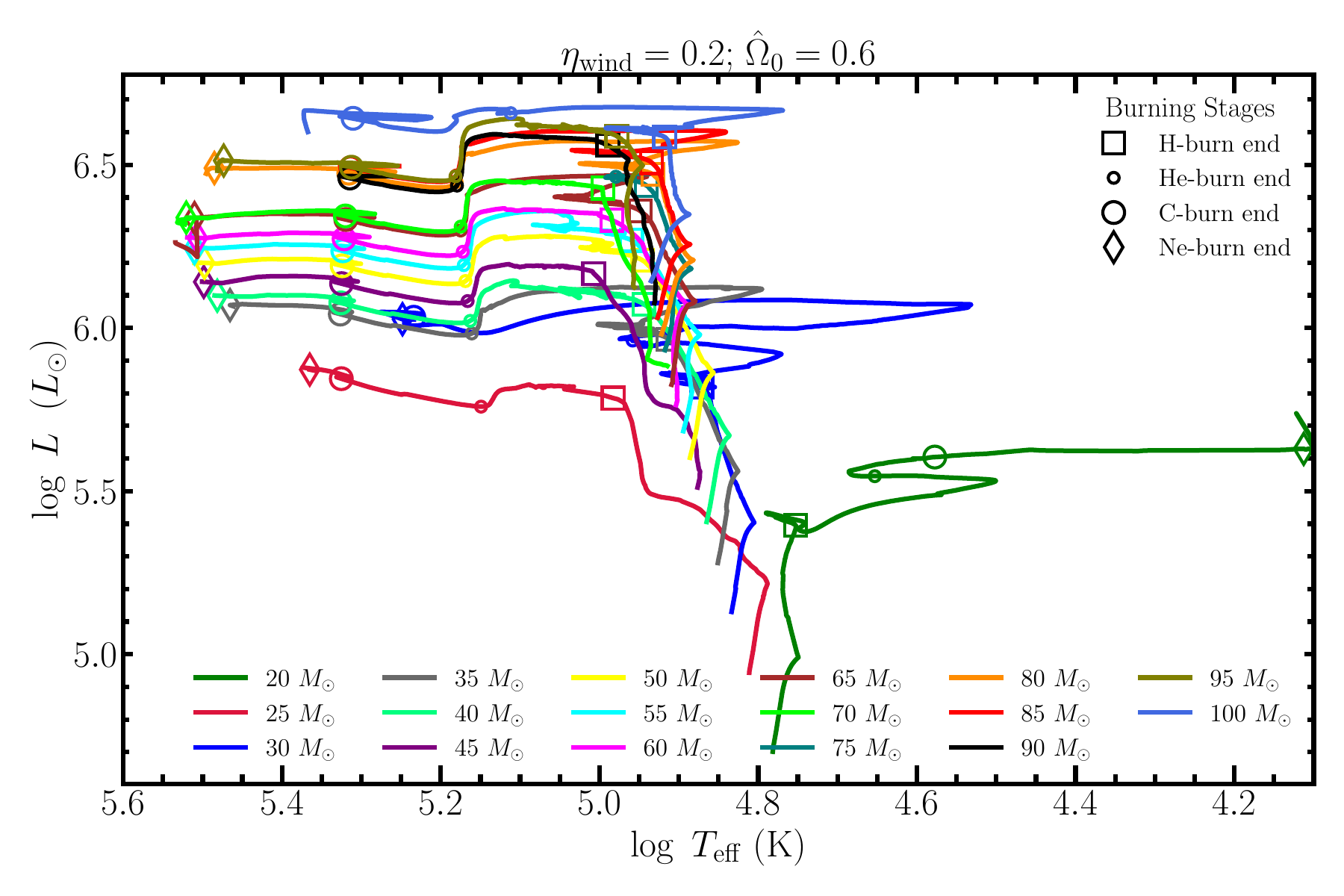}
    \caption{Evolutionary tracks in the HR diagram for stellar models with 
    initial masses ranging from 20 to 100 $M_{\odot}$, assuming an initial rotation fraction of 
    $\hat{\Omega}_0\equiv\Omega_0/\Omega_{\rm crit} = 0.9$ (left) and $\hat{\Omega}_0 = 0.6$ (right).  
    Each track shows luminosity evolution as a function of the effective temperature from the 
    ZAMS to advanced core nuclear burning stages before core-collapse. These different nuclear burning 
    stages are highlighted with different symbols. 
    (\textit{Top}) Stellar evolution with Dutch wind scaling factor of $\eta_{\rm wind} = 1.0$, with 
    evolutionary tracks shown up to a central temperature of $\log T_{\rm core} < 9.6$. 
    (\textit{Bottom}) The panel presents the scenario for $\eta_{\rm wind} = 0.2$, where tracks are 
    presented up to $\log T_{\rm core} < 9.4$. 
    These temperature thresholds are imposed to prevent numerical instabilities that arise in the 
    late evolutionary stages.
  }
    \label{fig:HR_vertical}
\end{figure*}

\subsection{Stellar Evolution of Rapidly Rotating Massive Pop III Stars}
\label{sec:pop-3-evolution}
Stellar rotation has been shown to alter the evolution of massive stars \citep{Heger+00,Maeder-Meynet-00,Langer-12}, 
and this effect is even more pronounced in lower metallicity stars \citep{Meynet-Maeder-02,Yoon+12}. In 
comparison to their non-rotating counterparts, rapid rotation tends to produce significantly larger 
(by $\sim25\%$ in mass) He cores due to rotationally-induced chemical mixing that supplies fresh fuel to 
the core from the outer stellar layers. Since the He-core mass is a proxy for the remnant mass, rotation 
reduces the limit on the zero-age-main-sequence (ZAMS) mass that can produce core-collapse supernovae. 
Rotationally-induced chemical mixing also leads to chemically homogeneous evolution \citep[CHE;][]{Maeder-87}, 
that produces dramatic changes in the evolutionary tracks on the HR diagram. As a result, massive metal-poor stars 
show a blueward evolution if they are rotating sufficiently fast as compared to a redward one for stars 
with lower rotation \citep{Maeder-87,Langer-92,Brott+11,Yoon+12}. 

As demonstrated in \citet{Brott+11}, CHE is sensitive to the 
initial stellar mass, rotation, and metallicity. In general, rotation tends to lower the local effective 
gravity due to the centrifugal acceleration, which affects the radiative energy flux \citep{vonZeipel-24} at 
the stellar surface. This leads to cooler and less luminous stars, a latitude dependence of the surface temperature 
and luminosity, as well as polar and equatorial outflows \citep{Maeder-99}. However, these effects are canceled 
in rapidly rotating massive stars (e.g. $\gtrsim15M_\odot$) due to the development of relatively more massive 
cores that yield higher luminosities when compared to non-rotating stars of the same mass. 

\begin{figure*}
    \includegraphics[width=0.48\textwidth]{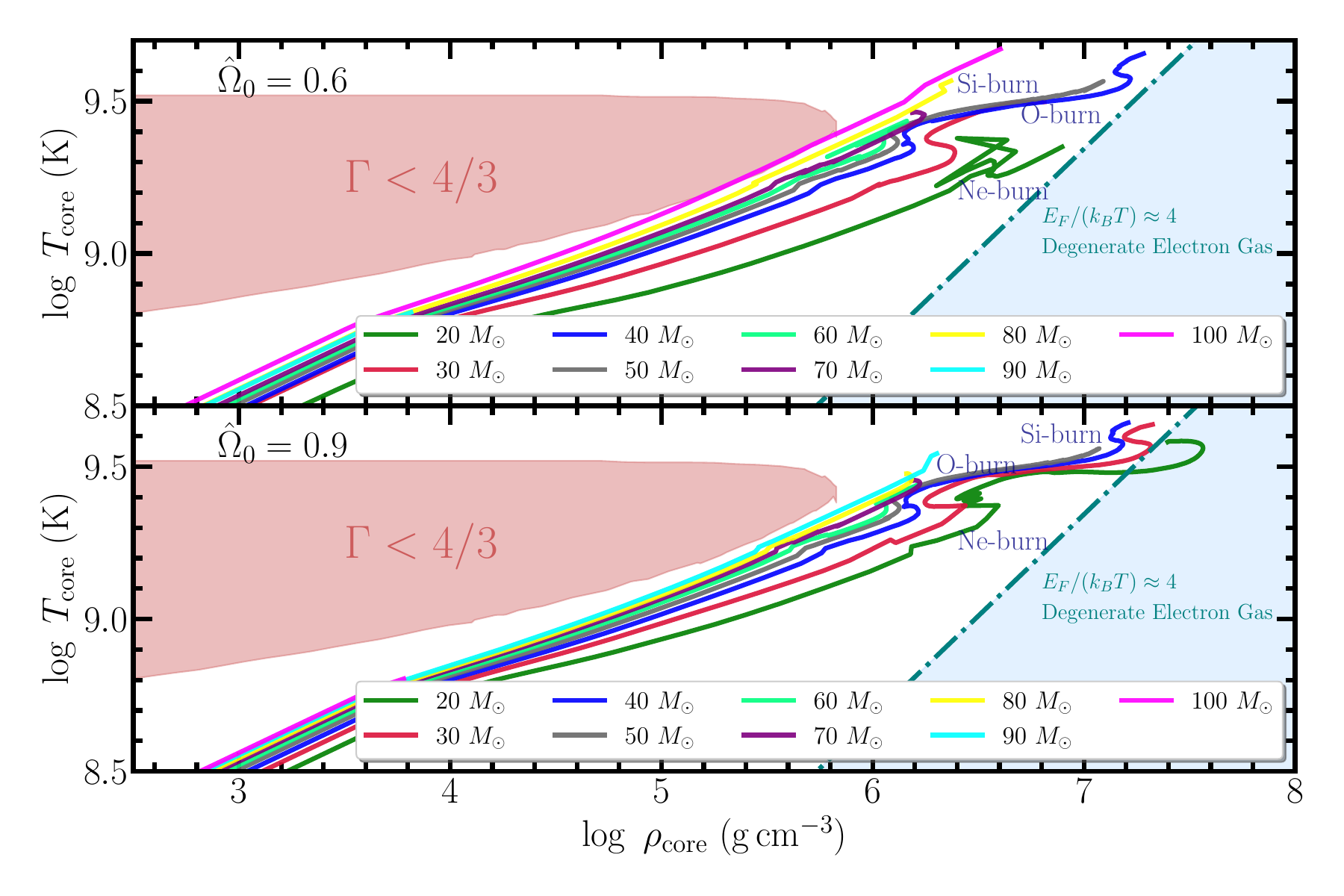}
    \includegraphics[width=0.48\textwidth]{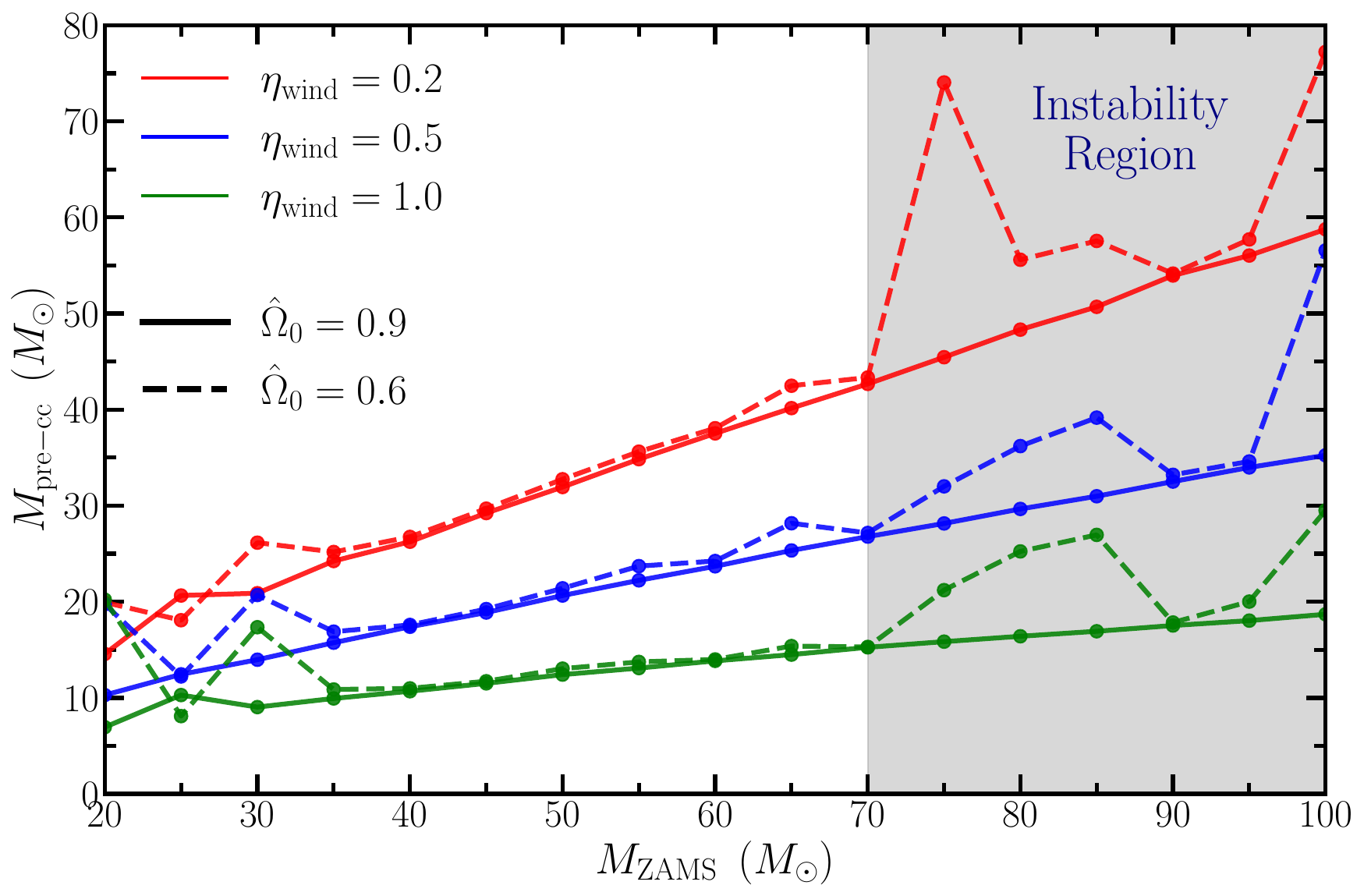}
\caption{ (\textit{Left})
    Evolution of the core temperature as a function of core density for stellar models 
    with masses ranging from $20M_{\odot}$ to $100M_{\odot}$ for $\eta_{\rm wind} = 0.2$ 
    and for an initial rotation of $\hat{\Omega}_0=0.6$ (top) and  $\hat{\Omega}_0=0.9$ (bottom).  The tracks 
    illustrate the progression through different nuclear burning stages (C, Ne, O, Si) 
    as the core evolves. The red shaded area indicates the $\Gamma < 4/3$ instability 
    criterion, valid particularly for non-rotating stars, where the stellar core becomes 
    dynamically unstable due to the creation of electron-positron pairs that reduce the 
    adiabatic index below the critical threshold, while the blue region corresponds to the 
    criteria for a degenerate electron gas. (\textit{Right}) Evolution of the pre-core-collapse 
    mass  as a function of the initial mass ($M_{\rm{ZAMS}}$), for different values of 
    $\eta_{\rm wind}$ and initial rotation of $\hat{\Omega}_0 = 0.9$ (solid) and $\hat{\Omega}_0 = 0.6$ (dashed). 
    Slower rotating massive stars ($M_{\rm ZAMS}>70M_\odot$) enter the (pulsational) pair-instability 
    region and their evolution was not followed all the way to core-collapse. Hence, the 
    non-monotonic behavior in the pre-core-collapse mass. 
}
    \label{fig:Tc_Rhoc_vertical}
\end{figure*}

Rotational mixing becomes more efficient at lower metallicity \citep{Yoon+06,Brott+11}. Stars with higher metallicity are prone 
to stronger line-driven winds that drive mass loss and along with it the loss of angular momentum \citep{Langer-98}. This 
slows down the star and reduces the efficiency of rotational mixing. In slow rotators, chemical mixing is 
inhibited due to the emergence of a chemical gradient between the convective core and radiative envelope \citep{Maeder-87}, 
when the mixing timescale becomes longer than the thermonuclear time. 
In contrast, very metal-poor rapid rotators, and in particular Pop III stars, do not lose angular momentum to 
line-driven winds and therefore become fully chemically mixed and show a blueward rise in luminosity in the 
HR diagram.

Rotation also alters the limits on the ZAMS mass of metal-free stars that are capable of producing core-collapse and 
pair-instability supernovae. In non-rotating stars, a type IIP supernova is expected where a star with mass 
$8\lesssim M_{\rm ZAMS}/M_\odot \lesssim 25M_\odot$ undergoes core-collapse and forms a neutron star \citep{Heger+03}. 
Above this limit and in the initial mass range of $25\lesssim M_{\rm ZAMS}/M_\odot\lesssim140$ the remnant collapses 
to a BH, either by fallback accretion ($M\lesssim 40M_\odot$) or directly \citep[][]{Fryer-99}. 
When the core collapses directly to a BH, there is no accompanying type II supernova. In the mass range of 
$140\lesssim M_{\rm ZAMS}/M_\odot\lesssim260$, the core suffers from pulsational pair-instability and is disrupted 
entirely, leaving no remnant \citep{Heger-Woosley-02}. Stars more massive than $\sim260M_\odot$ again form BHs 
directly as their cores implode due to photodisintegration \citep{Fryer+01,Heger-Woosley-02}. \citet{Yoon+12} carried 
out a comprehensive study of rotating massive Pop III stars and showed that the above limits are modified in rapid 
rotators. Stars born with initial surface angular velocity $\Omega\gtrsim 0.5\Omega_k$, where $\Omega_k = \sqrt{GM/R_\star^3}$ 
is the Keplerian angular velocity at the stellar surface, do not produce type II supernovae in the mass range 
$13\lesssim M_{\rm ZAMS}/M_\odot\lesssim190$. Instead, they experience CHE and end their lives in type Ib/c supernovae, 
with some in the mass range $13\lesssim M_{\rm ZAMS}/M_\odot\lesssim84$ producing GRBs and hypernovae.

\section{Simulations with \texttt{MESA}}\label{sec:simulations}

\subsection{Model Setup}\label{sec:setup}

\begin{figure*}
	\centering
	\includegraphics[width=0.48\textwidth]{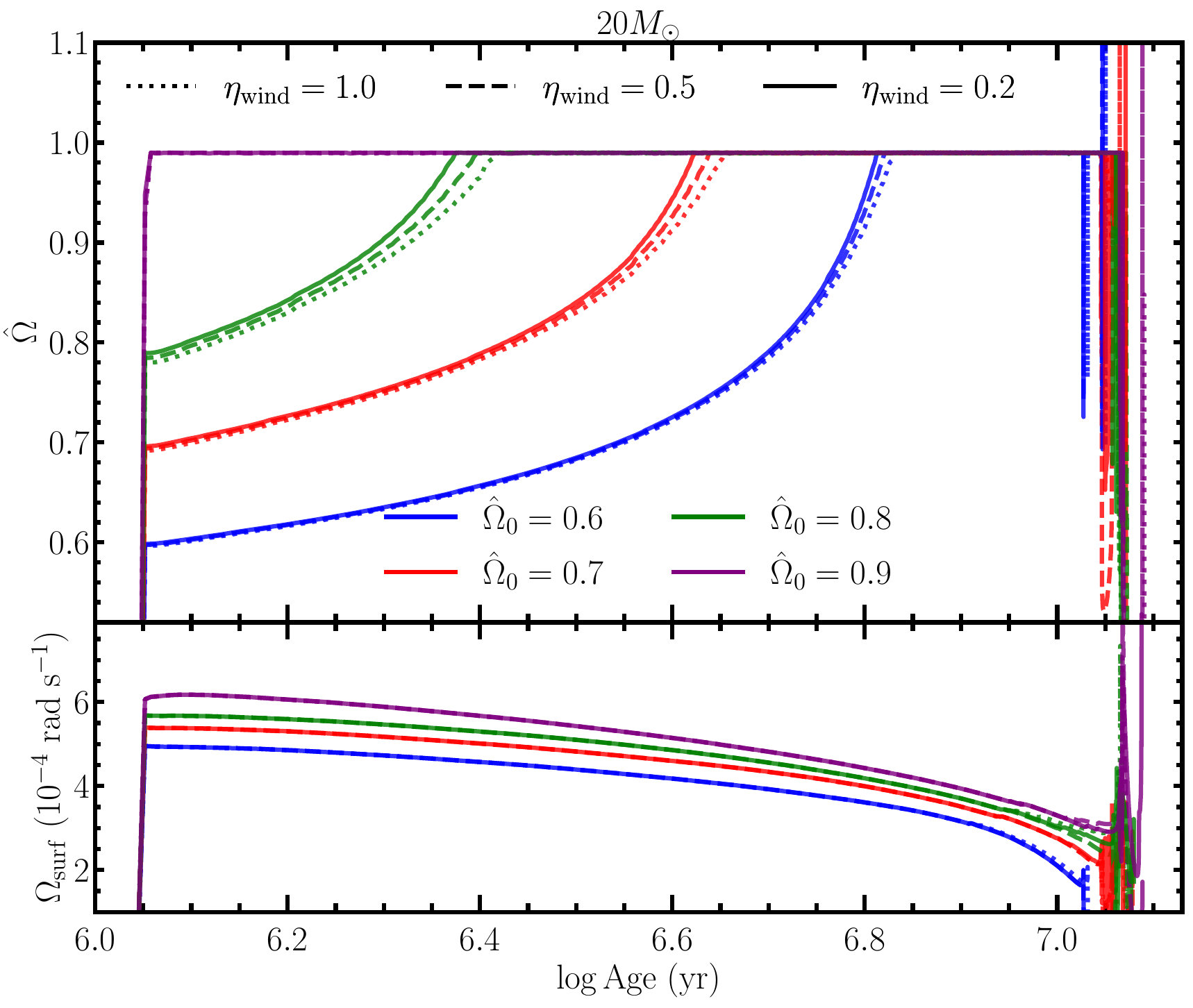}
        \includegraphics[width=0.48\textwidth]{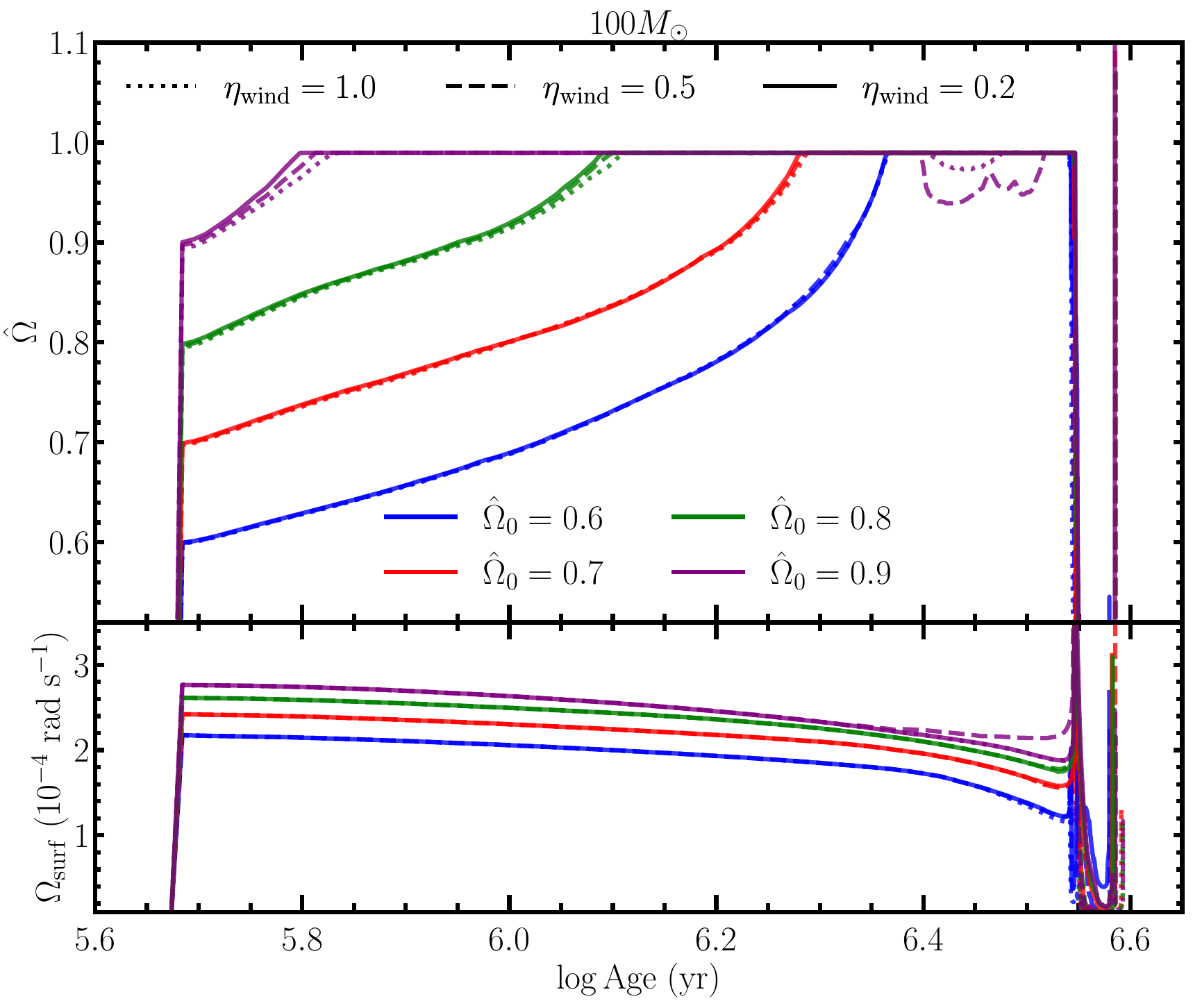}
\caption{
    Temporal evolution of the stellar surface rotation for $20\,M_\odot$ (\textit{left}) 
    and $100\,M_\odot$ (\textit{right}) stars. 
    (\textit{Top}) Normalized surface angular velocity $\hat{\Omega} \equiv \Omega / \Omega_{\rm crit}$, 
    as a function of stellar age, for different wind scaling factors: $\eta_{\rm wind} = 1.0$ 
    (dotted lines), $\eta_{\rm wind} = 0.5$ (dashed lines), $\eta_{\rm wind} = 0.2$ (solid lines), 
    and for different initial rotation rates: $\hat{\Omega}_0 = 0.6$ (blue), $\hat{\Omega}_0 = 0.7$ (red), 
    $\hat{\Omega}_0 = 0.8$ (green), and $\hat{\Omega}_0 = 0.9$ (purple). 
    (\textit{Bottom}) Corresponding evolution of the surface angular velocity.
}
	\label{fig:Omega_Age_eta_variation}
\end{figure*}

We present a grid of simulations covering the stellar evolution of massive zero-metallicity ($Z=0$) stars in the 
mass range of $20\leq M_{\rm ZAMS}/M_\odot\leq 100$. These were performed using the public stellar evolution 
code \texttt{MESA}  (version 22.05.1), which is a one-dimensional code that includes the effects of convection, 
rotation, and mass loss \citep{Paxton+11,Paxton+13,Paxton+15,Paxton+18,Paxton+19}. Our models follow the stellar 
evolution of rapidly rotating stars from the pre-main sequence (pre-ZAMS) to the advanced burning stages (e.g. 
core silicon burning) just before core-collapse. 

All stars are initialized at the ZAMS with solid-body 
rotation at an angular velocity $\Omega$ which is some fraction of the critical angular velocity 
$\Omega_{\mathrm{crit}} = \left(1 - \frac{L}{L_{\mathrm{Edd}}}\right)^{1/2}\Omega_k$, where the ratio 
$L/L_{\rm Edd} = \kappa L/4\pi cGM$ is the Eddington factor, $\kappa$ is the opacity to electron 
scattering, $L$ and $M$ are the stellar luminosity and mass, and $G$ is the gravitational constant. 
The critical angular velocity, evaluated at the stellar equator, represents the break-up limit and 
includes the effect of radiative acceleration via the Eddington factor. Here we consider a range of 
initial rotation rates, with $0.6\leq \hat{\Omega}_0\equiv \Omega_0/\Omega_{\rm crit} \leq 0.9$, 
that are subcritical. However, as the star evolves, the ratio 
$\Omega/\Omega_{\rm crit}$ at the surface approaches unity which is then followed by substantial 
rotationally-driven mass loss until the surface angular velocity becomes subcritical.

Line-driven mass loss is initially unimportant for Pop III stars due to the complete absence of metals. 
However,  rapid rotation drives the stars towards CHE that mixes the metals produced during core He burning and 
at later stages into the stellar envelope. At that point, stellar winds due to non-zero surface metallicity become 
important. For the treatment of stellar winds, we adopt the \textit{Dutch} wind scheme as described in \citet{Glebbeek+09} 
for both cool and hot phases, setting temperature thresholds of $0.8 \times 10^4 \text{ K}$ and $1.2 \times 10^4 \text{ K},$ 
respectively, to switch between mass-loss prescriptions. Specifically, we employ the wind prescription by \citet{Dejager+18} 
in the low-temperature regime, which is well-suited for cool, extended stellar envelopes. On the other hand, 
for the high-temperature regime and  given the high helium mass fraction at the surface, the mass-loss rate 
transitions to the \citet{Nugis-Lamers-00} Wolf-Rayet prescription. In both regimes, the wind mass-loss rate,  
$\dot{M}_{\rm wind} = \eta_{\rm wind}\dot M_{\rm wind}^{\rm model}$, is obtained by scaling the model mass-loss 
rate by a dimensionless parameter $0<\eta_{\rm wind}\leq1$. Different wind schemes and 
their efficiencies are calibrated by comparing theoretical expectations with observations of stars in our Galaxy and 
those in the small and large Magellanic clouds. Since there are no direct observations of Pop III stars, it 
is difficult to prescribe a certain wind efficiency $\eta_{\rm wind}$ in removing stellar mass. 
Earlier works discussing stellar evolution of GRB progenitors used $\eta_{\rm wind}\sim0.1 - 0.3$ 
\citep[e.g.][]{Woosley-Heger-06, Yoon+12} to retain sufficient angular momentum to launch relativistic 
jets. We employ the wind schemes in a parameterised way and consider a range of efficiencies with 
$0.2\leq\eta_{\rm wind}\leq 1.0$. These choices allow us to model a range of mass-loss regimes, from fully 
efficient winds to scenarios with significantly reduced mass-loss rates, thereby capturing the potential 
diversity in the evolution of metal-free massive stars. 

The effects of rotation are incorporated with the explicit inclusion of magnetic fields following the Spruit-Taylor 
dynamo \citep{Spruit-02} and fully accounting for the Solberg–Høiland instability, the secular shear instability, 
Eddington–Sweet circulation, and the Goldreich–Schubert–Fricke instability \citep{Paxton+13}. 
Convective and mixing processes are treated 
via mixing-length theory (MLT), where we set the mixing-length parameter to $\alpha_{\rm MLT}=1.5$ and adopt the Henyey 
option, which yields stable convergence in convective regions. The convective boundaries are determined using the Ledoux 
criterion, ensuring that the chemical composition gradients are taken into account. Semiconvection is incorporated using 
the \citet{Langer+85} mixing scheme with an efficiency parameter of $\alpha_{\rm sc}=0.02$, allowing moderate mixing in 
semiconvective zones. It is worth noting that in the late stages of evolution, the time scales associated with 
semiconvective mixing become longer than those for nuclear reaction processes. As a result, semiconvection is unlikely to have 
a considerable impact on evolution. Its induction causes numerical instabilities due to rapid changes in the stellar structure, 
and therefore we set $\alpha_{\rm sc}=0$ during advanced burning before core-collapse. A step overshoot scheme during 
core H burning is used with overshoot parameter $f = 0.345,$ while the scale length at the convective boundary is set to 
$f_0 = 0.01$. 

In most of our models, stellar evolution is followed up to late stages before core-collapse when the core Silicon ($^{28}$Si) 
fraction is $X_{\rm Si28} < 10^{-3}$. In some models, the evolution is terminated earlier than this limit due to onset of 
instabilities. However, in all cases, the stars are evolved up to a point when mass loss has terminated and the stellar structure, 
including the radial density and angular momentum profiles, remain unchanged in the subsequent stellar evolution before the 
instabilities set in. This ensures that the total angular momentum remains unchanged and profiles obtained at this stage are 
very similar to what would have been obtained just before core-collapse.

\subsection{Stellar Evolution Models}
Here we present the results of these simulations, focusing primarily on their evolutionary tracks and an exhaustive 
compilation of their physical properties throughout their evolution, but particularly during the final stage of stellar 
evolution that precedes core-collapse and in which, in most cases, the iron core has already formed.  

Figure \ref{fig:HR_vertical} shows the evolutionary track in the HR diagram for all masses in our grid 
from the ZAMS to advanced stages prior to core-collapse. The models are initialized with a rigidly rotating 
star with angular velocity $\hat{\Omega}_0 = \{0.6,\,0.9\}$. These diagrams 
are shown up to central temperatures of $\log T_{\rm c} < \{9.6,\ 9.4\}$ for $\eta_{\rm wind} = \{1.0,\  0.2\}$ 
to avoid numerical difficulties in resolving the outer stellar layers that cause erratic behaviour in the HR 
diagram. In contrast, the core remains stable and the models are in fact evolved to higher temperatures. 
Key nuclear burning stages are marked with distinct symbols, as indicated in the legend, corresponding to the 
end of hydrogen, helium, carbon, neon, oxygen, and silicon burning, 
when their respective central mass fractions are $X<0.01$. 

When $\hat{\Omega}_0=0.9$, all stellar models undergo CHE due to rapid rotation \citep{Brott+11,Yoon+12}, 
which causes the stars to evolve bluewards where they become hotter and more luminous due to an increase in the 
helium fraction at the surface, which in turn causes the surface to contract and heat up rather than expand towards 
cooler temperatures. This trend is different from that obtained in the non-rotating counterparts \citep{Marigo+01}, 
where the evolution is redwards towards cooler effective temperatures.

Left panels in Figure \ref{fig:Tc_Rhoc_vertical} show the evolutionary tracks in the $\log T_{\rm core}-\log \rho_{\rm{core}}$ 
plane for the cores of stars in our grid. 
The different rows show the evolution for different wind mass-loss efficiencies. These evolutionary 
tracks reflect the typical behaviour of the stars in our model, beginning with low core densities and temperatures 
(due to hydrogen burning) and progressing to higher densities and temperatures as the core contracts and 
heavier isotopes burn. The high initial rotation imposed in our models also favours rotational mixing, incorporating 
fresh hydrogen into the core during the main sequence and increasing the convective core mass \citep{Ekstrom+08}. 
The right panel of this figure shows the evolution of the pre-core-collapse mass 
($M_{\rm pre-cc}$) as a function of the initial mass ($M_{\rm ZAMS}$), considering 
different stellar wind efficiencies ($\eta_{\rm wind} = 0.2,\,0.5,\,1.0$) and initial 
rotation rates ($\hat{\Omega}_0 = 0.9$ for solid lines and $\hat{\Omega}_0 = 0.6$ for dashed lines). 
Here we can notice the effect of wind efficiency, with a low $\eta_{\rm wind}$ value resulting in minimal 
mass loss, thus allowing the stars to retain a more significant fraction of their initial mass.
The relation between the two masses shows a linear trend, where lower wind efficiencies 
allow for greater mass retention. In the mass range of $35\leq M_{\rm ZAMS}/M_\odot\leq70$, 
the initial rate of rotation produces minimal changes in the outcome. However, results 
diverge at both the low and high ends. At the high mass end, in particular, the slower 
rotators become susceptible to the pulsational pair instability (PPI), which causes thermonuclear 
pulsations that eject part of the envelope without destroying the star \citep{Woosley-17, Yoon+12}. 
This effect is illustrated in the top-left panel of Fig.~\ref{fig:Tc_Rhoc_vertical}, corresponding 
to models with $\hat{\Omega}_0 = 0.6$. As rotation decreases, stellar evolution paths approach 
the instability region where the adiabatic index $\Gamma$ drops below $4/3$. In this regime, 
the formation of $e^\pm$ pairs softens the equation of state, reducing $\Gamma$ and weakening 
the pressure support against gravity. This loss of support leads to rapid core contraction.
 
\begin{figure*}
	\includegraphics[width=\columnwidth]{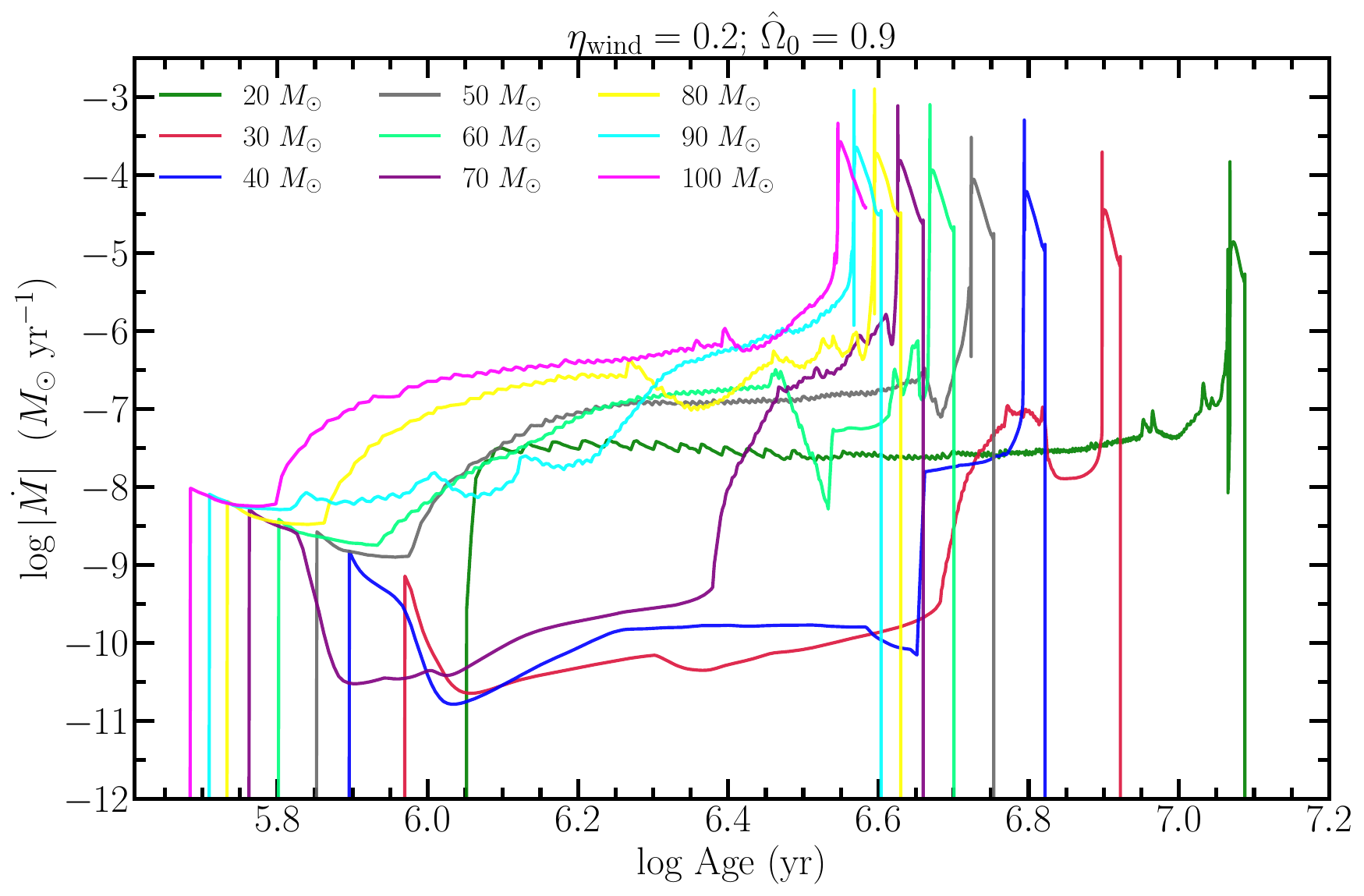}
    \includegraphics[width=\columnwidth]{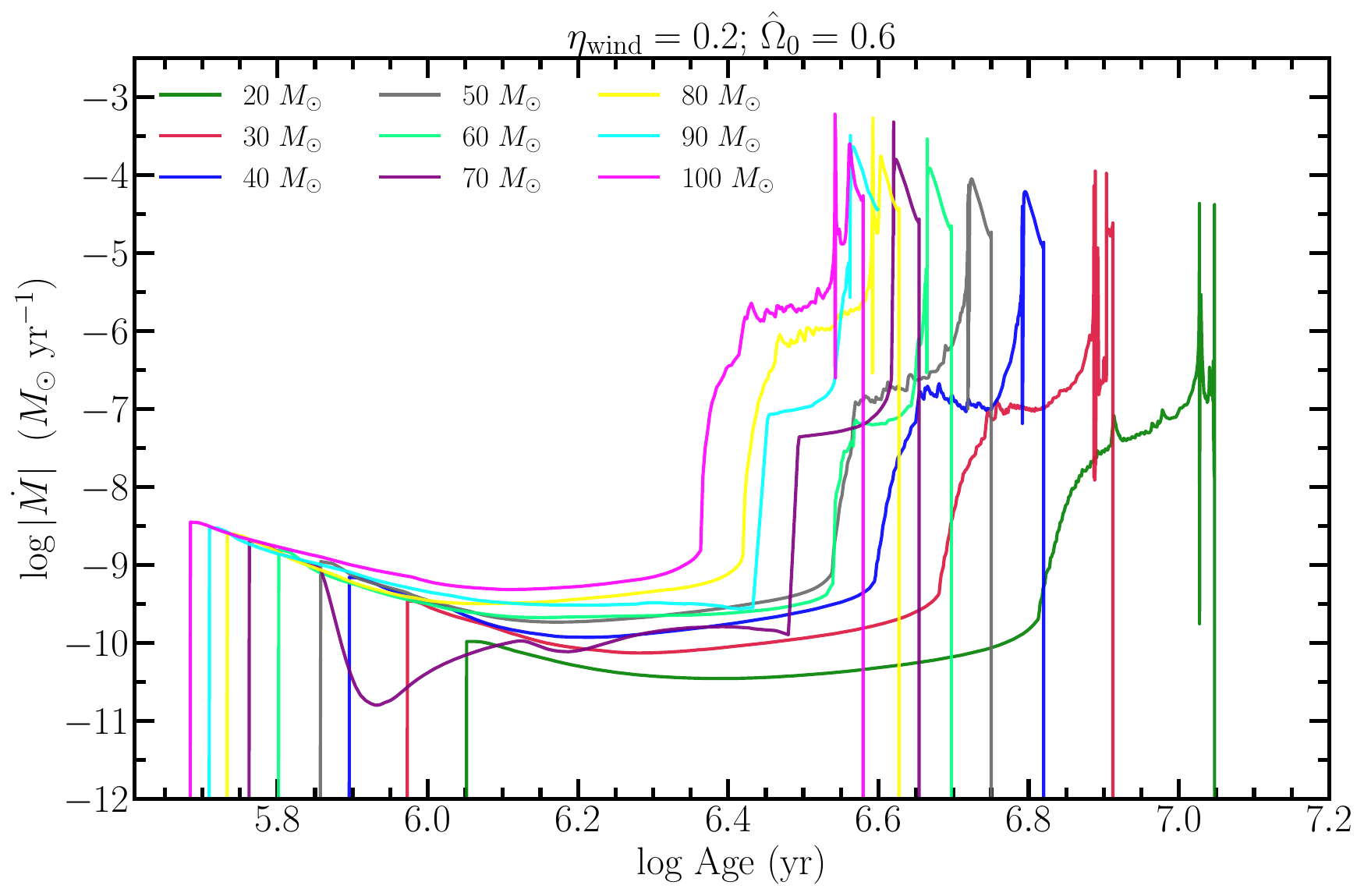}
	\caption{
    	Mass loss rates as a function of stellar age for the entire mass grid with wind efficiency 
        factor of $\eta_{\rm wind} = 0.2$, shown for two different initial rotation rates: $\hat{\Omega}_0=0.9$ 
        (\textit{left}), and $\hat{\Omega}_0=0.6$ (\textit{right}). Higher $\eta_{\rm wind}$ yields similar 
        mass loss rates. Models with higher initial $\hat{\Omega}_0$ reach critical rotation earlier 
        (see Fig.\,\ref{fig:Omega_Age_eta_variation}) and therefore show a sudden increase in mass loss 
        caused by centrifugal effects. 
	}
	\label{fig:Mdot_Age}
\end{figure*}

\subsection{Rotation and Mass Loss}
The evolution of rotational and mass-loss profiles in our Pop III star models is strongly 
influenced by the choice of initial parameters, such as initial mass $M_{\rm ZAMS}$, initial 
rotation rate $\hat{\Omega}_0$, and wind efficiency factor $\eta_{\rm wind}$. 
We examine these effects to understand the behaviour of each model. In general, we find that 
stars with high initial rotation and low $\eta_{\rm wind}$ (i.e., weak winds) retain more 
angular momentum in the core, which would eventually favor producing a successful GRB. 
In contrast, models with higher initial mass and high $\eta_{\rm wind}$ (i.e. strong winds) 
and low initial rotation experience significant loss of mass and angular momentum. Such stars will 
ultimately lack the requisite angular momentum to launch relativistic jets and will fail to 
produce GRBs.

Figure \ref{fig:Omega_Age_eta_variation} shows the evolution of the dimensionless surface angular 
velocity ($\hat{\Omega} \equiv \Omega/\Omega_{\rm crit}$) as a function of 
stellar age for 20 $M_\odot$ and 100 $M_\odot$ models, with varying $\eta_{\rm wind}$ 
factors across panels. As noted in prior studies \citep{Ekstrom+08, Chatzopoulos+12, Yoon+12, Murphy+21}, 
$\hat{\Omega}$ approaches critical rotation ($\hat{\Omega} \to 1$) over time, 
regardless of the initial rotation rate, due to the growth in stellar radius, which 
reduces $\Omega_k$ and consequently $\Omega_{\rm crit}$. For both cases with $\hat{\Omega}_0 = 0.9$ 
(purple curve), critical rotation is reached early, and then it is progressively delayed for 
slower rotators. When a star reaches critical rotation, it is accompanied by a 
significant increase in the rate of mass loss, and angular momentum, that allows the 
star to remain at $\hat\Omega=1$ and not exceed it. This is the scenario that is 
characterised as the \textit{forbidden region} in Fig.\,12 of \citet{Yoon+12}. Different 
wind efficiencies only have a mild effect on the time when the star reaches critical 
rotation.

In Figure \ref{fig:Mdot_Age}, we present the temporal evolution of the mass-loss rate from 
our grid of simulations, considering various initial masses ($M_{\rm ZAMS}$), initial rotation 
rates ($\hat{\Omega}_0$), and wind efficiencies ($\eta_{\rm wind}$). In our simulations, mass loss 
is driven by both rapid rotation and mixing of nuclear products from the core into the outer 
stellar layers due to CHE, which again is facilitated by rapid rotation. When comparing the two 
panels, showing the rate of mass loss for two different initial rotation rates, the 
onset of significant mass loss at early times can be seen for $\hat{\Omega}_0=0.9$. For example, in 
the $M_{\rm ZAMS}=100M_\odot$ case, the onset of rapid mass loss is triggered by the star achieving 
super-critical rotation, i.e. when $\hat\Omega>1$, as discussed above and shown in the right panel 
of Fig.\,\ref{fig:Omega_Age_eta_variation}. The same temporal coincidence can be seen for all of the 
stars in our grid. The rate of mass loss becomes even more extreme near the terminal stages of 
the star, when due to the CHE, the metal abundance in the stellar envelope rises greatly. 
This results in the removal of the outer stellar layers due to powerful line-driven winds. 
Mass loss is explicitly disabled when the central temperature reaches $T_{\rm core} = 1.1\times10^9$ K, 
during the terminal stage of the star, to maintain numerical stability. This is the reason for the very sharp drop in 
the rate of mass loss at late times. 

\begin{figure*}
    \centering
    \includegraphics[width=0.48\textwidth]{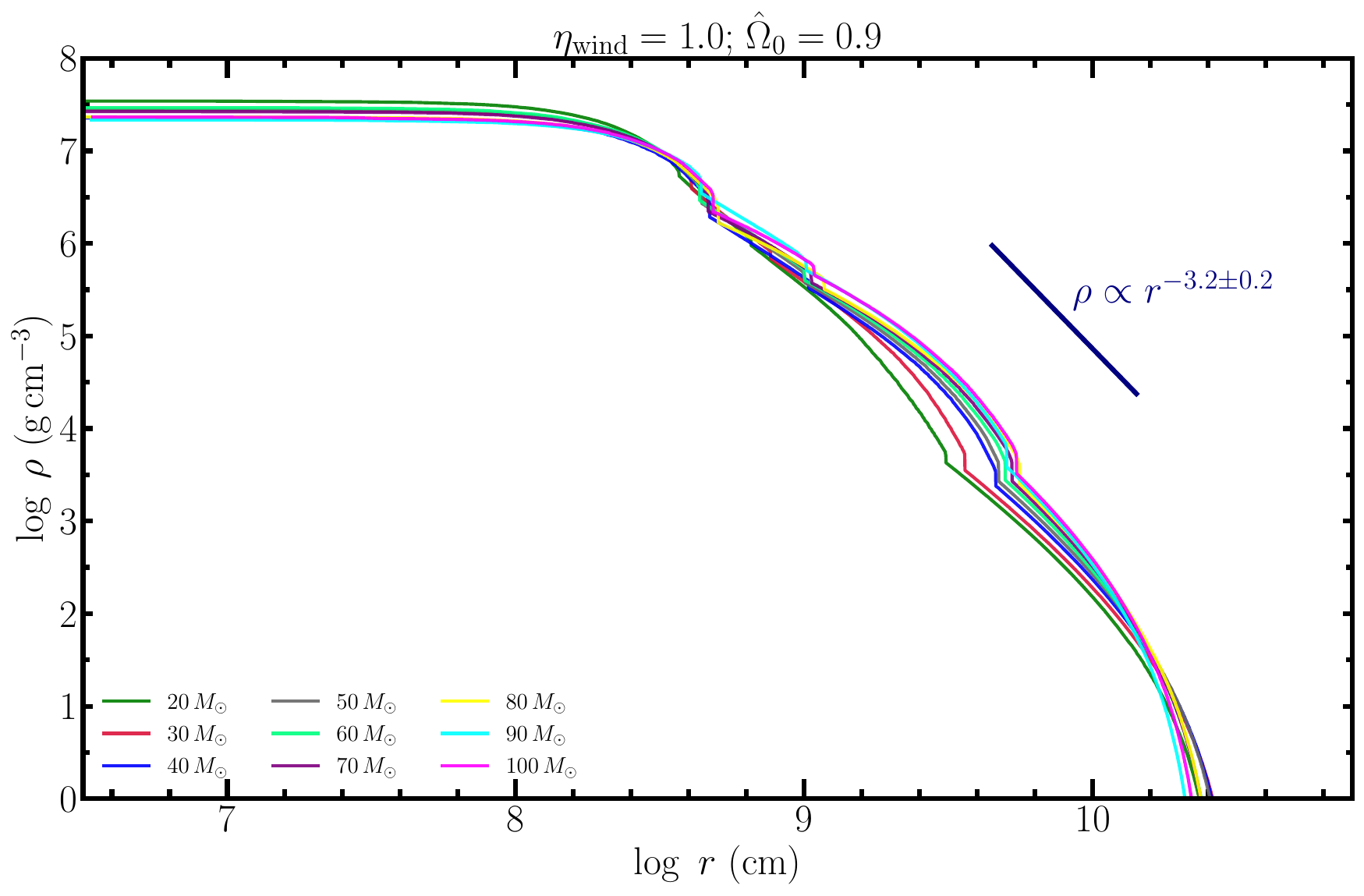}
    \includegraphics[width=0.48\textwidth]{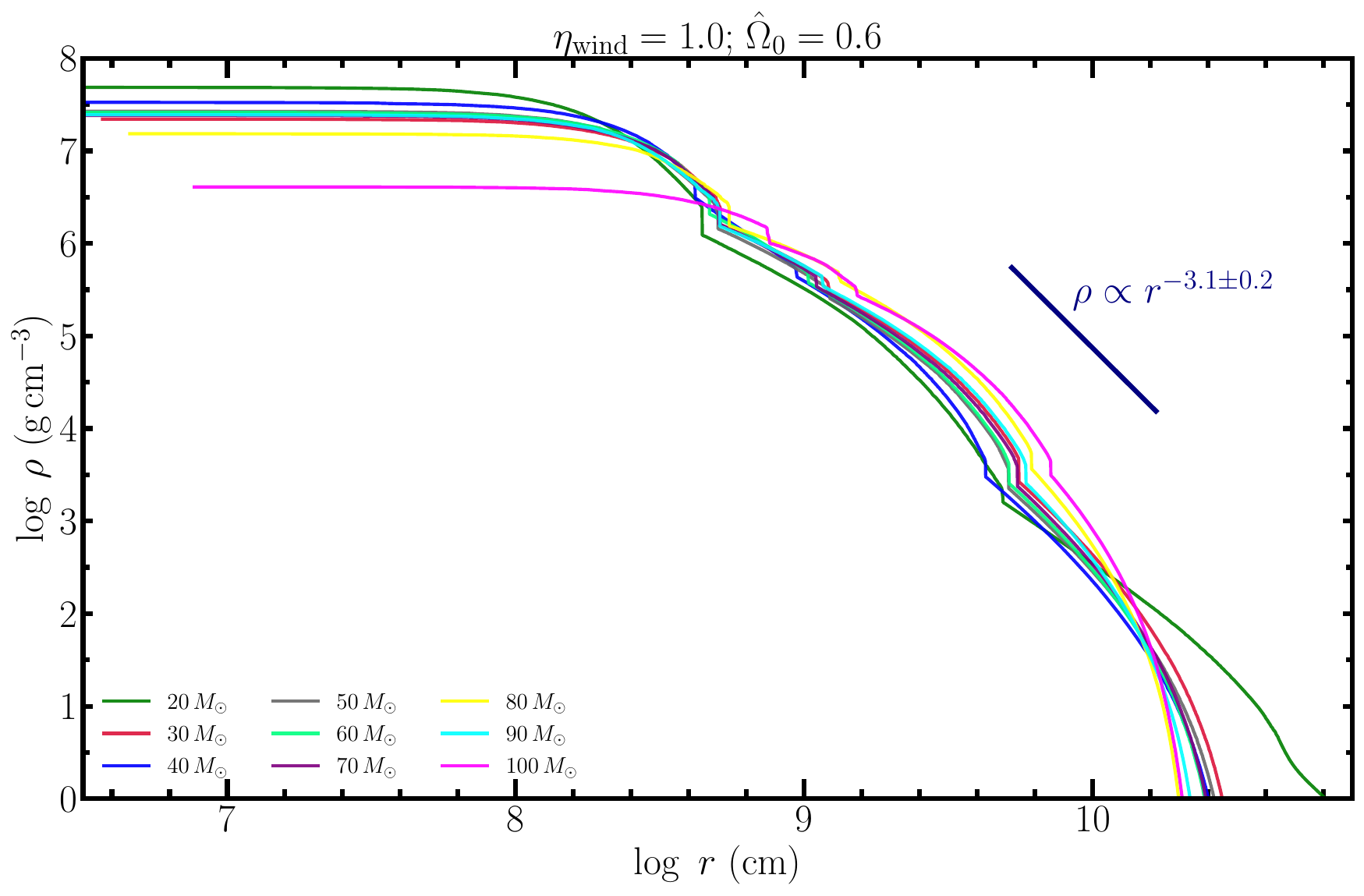}
    \includegraphics[width=0.48\textwidth]{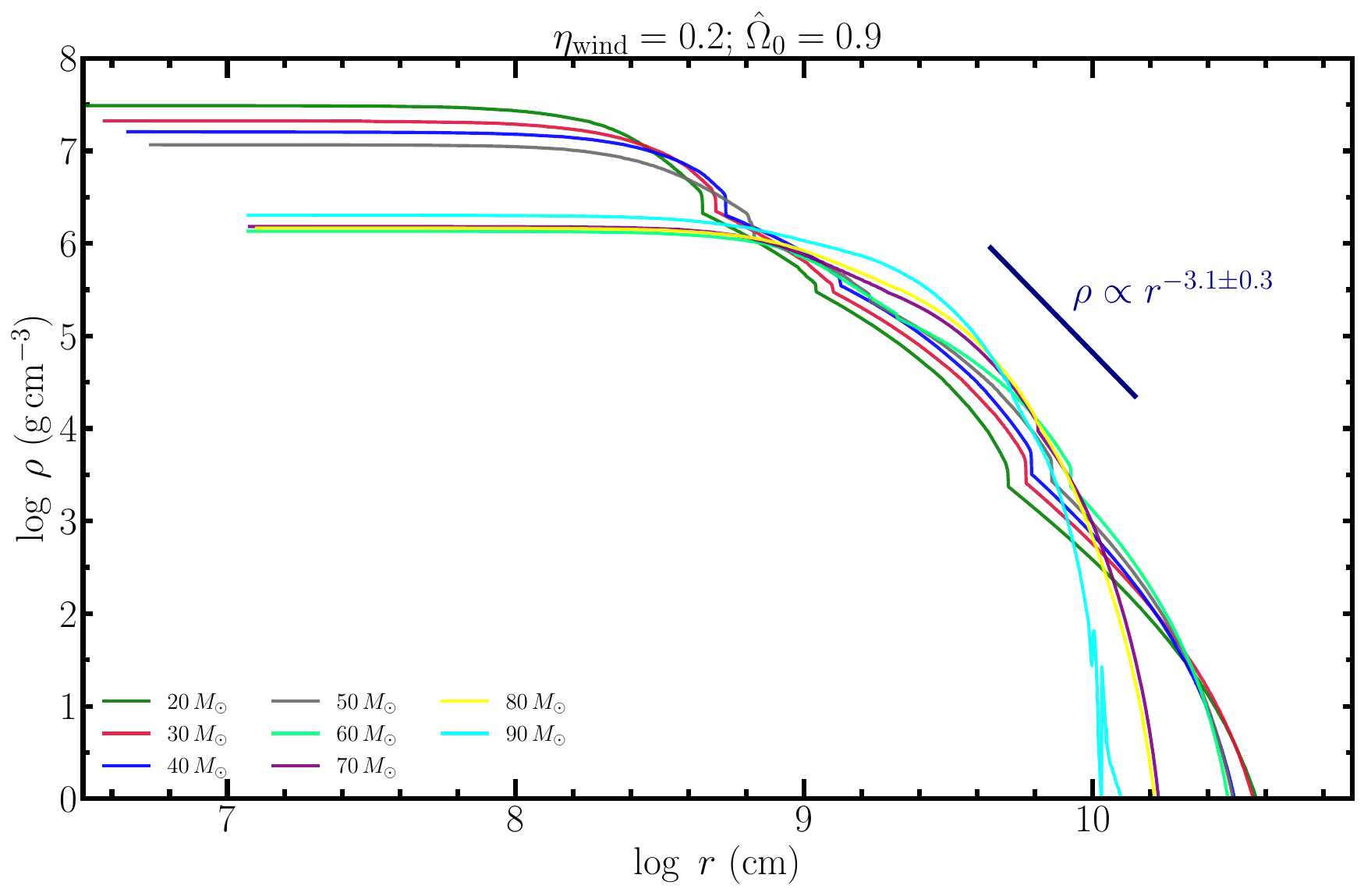}
    \includegraphics[width=0.48\textwidth]{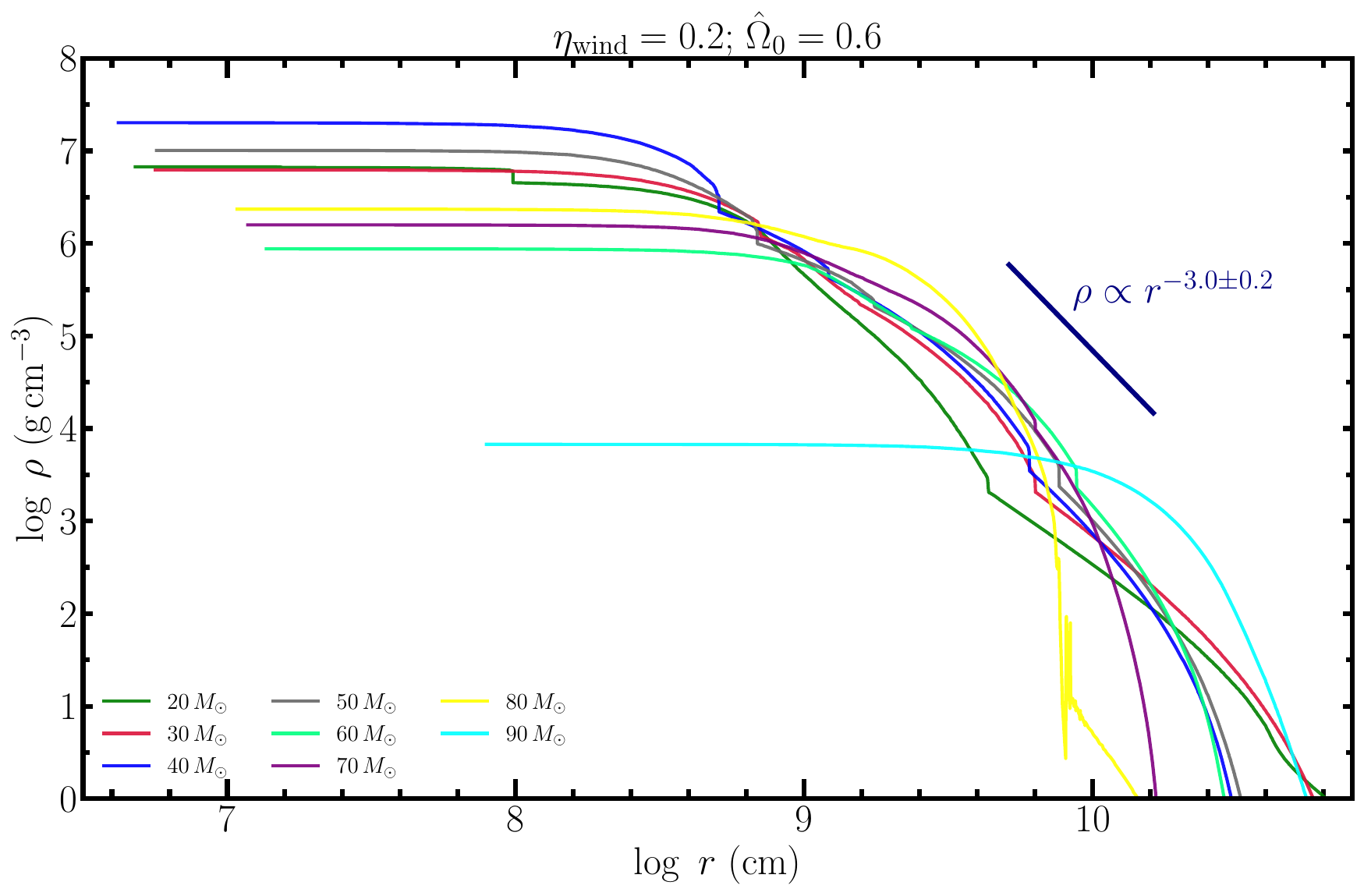}
    \caption{
        Pre-core-collapse density profiles as a function of radius for stellar models with initial masses ranging 
        from $20M_{\odot}$ to $100M_{\odot}$, shown for initial $\hat{\Omega}_0 = 0.9$ (\textit{left}) 
        and $\hat{\Omega}_0 = 0.6$ (\textit{right}) and wind scaling factor of $\eta_{\rm wind} = 1.0$ 
        (\textit{top}) and $\eta_{\rm wind} = 0.2$ (\textit{bottom}). 
    }
    \label{fig:Rho_R_grid}
\end{figure*}

\begin{figure*}
    \centering
    \includegraphics[width=0.48\textwidth]{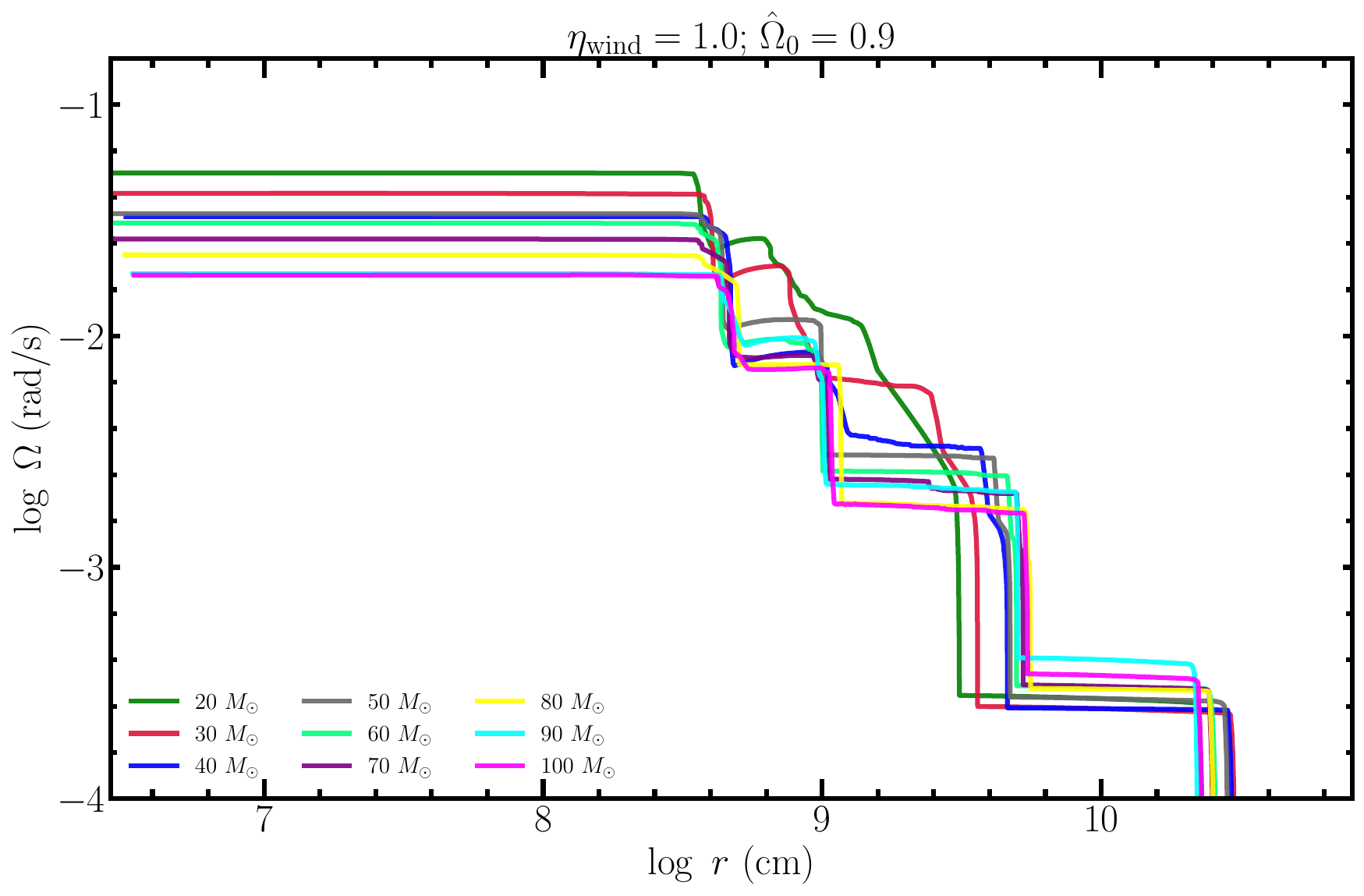}
    \includegraphics[width=0.48\textwidth]{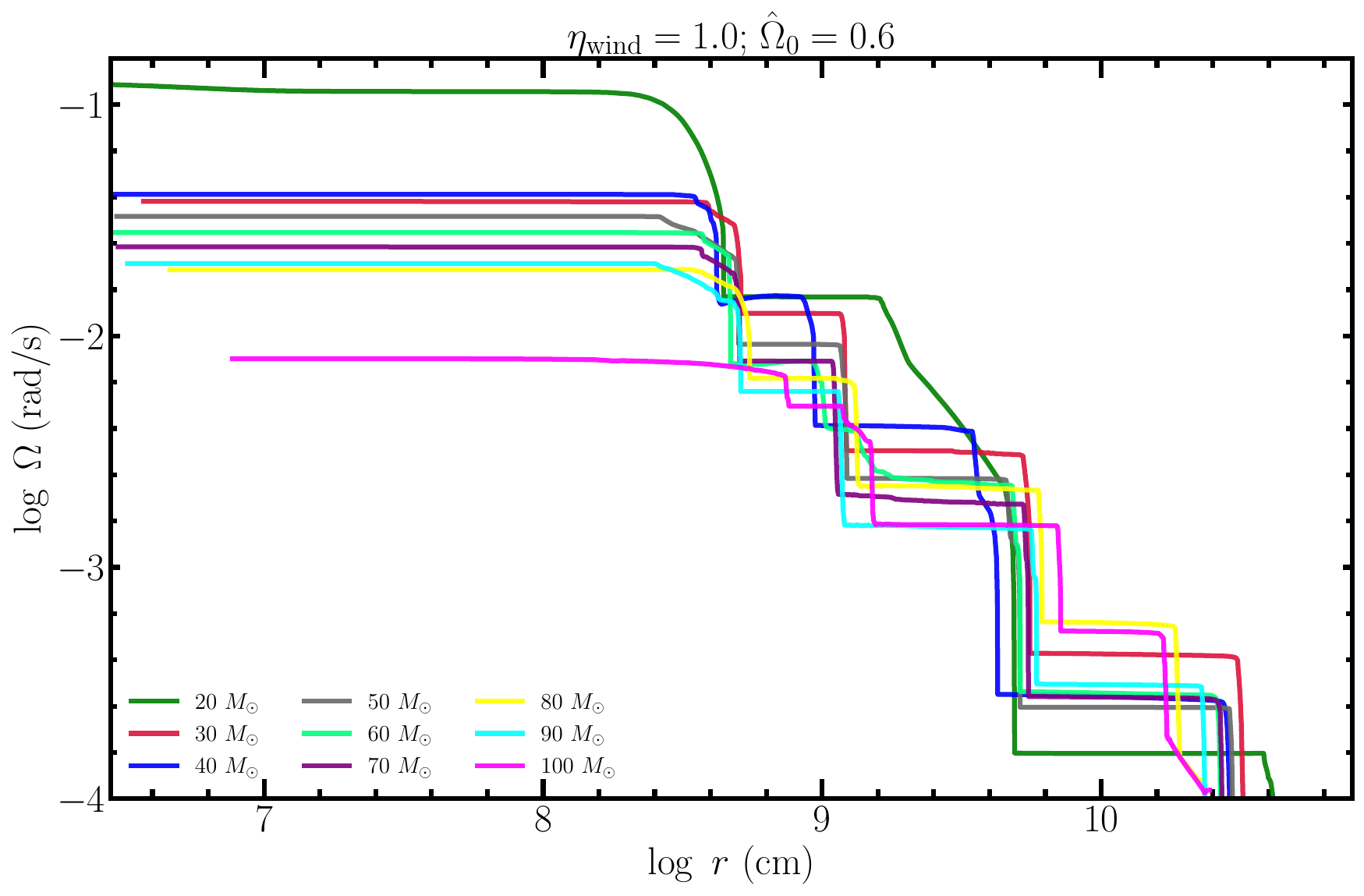}
    \includegraphics[width=0.48\textwidth]{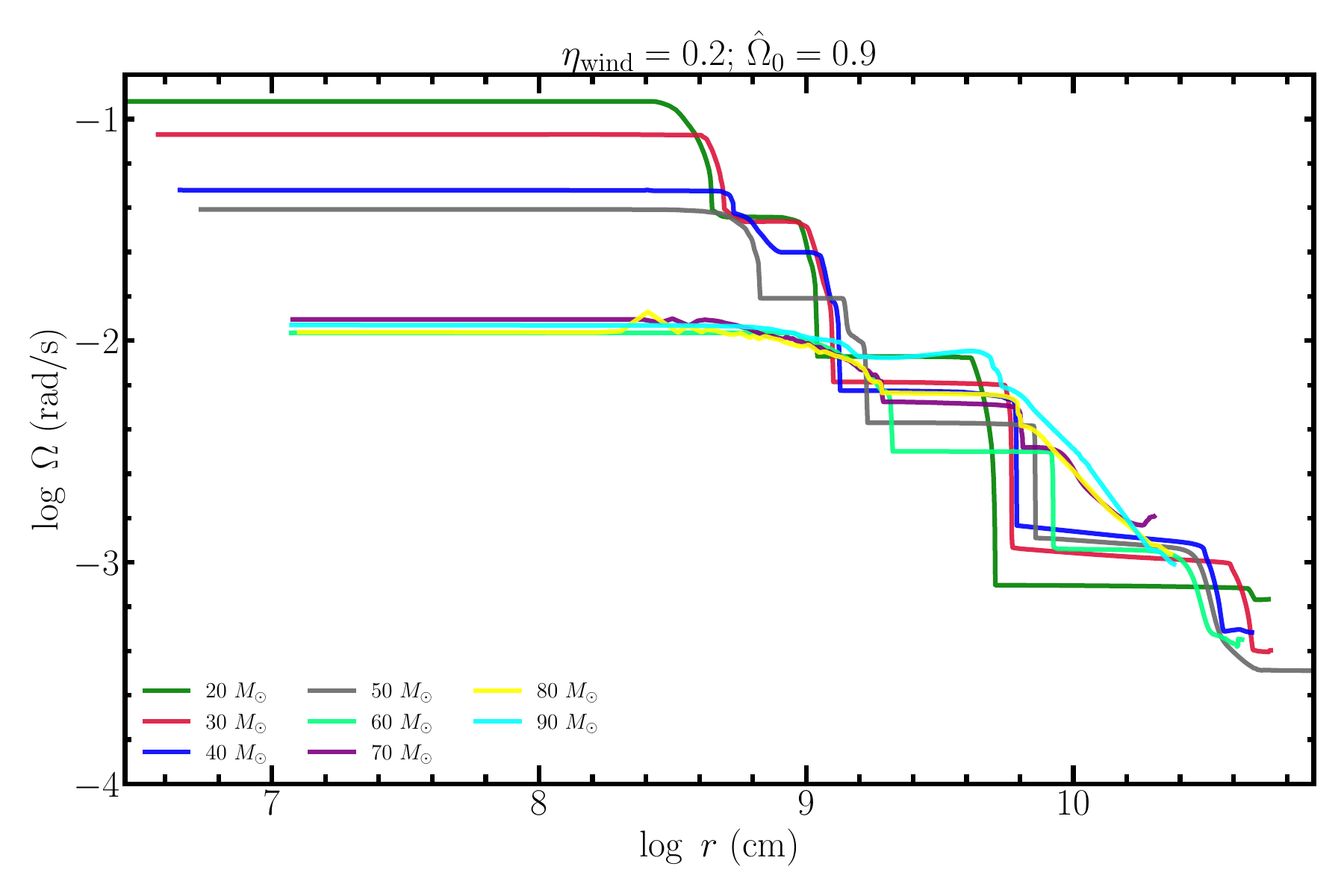}
    \includegraphics[width=0.48\textwidth]{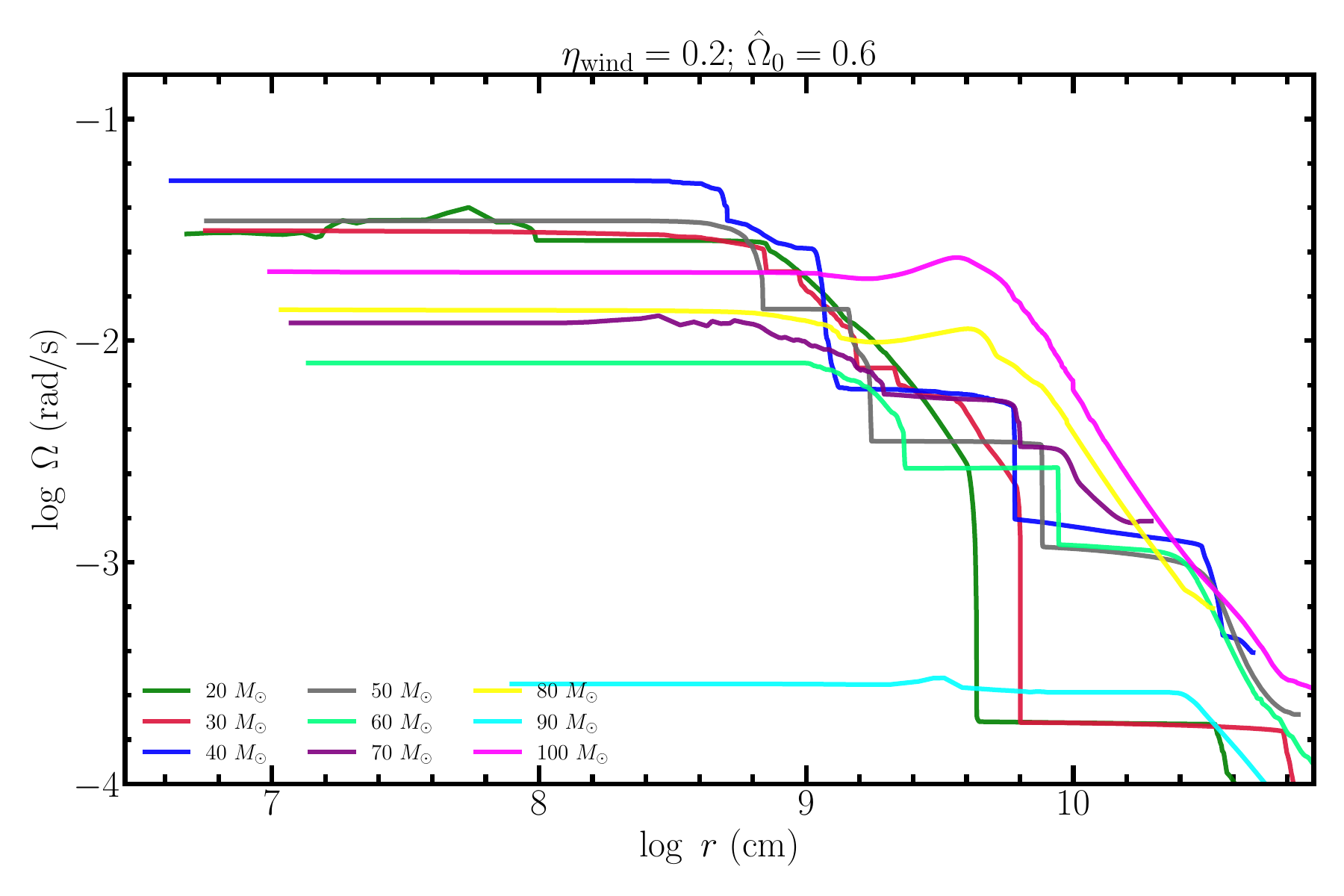}
    \caption{
        Pre-core-collapse angular velocity profiles as a function of radius for stellar models with initial 
        masses ranging from 20 to 100 $M_{\odot}$, shown for initial $\hat{\Omega}_0 = 0.9$ 
        (left) and $\hat{\Omega}_0 = 0.6$ (right) and with $\eta_{\rm wind} = 1.0$ (\textit{top}) 
        and $\eta_{\rm wind} = 0.2$ (\textit{bottom}). The staircase pattern arises due to the 
        shellular assumption in \texttt{MESA} that evolves the radial profile over several 
        isobaric shells of stellar material.
        }
    \label{fig:Omega_R_vertical}
\end{figure*}

\begin{figure*}
    \centering
    \includegraphics[width=0.48\textwidth]{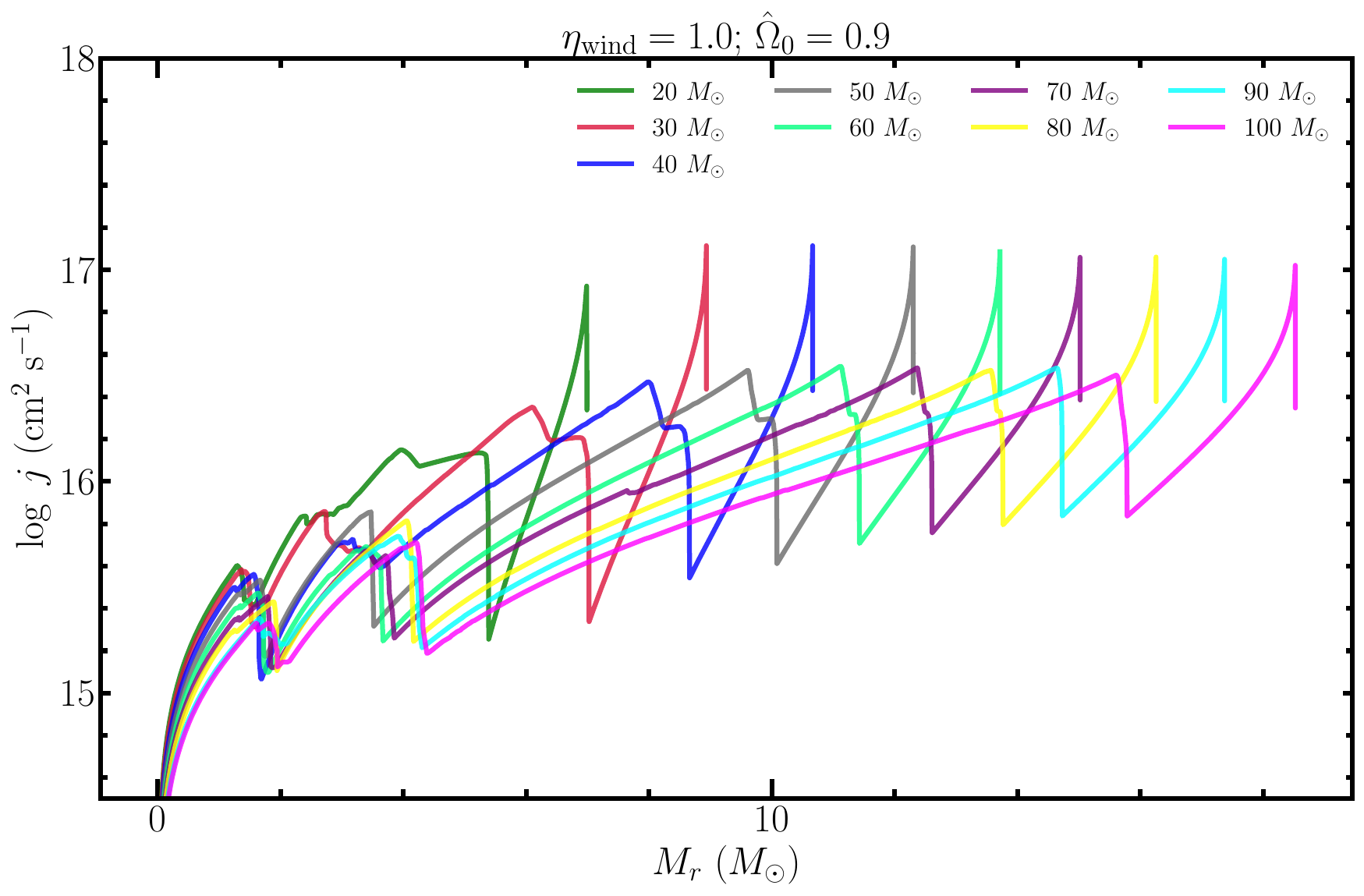}
    \includegraphics[width=0.48\textwidth]{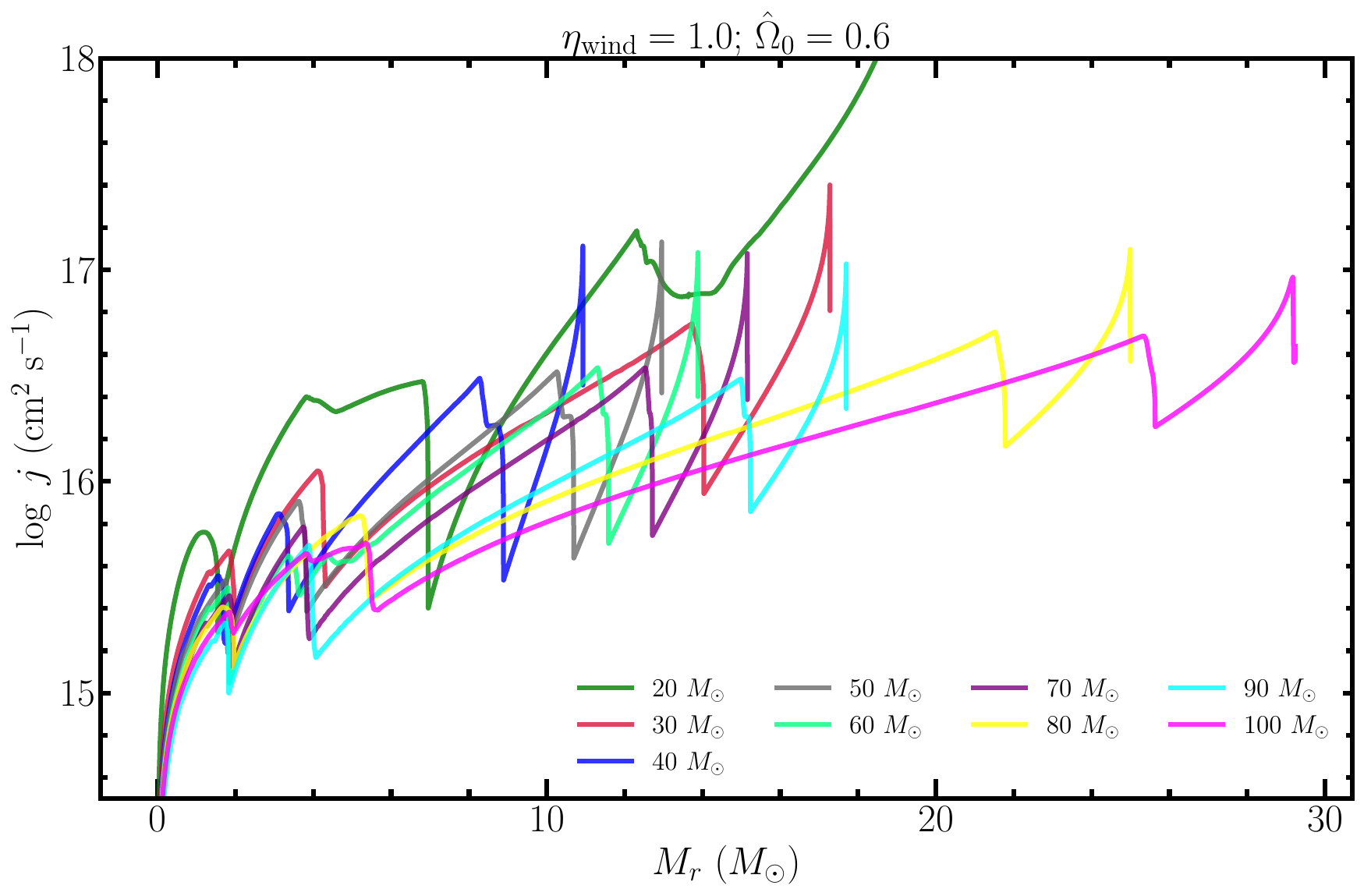}
    \includegraphics[width=0.48\textwidth]{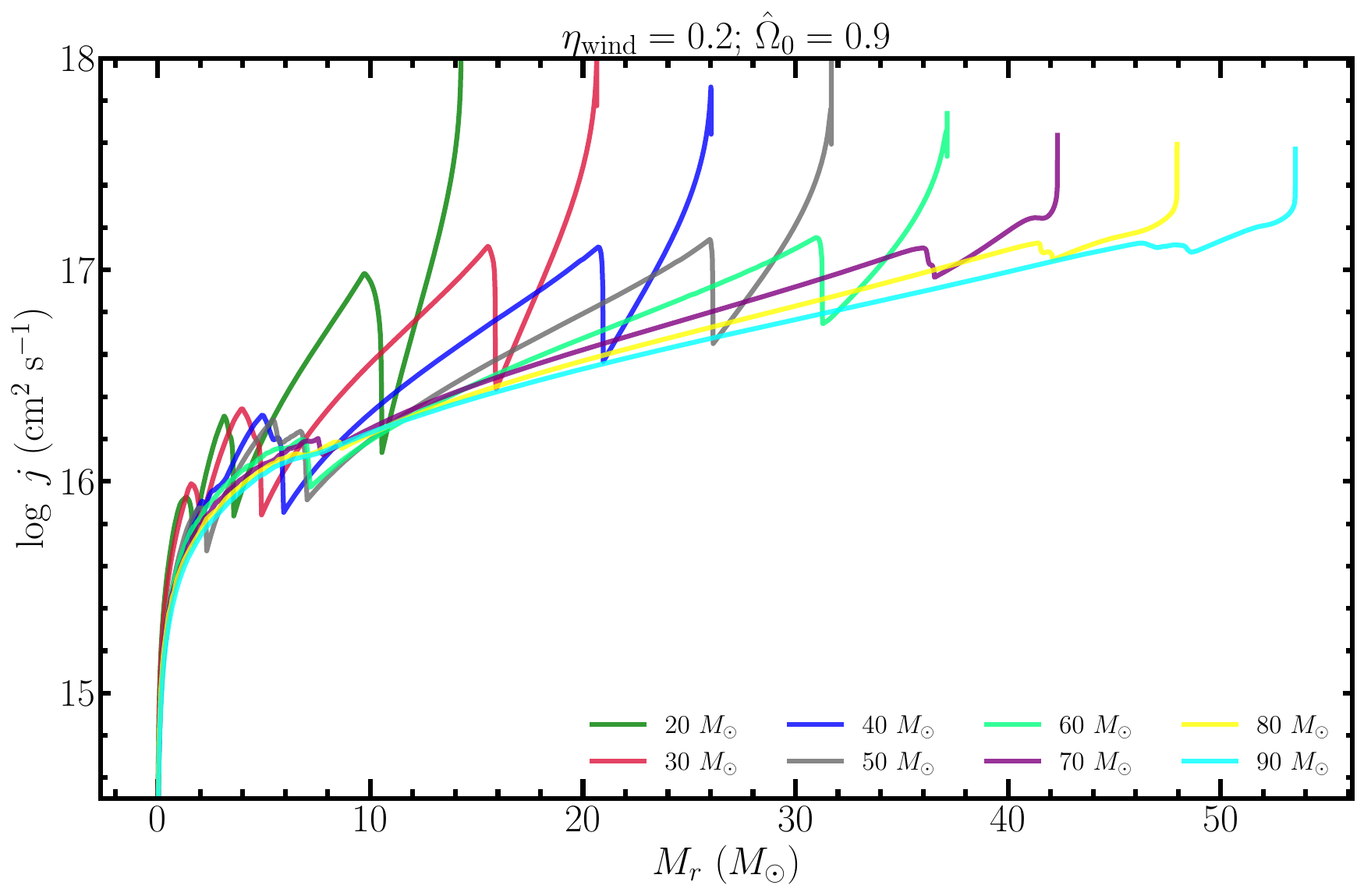}
    \includegraphics[width=0.48\textwidth]{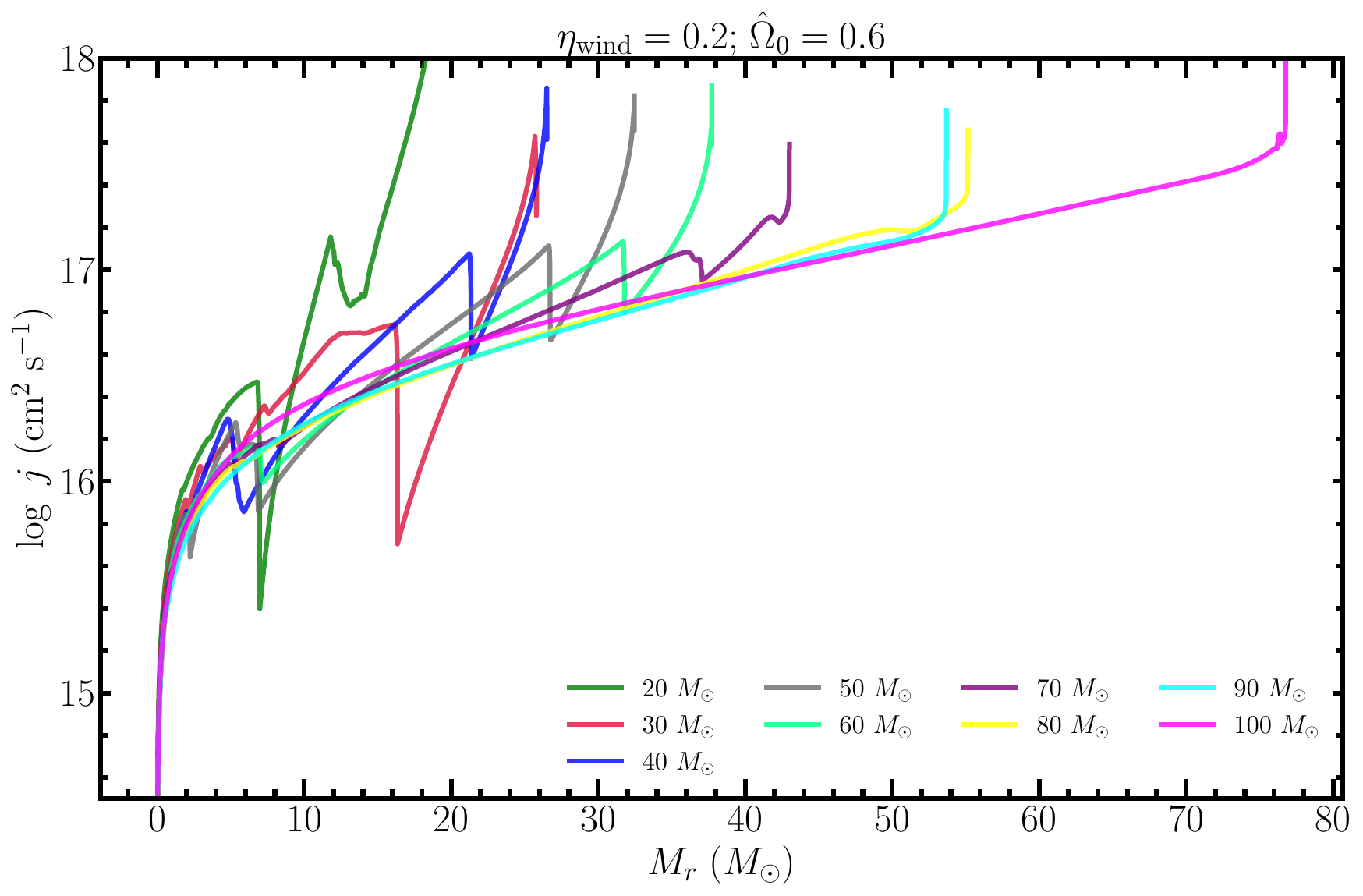}
    \caption{
        Pre-core-collapse specific angular momentum profiles as a function of enclosed mass 
        for stellar models with initial masses ranging from 20 to 100 $M_{\odot}$, with initial 
        rotation $\hat{\Omega}_0 = 0.9$ (left) and $\hat{\Omega}_0 = 0.6$ (right) and 
        $\eta_{\rm wind} = 1.0$ (\textit{top}) and $\eta_{\rm wind} = 0.2$ (\textit{bottom}). 
        The sawtooth profile corresponds to the staircase profile seen in Fig.\,\ref{fig:Omega_R_vertical} 
        that results from the shellular approximation in \texttt{MESA}.
    }
    \label{fig:J_Mr_vertical}
\end{figure*}

\subsection{Stellar Structure at Core-Collapse}
In this section, we examine the stellar structure of our Pop III star models at the 
onset of core collapse, first focusing on their density profiles, as shown in 
Figure \ref{fig:Rho_R_grid}. These profiles are extracted from the final snapshot of 
\texttt{MESA} simulations and they capture the state of the star immediately before collapse. The 
density distribution reveals two distinct regions: (a) the stellar core characterised 
by high and nearly constant density and (b) the outer layers that display a near power-law 
decline in density, with $d\ln\rho/d\ln r\sim -3.1$. Near the stellar surface the density 
declines very rapidly. We model the density profile as a piecewise function, 
\begin{equation}\label{eq:density}
\rho(r) = \begin{cases} 
\rho_{\rm{core}} & \text{if } r \leq R_{\rm{core}}, \\
\rho_{\rm core} \left( \frac{r}{R_{\rm{core}}} \right)^{-m} 
\left( 1 - \frac{r}{R_*} \right)^3 & \text{if } R_{\rm{core}} < r \leq R_*, \\
0 & \text{if } r > R_*,
\end{cases}
\end{equation}
where the core density varies over a large range for the different masses in our model, 
with $10^4\lesssim\rho_{\rm{core}}/(\rm{g\ cm}^{-3})\lesssim 10^8$. Likewise, the 
core radius varies in the range $1\lesssim R_{\rm core,8}\lesssim200$ and the stellar 
radius in the range $1\lesssim R_{*,10}\lesssim 6$. The density power-law index 
appears to show sensitivity to $\hat\Omega$, $\eta_{\rm wind}$, and $M_{\rm ZAMS}$, 
however, in many cases the power-law decline does not have enough dynamic range 
over which the asymptotic value of $m$ can be determined. When this is not the case, 
$m\sim3.1$.

We use these density profiles later to calculate the rate of accretion and jet propagation 
along the rotation axis of the star. In this way, the density of material in the path of the 
jet is over-estimated. Since the stellar material near the rotational poles of the star has very 
little angular momentum, it readily plunges towards the center upon losing pressure support, 
thus creating a lower density funnel. To what extent the density is diluted along the pole when 
the jet is launched can be understood with core-collapse numerical simulations that allow proper 
accounting for free-fall and hydrodynamic effects. This was done in \citet{Halevi+23}, where they 
found that pre-core-collapse density profiles, with $d\ln\rho/d\ln r \sim -2.5$, flatten during 
the collapse to $d\ln\rho/d\ln r\sim-1.5$ before BH formation. 

Figure \ref{fig:Omega_R_vertical} illustrates the radial variation of angular velocity 
in the star prior to core collapse, shown for $\hat\Omega_0=\{0.6,\,0.9\}$ and 
$\eta_{\rm wind}=\{0.2,\,1\}$. 
In most 1D stellar evolution codes, like \texttt{MESA}, rotational effects are included using the 
\textit{shellular approximation} \citep[e.g.][]{Meynet-Maeder-97} that makes the assumption 
of a constant $\Omega(r)$ over isobars (constant pressure) which is valid in the presence 
of strong anisotropic turbulence acting along isobars \citep{Paxton+13}. This is the 
reason behind the discontinuous behavior seen in the $\Omega(r)$ profile. 
It is evident that in the absence of strong winds ($\eta_{\rm wind}=0.2$), the models with 
higher rotation ($\hat\Omega_0 = 0.9$) show greater angular velocities. Some models at the 
high mass end, particularly with $M_{\rm ZAMS}>70M_\odot$, show a smooth $\Omega(r)$ profile 
when $\eta_{\rm wind}=0.2$. These stars suffer from the PPI, as discussed above, making 
it numerically challenging to evolve them to advanced nuclear burning stages close 
to core-collapse. Since the evolution was stopped at an earlier time in their evolutions, 
the angular velocity profiles do not yet show the shellular structure which is apparent in 
other models that were evolved to near core-collapse. 

Figure \ref{fig:J_Mr_vertical} shows the distribution of the specific angular momentum 
as a function of the enclosed mass ($M_r$) for our stellar models. Like the angular velocity 
profiles, the angular momentum profiles also show the discontinuous behavior due to the 
shellular approximation. The star is able to retain a larger angular momentum when the wind efficiency is lowered. 
For example, when comparing the $\eta_{\rm wind}=1.0$ with $\eta_{\rm wind}=0.2$ case for the 
$20\,M_\odot$ model with $\hat{\Omega}_0=0.9$, we find that the specific angular momentum is 
around a factor of 3 higher for a $3M_\odot$ core (see Table\,\ref{tab:eta_d0.2} and \ref{tab:eta_d1.0}) 
when winds are suppressed. Furthermore, the total angular momentum of the entire star is an order of magnitude larger 
when $\eta_{\rm wind}=0.2$ over that when $\eta_{\rm wind}=1.0$. Comparison of these profiles with 
previous results \citep{Yoon+12} reveals consistency in the observed trends. In particular, our models 
with reduced values of $\eta_{\rm wind}$ resemble the more favourable models for GRBs previously 
reported by \citet{Woosley-Heger-06}. For a thin accretion disc, the requisite specific angular momentum 
at the inner-most stable circular orbit (ISCO) is given by \citep{Bardeen+72}
\begin{equation}
    j_{\rm ISCO} = \frac{2GM_{\rm BH}}{3^{3/2}c}\left[1 + 2(3\hat R_{\rm ISCO} - 2)^{1/2}\right]\,,
\end{equation}
where $\hat R_{\rm ISCO} = R_{\rm ISCO}/R_g$ and $R_g = GM_{\rm BH}/c^2$ is the gravitational radius of 
a BH with mass $M_{\rm BH}$. To launch a successful jet, the basic requirement of $\hat R_{\rm ISCO}>1$ 
must be met. When considering, e.g., a BH with mass $M_{BH}=5M_\odot$, this condition is met for 
$j_{\rm ISCO}\gtrsim3\times10^{16}\,{\rm cm}^2\,{\rm s}^{-1}$ \citep[e.g.][]{MacFadyen-Woosley-99}. 
When taking this to be the fiducial scenario, we find that stars with $\hat{\Omega}_0=0.9$ and higher 
wind efficiencies will not have enough angular momentum to form an accretion disc. Only in the case 
with $\eta_{\rm wind}=0.2$ there is enough material with angular momentum larger than the minimum 
needed to have a successful accretion disc and even launch a relativistic jet.

For low $\eta_{\rm wind}$ values, such as $\eta_{\rm wind} = 0.2$, reaching core collapse in 
\texttt{MESA} becomes particularly challenging due to numerical instabilities, which leads to 
the difficulty of obtaining specific angular momentum profiles for high-mass, low-rotation models 
($\hat{\Omega}_0 = 0.6$). Consequently, profiles are presented only for the mass range of 20--70 
$M_\odot$ for this case.

\begin{figure*}
    \centering
    \includegraphics[width=0.95\textwidth]{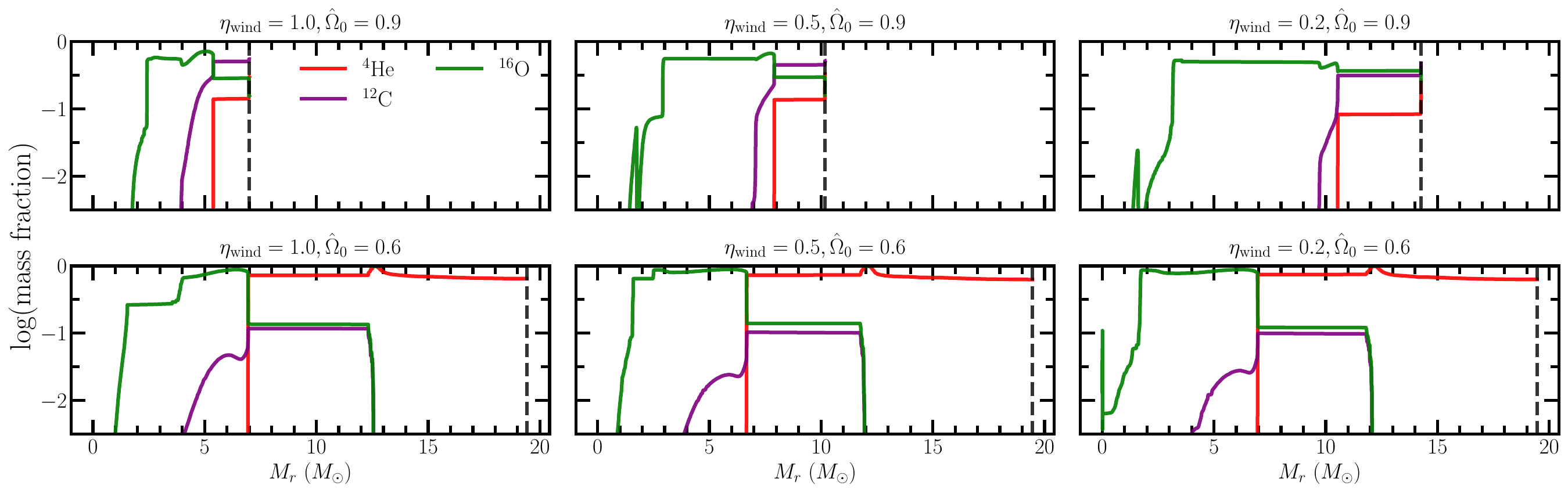}
    \caption{Mass fraction of  key elements ($^4$He, $^{12}$C, $^{16}$O) as a function of the mass 
    coordinate ($M_r$) for a model with an initial mass of $M_{\rm ZAMS} = 20\ M_\odot$, 
    shown for initial $\hat{\Omega}_0 = 0.6$ (\textit{top}) and $\hat{\Omega}_0 = 0.9$ (\textit{bottom}), 
    and $\eta_{\rm wind} = \{1.0$, $0.5$, $0.2$\} (left to right). The vertical dashed line in each subplot indicates the pre-core-collapse mass at 
    the final stage prior to core-collapse. The bottom row represents a special case, as the 
    evolution of this model at  $\hat{\Omega}_0=0.6$ is the only one lacking C and O in the envelope, 
    likely resulting in a type Ib SN, while all other models (showing fractions similar to 
    the top row of our grid) exhibit C and O in the envelope, leading to a type Ic SN.
    }
    \label{fig:fractions_20Msun}
\end{figure*}

\subsection{Remnant Properties and Associated Supernovae}
Here we analyze pre-core-collapse (pre-cc) stellar parameters such as the final 
pre-cc mass ($M_{\rm{pre-cc}}$), the specific angular momentum of the central 3\,$M_{\odot}$ 
core ($\log j_{3M_{\odot}}$), the compactness parameter $\xi_{2.5}$, and the carbon-oxygen 
core mass ($M_{\rm{CO}}$). These parameters are presented in detail in Tables 
(\ref{tab:eta_d0.2}--\ref{tab:eta_d1.0}). We use these data 
to infer the types of supernovae (SNe) and resulting remnants. The compactness parameter, 
valid particularly for non-rotating stars, is defined as \citep{OConnor+11}
\begin{equation}
    \label{eq:xi_2.5}
    \xi_{2.5} \equiv \dfrac{M_{2.5}/M_\odot}{R(M_{2.5})/10^3\ {\rm km}}\,,
\end{equation}
where $M_{2.5}\equiv 2.5\ M_\odot$ and $R(M_{2.5})$ is the radial coordinate that includes 
that mass, indicates whether the core is compact enough to collapse to a BH. 
When $\xi_{2.5}\gtrsim 0.45$, it becomes very difficult to explode the star, as in a 
supernova, and BH formation via collapse is the most likely option. Here we use the 
pre-collapse properties of the progenitor 
to calculate the compactness, but its actual value should be strictly determined at the 
time of core bounce. The latter can only be obtained using an advanced numerical code, which 
is outside the scope of this work. Consequently, the values that we obtain are approximative. 
Our simulation grid finds $\xi_{2.5} \sim 0.3$, a value typically associated with neutron star 
formation in non-rotating models for some equations of states. When factoring in the uncertainty in 
reliably calculating the compactness parameter from stellar evolution codes alone, we make the 
assumption in later sections that all of our models do not explode and collapse directly to a 
BH remnant.

The chemical composition of the stellar envelope at the pre-core-collapse stage, as illustrated in 
Figure \ref{fig:fractions_20Msun}, is crucial for classifying the engine-powered supernova types. 
These figures present mass fraction profiles of key elements ($^4\rm He$, $^{12}\rm C$, $^{16}\rm O$) for 
initial rotation rates of $\hat{\Omega}_0 = 0.6$ and $0.9$, with different wind efficiencies. 
The complete absence of hydrogen in all models, resulting from rotation-induced mass loss, 
prevents the formation of Type II supernovae (SN II), which require a significant hydrogen envelope 
\citep{Heger+03}. Initially, the surface metal abundance decreases gradually due to dilution from rotational mixing 
that transports hydrogen from the surface inward, but once critical rotation is reached 
it increases abruptly as the hydrogen-rich envelope is ejected, exposing the elements resulting 
from nuclear reactions in the core. Consequently, the outer envelopes become dominated by helium, 
while the cores exhibit substantial fractions of $^{12}\rm C$ and $^{16}\rm O$, influencing the final 
remnant and favouring Type Ib or Ic supernovae over SN II. In fact, several SNe associated 
with observed population of long-duration GRBs, that are produced by more metal-rich progenitors, 
are of type Ic \citep{Woosley+06b, Cano+17}. This provides additional motivation to relate GRBs with 
rapidly rotating massive stars that may have undergone CHE.

The SN outcome is sensitive to initial mass and rotation. For example, the model with $M_{\rm ZAMS}=20\,M_\odot$ 
and $\hat{\Omega}_0=0.6$ shows no $^{12}$C or $^{16}$O on the surface, as can be 
seen in the bottom row of Figure \ref{fig:fractions_20Msun}. Instead, the surface is dominated by $^4$He. Unlike all 
other models undergoing CHE, this star evolves toward the red giant phase in the HR diagram (Figure \ref{fig:HR_vertical}), 
indicating non-CHE. The combination of low rotation and minimal mass loss, driven by the low 
$\eta_{\rm wind}$, favours the formation of a Type Ib supernova, characterised by a helium-rich 
envelope devoid of hydrogen. In contrast, all other models with higher rotation ($ \hat{\Omega}_0 = 0.9$) 
or initial masses ($M_{\rm ZAMS} \geq 25 \, M_{\odot}$) develop massive CO cores with significant surface fractions of $^{12}\rm C$ and $^{16}\rm O$, leading to a direct core-collapse.

\section{GRB Production Efficiency}

\subsection{Core Collapse and Accretion}\label{sec:core-collapse-accretion}

In contrast to the type II core-collapse that delivers a successful supernova explosion in massive 
and non-rotating stars, rapidly rotating such stars may form an accretion disc after the iron-core 
collapses into a black hole \citep{MacFadyen-Woosley-99}. In the collapsar model of long-GRBs, 
infalling stellar material with specific angular momentum in the range $3 \lesssim j/(10^{16}\,{\rm cm}^2\,{\rm s}^{-1})\lesssim 20$ 
will form a compact disc at the optimal radius away from the Kerr BH to power relativistic outflows 
that produce GRBs. To calculate the rate of accretion in this scenario, we use the model of \citet{Kumar+08} 
that considers the infall of stellar material from an axisymmetric, rotating star onto an accretion 
disc after the star has lost pressure support. In our formulation below, many quantities that depend 
on the polar angle $\theta$ have been averaged since \texttt{MESA}, being a 1D code, only provides stellar profiles 
as a function of radius.

After a temporal delay given by the sound crossing time, which can be approximated using the free-fall time, 
$t_s(r)\sim t_{\rm ff}(r) = \sqrt{3\pi/32G\bar\rho(r)}$, where $\bar\rho(r) = 3M_r/4\pi r^3$ is the 
mean density of material within radius $r$, the material at a spherical coordinate $(r,\theta,\varphi)$ 
starts to free-fall. Here $M_r$ is the mass enclosed within radius $r$, 
\begin{equation}
M_r = \int_0^r 4\pi r'^2\,\rho(r')\,dr'. \label{eq:Mr}
\end{equation}
and $\rho(r)$ is the local density. Given its angular velocity $\Omega(r,\theta)$ the material follows 
an elliptical trajectory, with eccentricity $e = 1 - \Omega^2(r,\theta)\sin^2\theta/\Omega_k^2$, 
that intersects the equatorial plane after duration
\begin{equation}\label{eq:teq}
t_{\rm eq}(r,\theta)= \frac{1}{\Omega_k}\left(\arccos(-e)+e\sqrt{1-e^2}\right)(1+e)^{-3/2}+t_s(r)\,, 
\end{equation}
and then circularises at the fallback radius in the equatorial plane at 
\begin{equation} 
    R_{\rm fb} \approx r\,\frac{\Omega^2(r)}{\Omega_k^2(r)}\,.
    \label{eq:isco_condition} 
\end{equation}
This radius is approximated using the polar-angle averaged angular velocity of the infalling material from 
a given radius $r$ and where $\Omega_k = \sqrt{G\,M_r/r^3}$ is the Keplerian angular velocity of that material. 
Mass falls at the equatorial plane at the rate
\begin{equation}
    \dot M_{\rm fb}\equiv\frac{dM_{\rm fb}(r)}{dt} = \frac{dM(r)}{dr}\left[\frac{d\langle t_{\rm eq}(r)\rangle}{dr}\right]^{-1}\,,
\end{equation}
where $dM/dr = 4\pi\rho(r)r^2$ and 
\begin{equation}
    \langle t_{\rm eq}(r)\rangle = \frac{1}{2}\int t_{\rm eq}(r,\theta)\sin\theta\,d\theta
\end{equation}
is the angle averaged equatorial fallback time. It is possible that regions close to the rotational 
axis of the star might experience an outflow instead, in which case the above estimate needs to be 
appropriately modified considering the solid angle of the material that falls back instead of the $4\pi$ 
factor. Here we do not consider such possibility due to its inherent uncertainty and 
instead consider material from all polar angles to contribute either to the BH mass or the accretion disc. 
{In what follows, we also do not consider general relativistic effects near the newly formed BH and our treatment remains non-relativistic \citep[see, however,][for a formalism including a dynamical spacetime]{Ghodla-Eldridge-24}.}

\begin{figure}
\centering
\includegraphics[width=\columnwidth]{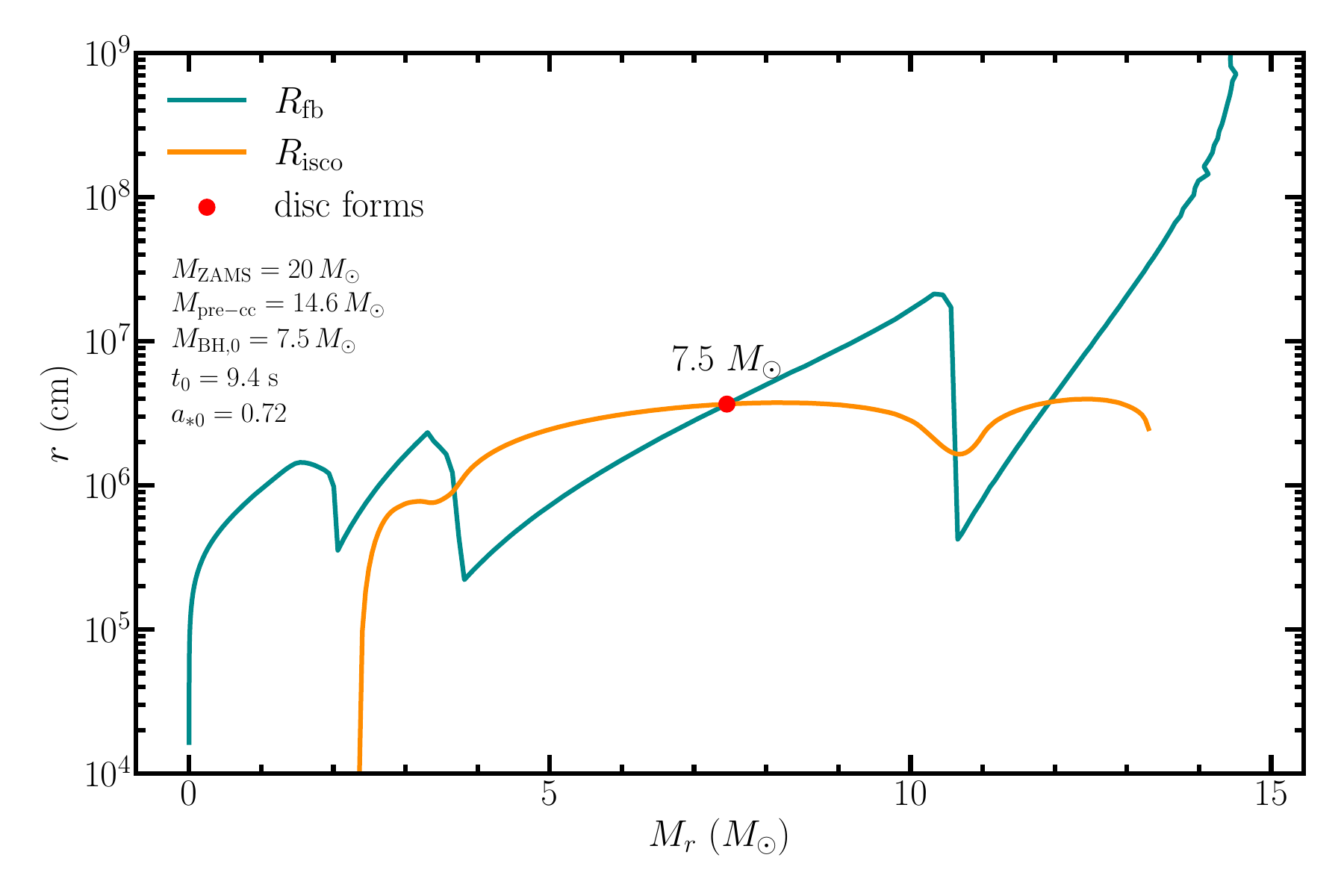}
\caption{Intersection of the fallback radius profile, $R_{\rm fb}$, with $R_{\rm ISCO}$, shown for 
a $M_{\rm ZAMS}=20M_\odot$ star post core-collapse as a function of the mass coordinate $M_r$. 
This model assumes a wind scaling factor of $\eta_{\rm wind}=0.2$ and an initial rotation rate of $\hat{\Omega}_0 = 0.9$. 
All mass below the intersection point (red dot) falls directly into the BH, while that above it goes into 
an accretion disc. The mass and spin of the BH at the critical time ($t_0=9.4$ s) when the disc forms are 
$M_{\rm BH,0}=7.5M_\odot$ and $a_{*0}=0.72$.
}
\label{fig:Rfb_Risco_20M_W0.9_D0.2}
\end{figure}

\begin{figure}
    \centering
    \includegraphics[width=0.95\columnwidth]{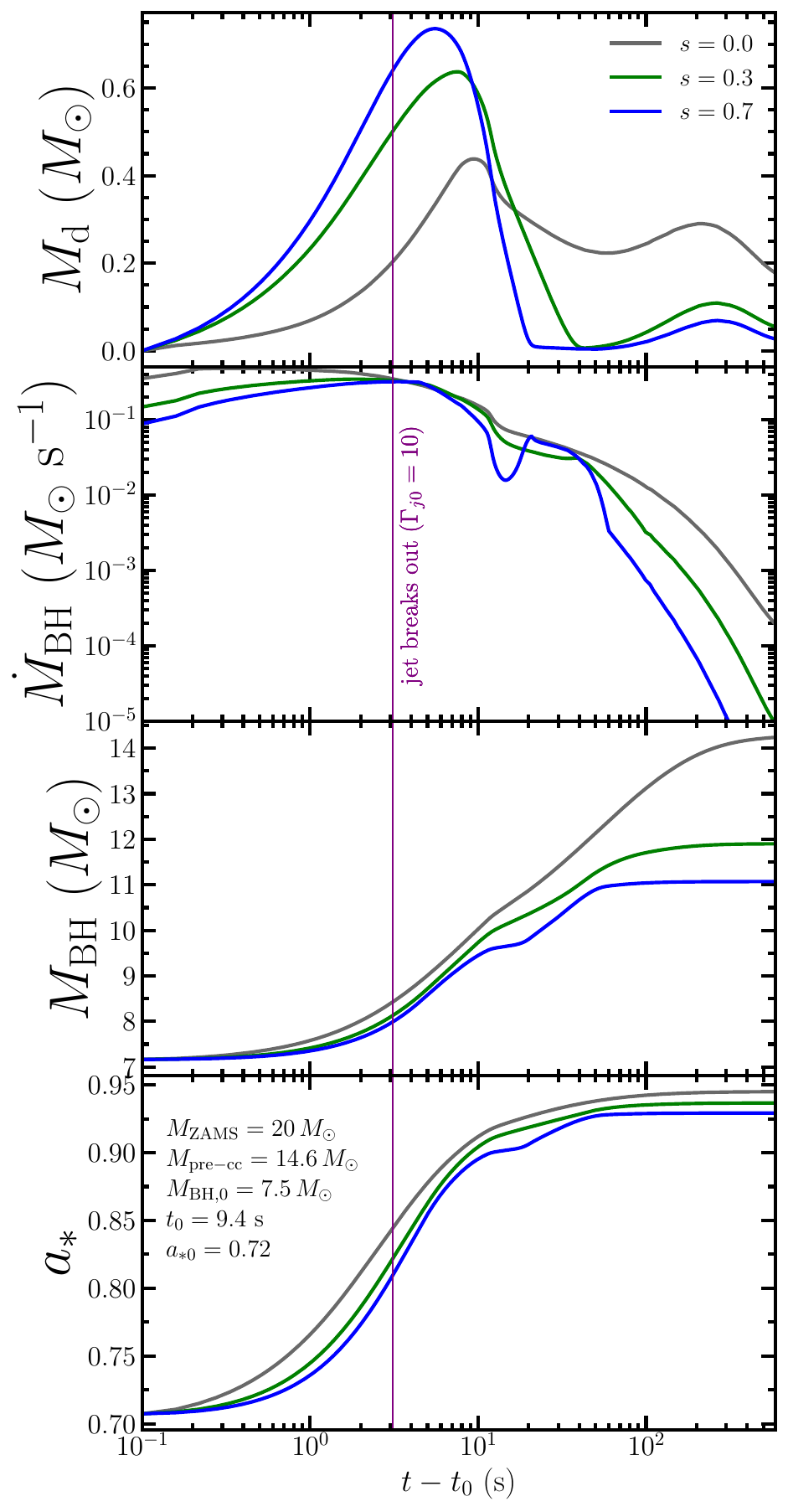}
    \caption{
    Temporal evolution of key quantities describing the accretion disc and central black 
    hole properties in the 20R9W2 model. From top to bottom, the panels show the temporal 
    evolution of i) the disc mass, ii) the accretion rate $\dot{M}_{\rm BH}$, iii) the 
    black hole mass, and iv) dimensionless spin parameter $a_*$ for different values of the 
    parameter $s = \{0.0,\,0.3,\,0.7\}$. The vertical dashed line 
    represent the moment when the jet breaks out of the surface of the progenitor star considering 
    an initial Lorentz factor of $\Gamma_{j0}=10$ with jet opening angle $\theta_{0}=0.1$. 
    See Fig.\,\ref{fig:Rfb_Risco_20M_W0.9_D0.2} for further details. 
    }
    \label{fig:MJA}
\end{figure}

\begin{figure*}
    \centering
    \includegraphics[width=0.45\textwidth]{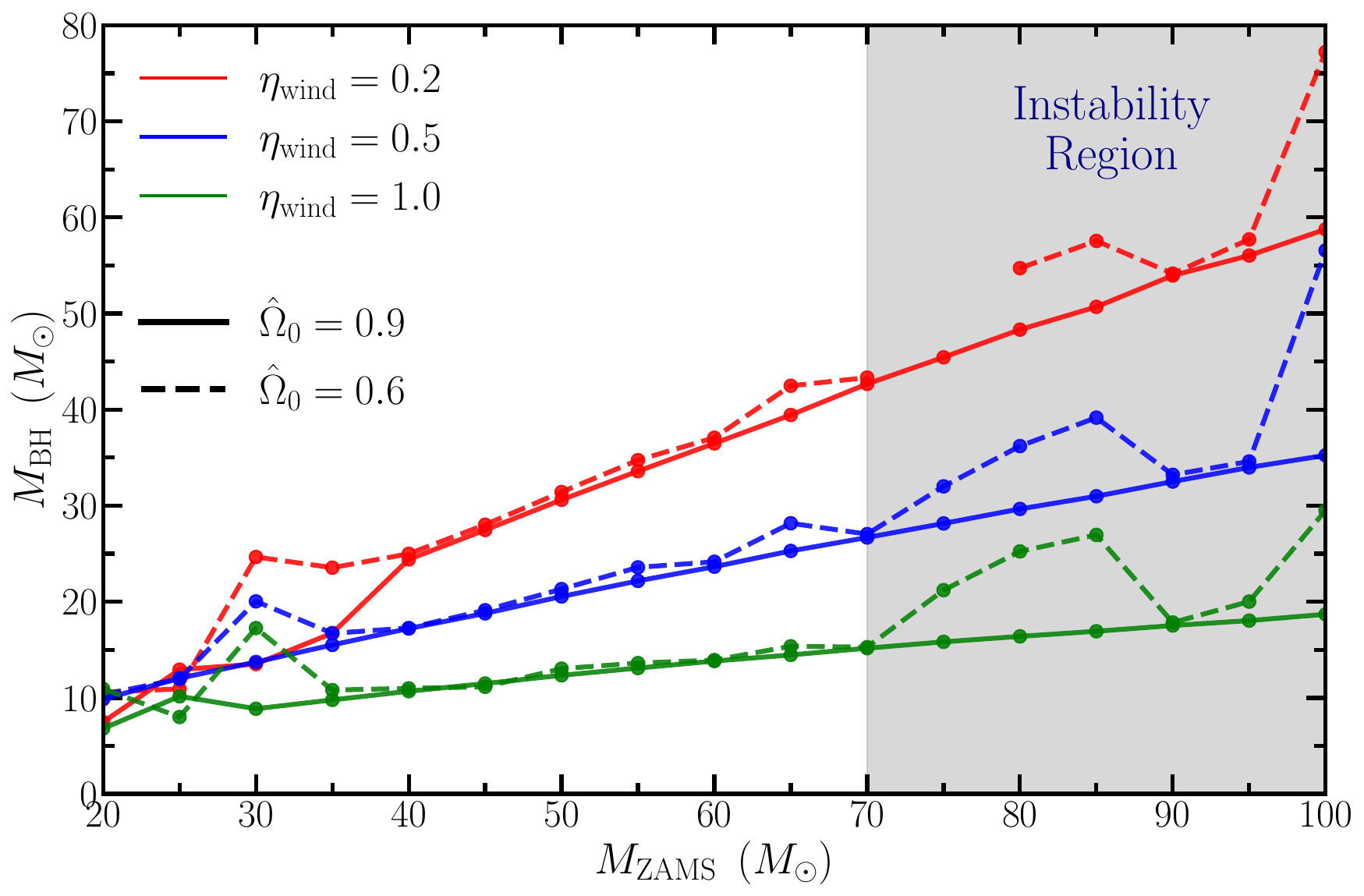}\quad\quad
    \includegraphics[width=0.45\textwidth]{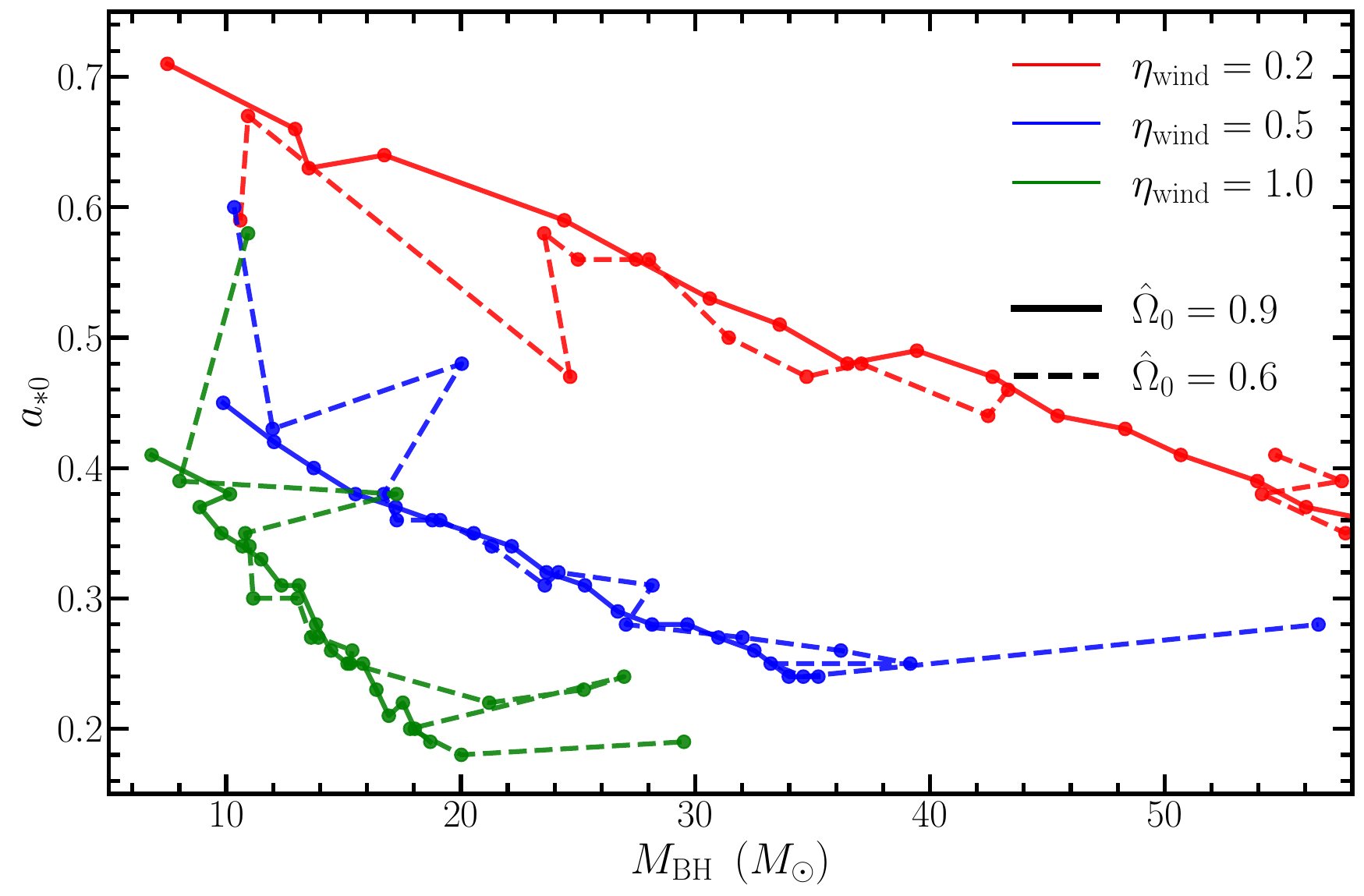}
    \caption{
    (Left) Evolution of the initial black hole mass ($M_{\rm BH,0}$) as a function 
    of the initial stellar mass ($M_{\rm ZAMS}$). Results 
    are shown for different values of $\eta_{\rm wind}$. Solid lines correspond to 
    models with an initial rotation rate of $\hat{\Omega}_0 = 0.9$, while dashed lines 
    represent those with $\hat{\Omega}_0 = 0.6$. 
    (Right) Evolution of the dimensionless spin parameter ($a_{*0}$ when disc forms) 
    as a function of $M_{\rm ZAMS}$. 
    }
    \label{fig:MBH0_vs_MZAMS}
\end{figure*}

Initially, material in the central regions falls freely onto the proto-neutron-star and turns it into a BH {when the mass of the compact remnant grows beyond the threshold mass of $M_{\rm th}\simeq1.2\,M_{\rm TOV}\simeq2.4\,M_\odot$ \citep[e.g.][]{Breu-Rezzolla-16}, for a maximum mass of a non-rotating NS $M_{\rm TOV}\simeq2\,M_\odot$ \citep[e.g.][]{Margalit-Metzger17,Shibata+17,Rezzolla+18}}. Subsequently, more mass with specific angular momentum less than that required 
to circularise continues to fall, and the mass and spin of the BH continues to grow, with
\begin{equation}
M_{\rm BH} = \int_0^{t_0} \dot{M}_{\rm fb}(t_{\rm eq})\,dt_{\rm eq}\,, \label{eq:M_BH}
\quad\quad
a_* = \frac{cJ_r}{GM_r^2}
\end{equation}
where $J_r$ is the angular momentum of the mass $M_r$ that fell into the BH until some 
critical time $t\leq t_0$. At this critical time, a disc begins to form of material that 
has enough specific angular momentum to launch it into orbit around the BH at radii beyond 
$R_{\rm ISCO}$, the inner-most stable circular orbit \citep{Bardeen+72},
\begin{equation} 
    R_{\rm ISCO}(M_{\rm BH},a_*) = \frac{GM_{\rm BH}}{c^2}\left\{ 3 + z_2 - \sqrt{(3-z_1)(3+z_1+2z_2)} \right\}, 
    \label{eq:isco_radius} 
\end{equation} 
with 
\begin{eqnarray} 
    z_1 &=& 1 + \left(1-a_*^2\right)^{1/3}\left[(1+a_*)^{1/3}+(1-a_*)^{1/3}\right], \\
    z_2 &=& \sqrt{3a_*^2+z_1^2}\,.
    \label{eq:z1-z2} 
\end{eqnarray} 
The critical time to disc formation is determined from the condition, 
$R_{\rm ISCO}(M_r,a_*)=R_{\rm fb}(r)$, when the fallback radius in the 
equatorial plane becomes comparable to $R_{\rm ISCO}$. Material with $R_{\rm fb}<R_{\rm ISCO}$ 
falls directly into the nascent BH and that with $R_{\rm fb}>R_{\rm ISCO}$ has sufficient angular 
momentum to circularise and form an accretion disc. 
We first obtain the critical radius $R_{\rm fb,crit}$ from the stellar profiles and then calculate 
the corresponding $t_0$ as well as the BH mass and spin at $t_0$. This is demonstrated 
in Fig.\,\ref{fig:Rfb_Risco_20M_W0.9_D0.2}, using a $20\,M_\odot$ model (referred to as 20R9W2 henceforth) 
with $\hat{\Omega}_0=0.9$ and $\eta_{\rm wind}=0.2$, that illustrates the intersection between the 
$R_{\rm fb}$ and the $R_{\rm ISCO}$ as a function of the mass coordinate $M_r$. 
The red point denotes the critical intersection at $M_r = M_{\rm BH,0} = 7.5\ M_\odot$, with spin 
$a_{*0} = 0.72$ at the critical time $t_0 = 9.4$ s after core-collapse. 
For the innermost part of the stellar core, the $R_{\rm ISCO}$ curve rises at a later mass coordinate 
as it is not well defined since $a_*$ exceeds unity there. As a result, we only show the range in mass coordinate 
where $R_{\rm ISCO}$ is well-defined, according to Equation \ref{eq:isco_radius}. 
The critical solution depends on the rotational profile of the star at core-collapse and 
it is not obtained in all models, in which we find $R_{\rm fb} < R_{\rm ISCO}$ for the majority 
of the stellar material lacking of centrifugal support and leaving very little for accretion. 
Such models are not capable of launching relativistic jets and do not produce GRBs. {We have verified the results of our model using the pre-collapse stellar profiles for the \texttt{16TI} model from \citet{Woosley-Heger-06}, for which we find $M_{\rm BH,0}\simeq3\,M_\odot$ and $a_{*0}\simeq0.7$. These are consistent with results from advanced numerical simulations of core-collapse for the same stellar model \citep[e.g.][]{Just+22,Coleman-Fernandez-24}.}

We consider the formation of a thick accretion disc that accretes matter onto the BH over the viscous timescale \citep{Narayan+01}. Accretion in such a disc may proceed in three 
different regimes, namely (i) NDAF (neutrino-dominated accretion flow; \citet{Popham+99}), 
(ii) ADAF (accretion-dominated accretion flow; \citet{Narayan-Yi-94}), and (iii) a mix of both 
(i) and (ii). The mass accretion rates, $\dot M_{\rm BH}$, onto the black in all three regimes 
are given in Eqs. 22 - 24 of \citet{Kumar+08}; the interested reader should look there for more details. 
When accretion occurs via an ADAF, a significant amount of disc mass is lost due to outflows, in 
which case the accretion rate at any given radius, $\dot M(r)$, is suppressed as a power-law, 
\begin{equation}
    \dot M(r) = \dot M_{\rm acc}(R_d)\left(\frac{r}{R_d}\right)^s\quad\quad{\rm for}\quad\quad 0\leq s\leq 1\,.
\end{equation} 
The exact value of the power-law index $s$ is unclear and numerical simulations in different 
works find different values \citep{McKinney+12,Narayan+12}. 
We keep it as a free parameter in what follows. Here $\dot M_{\rm acc} = M_d/t_{\rm acc}$ is the 
mass accretion rate out of the disc of mass $M_d$. The accretion proceeds over the standard 
characteristic viscous timescale $t_{\rm acc}\sim2/\alpha\Omega_k$, where $\alpha\sim10^{-2}-10^{-1}$ 
is the dimensionless viscosity parameter. The mass of the disc changes due to fallback and accretion, 
with $\dot M_d = \dot M_{\rm fb} - \dot M_{\rm acc}$ and some fraction of the disc mass is then 
accreted onto the BH.

The mass and spin of the BH grows over time as mass accretes from the disc. To calculate the temporal 
evolution of disc mass, and the mass and spin of the BH, we numerically solve the coupled Eqs.\,25 - 32 
from \citet{Kumar+08}. {One caveat is that this formalism does not include the effect of outflows at higher latitudes (i.e. near the rotational pole) as well as accretion of low-angular momentum stellar material from the same polar regions. The latter may influence the spin evolution of the BH, and that effect is not included here due to its complexity which requires numerical simulations.}

Figure \ref{fig:MJA} shows the temporal evolution of accretion and the properties of the BH 
for the 20R9W2 model. From top to bottom, the panels show (i) the temporal evolution of the disc mass ($M_{\rm d}$), 
that initially grows, reaches a peak, and then declines. The disc mass is controlled 
by the rate of mass fallback to the equatorial plane and that of accretion, where the 
former is governed by the angular momentum profile at core-collapse. The disc mass initial 
grows due to the infalling material having sufficient angular momentum to join the accretion 
disc, until it reaches the point (shown by the sharp drop in the green curve in 
Fig.\,\ref{fig:Rfb_Risco_20M_W0.9_D0.2}) where some material from larger radii falls directly 
into the BH, having insufficient angular momentum to join the disc. This manifests as a 
sudden drop in the disc mass. The disc mass starts to grow at later times as material from the 
outer stellar layers continues to join the disc. (ii) The temporal evolution of the accretion rate, 
which for the first ten seconds remains above $0.1\,M_\odot\,{\rm s}^{-1}$. 
(iii) and (iv) The temporal evolution of the BH mass and spin, 
where both are sensitive to the accretion rate radial profile power-law index $s$. The pre-core-collapse 
mass of the star in the 20R9W2 model is $14.6\,M_\odot$ and not all of it makes it into the 
BH as some is lost to disc-driven winds and outflows during the accretion process. The BH spin increases from 
its initial value $a_{*0}=0.72$ to $a_*>0.9$ as more angular momentum is brought in by the 
accreting material. When the BH launches any outflows, some of the angular momentum will be 
lost to those, and therefore, its spin will be regulated by gain and loss of angular momentum 
\citep{Lowell+24,Jacquemin-Ide+24,Wu+25}. How much angular momentum is lost to outflows depends 
on the jet launching process, which is unclear. 

Figure \ref{fig:MBH0_vs_MZAMS} (left panel) shows the initial black hole mass ($M_{\rm BH,0}$) 
that forms promptly upon core-collapse as a function of the ZAMS mass for our grid. The results 
are presented considering different stellar wind efficiencies ($\eta_{\rm wind}=0.2, 0.5, 1.0$) 
and two initial rotation rates, $\hat{\Omega}_0=0.9$ (solid lines) and $\hat{\Omega}_0=0.6$ (dotted lines). 
The right panel shows the initial spin parameter $a_{*,0}$ as a function of intial BH mass, 
it decreases with increasing BH mass and correlates well with decreased wind efficiency that retains 
larger angular momentum in the star at core-collapse. Both figures highlight a region of instability 
for $M_{\rm ZAMS} > 70,M_\odot$, previously discussed in Figure \ref{fig:Tc_Rhoc_vertical} where significant 
effects due to pulsational pair-instability are observed.

\subsection{Jet Power}\label{sec:jet-power}
The composition of GRB jets, i.e. whether they are kinetic-energy-dominated or Poynting-flux-dominated, remains 
unclear, and so is the launching mechanism. Therefore, the main problem is how to convert, and with what efficiency, 
the power brought in by the accreting matter, which is $\dot M_{\rm BH}c^2$. Two different jet launching mechanisms 
have been proposed thus far: neutrino annihilation \citep{MacFadyen-Woosley-99} and magnetohydrodynamic (MHD) processes 
in the vicinity of the rapidly rotating BH, e.g. the \citet{Blandford-Znajek-77} (BZ) process. Several works have 
explored the efficiency of both mechanisms \citep{Popham+99,Liu+15,Lei+17}, but it is still unclear which one dominates 
in GRBs. 

In the neutrino-annihilation framework, the jet is powered by the energy deposition from neutrino and anti-neutrino 
annihilation above the accretion disc. Following \citet{Leng-Giannios-14}, the jet power for a black hole with 
spin $a_{*}=0.95$ is approximated as 
\begin{equation}
    L_{\nu \bar\nu} \approx 1.3 \times 10^{52} \left( \frac{M_{\rm BH}}{3 M_\odot} \right)^{-3/2}
    \times \left( \frac{\dot{M}_{\rm BH}}{M_\odot\, {\rm s}^{-1}} \right)^{9/4} 
    \,{\rm erg~s^{-1}},
\label{eq:Lnunu}
\end{equation}
with a more detailed discussion given in \citet{Zalamea-Beloborodov-11}. As shown in \citet{Leng-Giannios-14}, 
the jet power from neutrino annihilation is typically limited to a maximum of $\sim5\times10^{51}\,{\rm erg\,s}^{-1}$ 
for an engine duration of $\sim10$\,s, and it starts to fall for longer lasting bursts. This is problematic 
for ultra-long bursts, with duration of $\sim10^3$\,s, where this scenario is unable to account for the 
radiated energy.

Alternatively, in the collapsar scenario, what is clear is that the jet is ultimately powered by accretion, 
and therefore its power can be parameterized using an efficiency parameter that is regulated by the spin of 
the BH, such that 
\begin{equation}
L_j = \eta_j(a_*) \dot{M}_{\rm BH} c^2~~~{\rm and}~~~ 
\eta_j(a_*) \approx 0.07\left[\frac{a_*}{1+\sqrt{1-a_*^2}}\right]^5\,,
\label{eq:Lj}
\end{equation}
where $\eta_j(a_*)$ is obtained from GRMHD simulations \citep{McKinney-05}. In this scenario, the spin of the BH 
is only allowed to increase, from an initial value of $0<a_{*0}<1$ to the maximal value of $a_*=1$, due to addition of 
angular momentum brought by the accreting matter.

However, torques due to large scale magnetic fields threading the accreting gas and the event horizon of the 
rapidly spinning BH, as posited in the BZ process, remove angular momentum from 
the BH and power a Poynting-flux-dominated outflow. As demonstrated by \citet{Wu+25} 
(also see \citealt{Lowell+24,Jacquemin-Ide+24}), 
the spin of the BH may thus be regulated to attain an equilibrium value of $a_{*\rm eq}\sim0.5$ when the accumulated 
magnetic flux at the event horizon ($\Phi_{\rm BH}$) in the BZ scenario is $\sim0.4\Phi_{\rm MAD}$, where 
$\Phi_{\rm MAD}$ is the magnetic flux when the accretion transitions to the \textit{magnetically arrested disk} 
(MAD) state \citep{Narayan+03,Tchekhovskoy+11}. In the MAD state, the accumulated magnetic flux is strong 
enough to disrupt the accretion of gas when the magnetic pressure exceeds the ram pressure of the accreting 
gas. At the equilibrium spin value, the accretion efficiency is maximized in powering the jet, 
which reaches $L_{j,\rm eq}/\dot M_{\rm BH}c^2\approx0.022$ \citep[][see their Fig.\,2]{Wu+25}.

\begin{figure}
    \centering
    \begin{subfigure}[b]{0.48\textwidth}
        \centering
        \includegraphics[width=\textwidth]{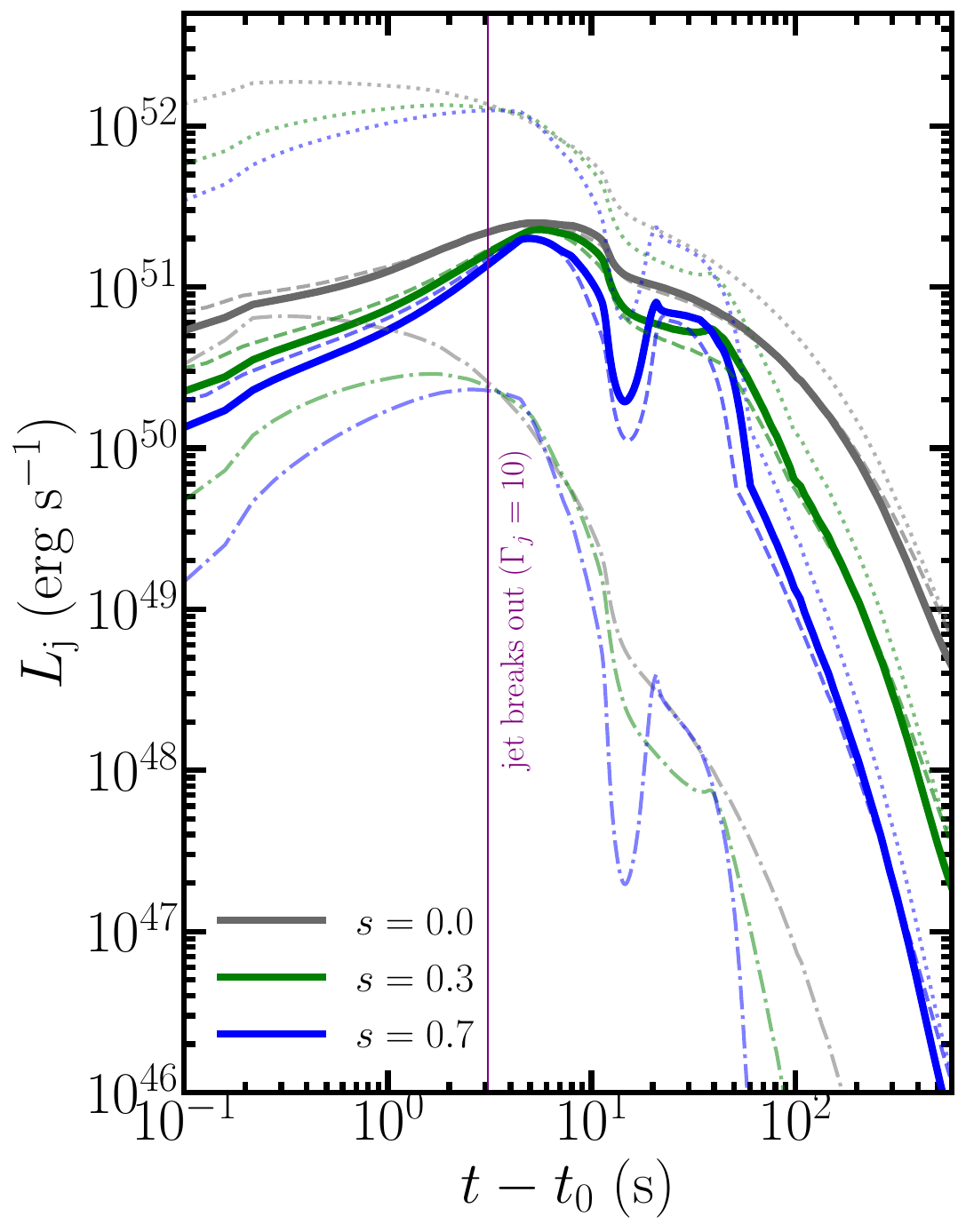}
    \end{subfigure}
    \caption{
        Temporal evolution of the true jet power $L_{\rm j}$ for our model, shown for 
        $s =\{0.0,\ 0.3,\ 0.7\}$. The linestyles correspond to the different ways to calculate the jet power:
         \textit{i) Solid lines:} Standard jet power obtained with Eq.\,(\ref{eq:Lj}).
         \textit{ii) Dash-dotted lines:} Considering the energy deposition rate due to neutrino annihilation (Eq.\,\ref{eq:Lnunu}).
       \textit{iii) Dotted lines:} Maximum jet power assuming that the system reaches equilibrium between the thin-disc accretion and MAD state.
        \textit{iv) Dashed lines:} Jet power for a new model 20R6W2 with higher metallicity ($M_{\rm ZAMS}=20\,M_\odot$, $\hat\Omega_0=0.6$, $\eta_{\rm wind}=0.2$, $Z=0.01\,Z_\odot$) and calculated with Eq.\,(\ref{eq:Lj}).
    The accretion disc forms at $t=t_0$ and it is assumed that a hydrodynamic jet is launched shortly thereafter at the injection radius of $z_{\rm inj} = 10\,r_g \approx 10^7$ cm. The initial bulk Lorentz factor is $\Gamma_{j0} = 10$ with jet opening angle $\theta_{0}=0.1$.
       For $s = 0.0$, the total injected energies after breakout for each case are:
(i) $E_{\rm j} \approx 1.00\times10^{53}$ erg,
(ii) $E_{\rm j} \approx8.32\times10^{50}$ erg,
(iii) $E_{\rm j} \approx2.33\times10^{53}$ erg,
(iv) $E_{\rm j} \approx9.98\times10^{52}$ erg.
        }
    \label{fig:Jet-Power}
\end{figure}

Figure \ref{fig:Jet-Power} shows the temporal evolution of the jet power obtained from the three different 
scenarios, as discussed above, for the 20R9W2 model and for different values of the parameter $s$. All three 
scenarios show different efficiencies, with the least efficient being the neutrino annihilation and the most 
efficient is in which the BH spin is regulated by the BZ process. 
The neutrino-annihilation-powered jets are up to an order of magnitude less luminous than those obtained from 
Eq.\,(\ref{eq:Lj}), especially at low accretion rates where neutrinos may not efficiently 
cool the disc. 

The different values of the $s$ parameter, that controls 
how much mass is lost to outflows when accretion occurs in the ADAF regime, has some affect on the 
jet power due to its effect on the accretion rate. What is particularly interesting is the dip in the 
jet power seen most prominently for $s=0.7$. This can lead to the ejection of two different mass shells 
that can then power distinct prompt emission episodes followed by afterglow emission with the potential 
for a re-brigtening due to refreshed shocks (however see, e.g., \citealt{Metzger+18} for an alternative 
scenario with fallback accretion onto a millisecond magnetar). The main parameters in the two-shell collision problem 
are the ratio of (kinetic) energies and that of the bulk Lorentz factors of the two shells. The former can be obtained 
from the jet power evolution, but the latter depends on the baryon loading of each shell, which is not 
so clear. This type of re-brightening has been seen in a small fraction of afterglows (see, e.g., Fig.\,9 of 
\citealt{Busmann+2025}) in long GRBs. Typically, both parameters are not known a priori and are instead constrained from fits to 
re-brightening episodes in the afterglow. Knowledge of the temporal evolution of jet power, and therefore 
the energy in the two shells reduces the unknown parameters by one.

We also compare the jet power (dashed lines), calculated using Eq.\,(\ref{eq:Lj}), for a metal rich ($Z=0.01\,Z_\odot$) 
progenitor of the same mass but with slower initial rotation ($\hat\Omega_0=0.6$) in Fig.\,\ref{fig:Jet-Power}. 
This model is similar to a Pop II progenitor and its comparison with the Pop III model shows that stars 
in our model grid may ultimately produce GRBs with properties, e.g. energetics and durations, similar to that 
expected from Pop II stars. This adds to the difficultly of distinguishing between Pop III and more metal rich progenitors 
of high redshift GRBs.

In what follows, we use Eq.\,(\ref{eq:Lj}) to calculate the jet power, which is intermediate between the 
neutrino-annihilation and BZ powered jets.

\subsection{Criteria for a Successful GRB: Jet Launching \& Breakout}\label{sec:criteria_GRB}
Given the expansive grid of simulations performed in this work, a variety of outcomes are possible, 
including successful and choked jets. In the former case, the jets are able to penetrate out of the 
star before the jet power declines significantly, and in the latter, the jet either does not have 
sufficient power or it is only powered for a duration shorter than needed for a successful breakout, causing 
it to choke inside a star. Below we quantify the different requirements to select cases that will 
yield a successful jet and a GRB. We find that a rapidly rotating stellar core, a sufficiently 
massive accretion disc, and a powerful, long-lived jet are all necessary conditions to successfully 
propagate through the progenitor star's envelope and produce a GRB without being choked. 

\subsubsection{Threshold Jet Power and Breakout Conditions}
The propagation of the jet inside the star depends mainly on the engine activity time, $t_{\rm eng}$, 
and the jet power, which must exceed a threshold value, $L_j^{\rm thr}$. This threshold is governed 
by the condition that the velocity of the jet head must exceed the local sound speed at the distance 
from the central engine where the jet is injected \citep{Aloy+18}. The local sound speed depends on 
the ambient density $\rho_a$ and pressure $p_a$, such that $c_{\rm s,a} = \sqrt{\hat\gamma p_a / \rho_a}$, 
where $\hat\gamma$ is the adiabatic index. The velocity of the jet head is affected by the jet's 
interaction with the surrounding medium, that produces a forward shock which shock heats the ambient medium 
and a reverse shock which shock heats the relativistic jet and slows it down to at most mildly relativistic 
speeds. By balancing the momentum flux density in the frame of the jet head, its velocity can be derived 
to yield $v_h \approx c / (1 + \tilde{L}^{-1/2})$ \citep{Matzner-03, Bromberg+11}, where $\tilde{L} = L_j / \Sigma_j \rho_a c^3$ 
gives the ratio of the jet energy density to that of the ambient medium. Here $\Sigma_j = z_h^2\Delta\Omega_j$ 
is the cross-sectional area of the jet head at a distance $z_h$ away from the central engine along the 
jet axis, and $\Delta\Omega_j$ is the solid angle subtended by it. Since we are interested in the conditions 
near the injection distance $z_h\approx z_{\rm inj}$, $\Delta\Omega_j=\pi\theta_0^2$ for a conical jet with 
initial jet half-opening angle of $\theta_0\ll1$. By enforcing the injection condition, 
$v_h > c_{\rm s,a}$, where the jet must move supersonically to avoid being choked, the threshold jet power 
can be obtained from \citep{Aloy+18}
\begin{equation}
    \label{eq:L_j;thr}
 L_j^{\rm thr} \approx 1.6 \times 10^{49} z_{\rm inj,9}^2 \theta_{0,-1}^2 p_{a,22}\,{\rm erg\,s}^{-1}\,.
 \end{equation}
 For the 20R9W2 model, the threshold jet power is estimated as $L_j^{\rm thr} \approx 10^{48}{\rm erg\,s}^{-1}$, 
 where we have assumed an opening angle $\theta_0 = 0.1$ rad, an injection radius of 
 $z_{\rm inj}=10 r_g \approx  10^7$ cm, with $r_g \approx  10^6$ cm for a 7.5 $M_\odot$ BH, and an 
 ambient pressure of $p_a(z_{\rm inj}) \simeq 6.5 \times 10^{24}$ dyne cm$^{-2}$ from our model.

\begin{figure}
    \centering
    \includegraphics[width=\columnwidth]{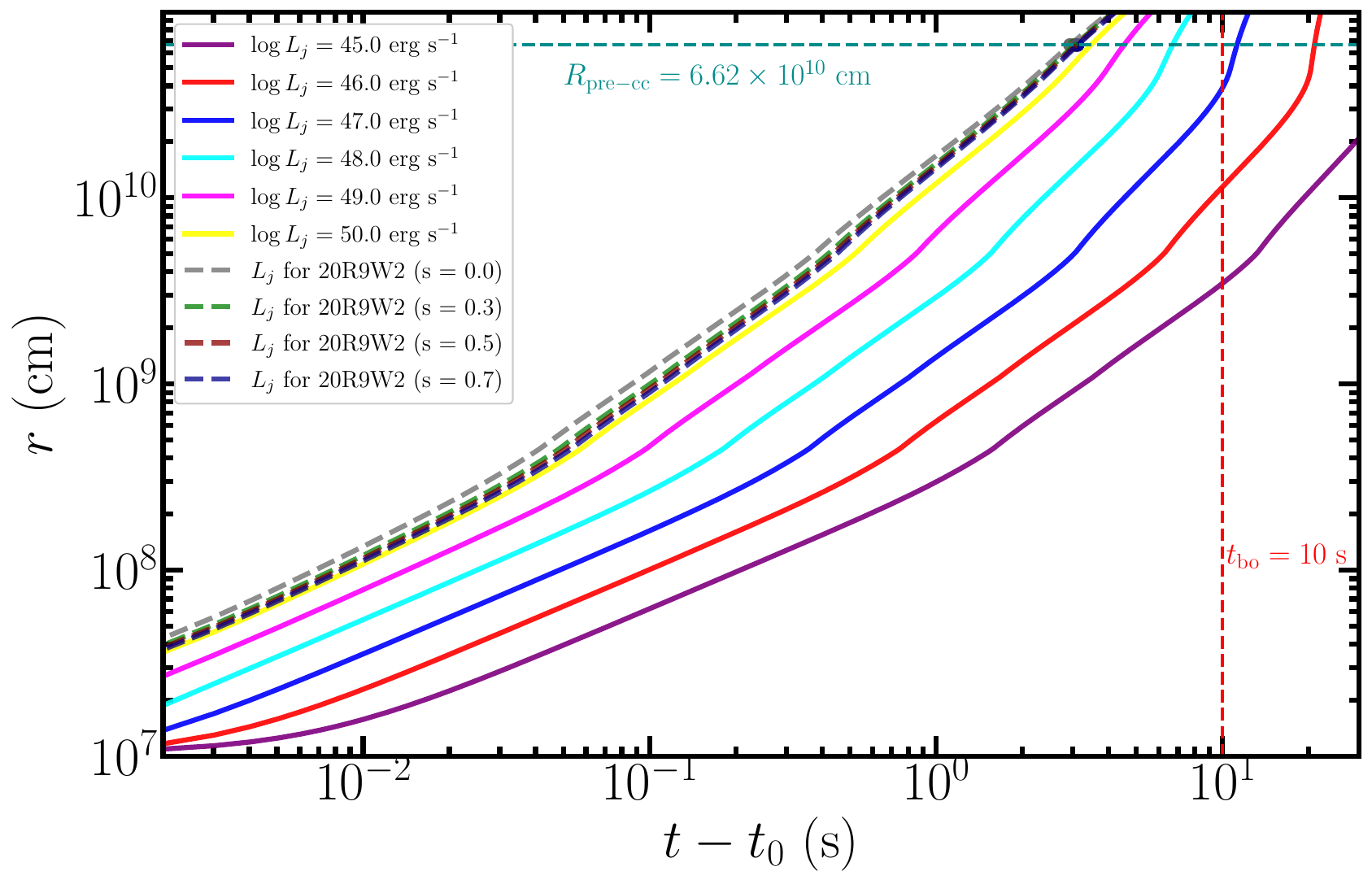}
    \caption{
        Temporal evolution of the jet head position $z_h$ for the 20R9W2 model. 
        The solid lines display the jet head trajectories for a range of constant $L_j$, 
        spanning from $\log L_j = 45$ to $50$ (in erg s$^{-1}$), computed within the 
        framework of \citet{Bromberg+11}. The stellar surface is shown by the horizontal 
        line at $R_{\rm pre-cc} = 6.62 \times 10^{10}$ cm,  and the typical breakout 
        time benchmark of $t_{\rm bo} \lesssim 10$ s, denoted by the vertical line. 
        The dashed lines represent the breakout times for the $L_j$ curves shown in Figure 
        \ref{fig:Jet-Power}, corresponding to different values of the parameter 
        $s$ ($s = 0.0 ,\ 0.3$, $0.5$, and $0.7$). The jet breakout times for this model 
        are $t_{\rm bo} = (2.94,\ 3.05,\  3.09,\ 3.12)$ s for $s = (0.0,0.3, 0.5, 0.7)$, 
        respectively. See the caption of Fig.\,\ref{fig:Jet-Power} for more details. 
    }
    \label{fig:breakup_combined}
\end{figure}

We ascertain the threshold jet power for the 20R9W2 model by analytically propagating 
the jet inside the star using the framework of \citet{Bromberg+11}. We inject a relativistic, 
cold jet at the base of the flow, which is set to a distance $z_{\rm inj} = 10R_g\approx10^7$\,cm 
along the jet symmetry axis. The jet is introduced with a bulk Lorentz factor of $\Gamma_{j0}=10$ 
with a half-opening angle of $\theta_0=0.1$\,rad, and it is powered with a time-dependent 
jet power $L_j(t)$ for an engine activity time $t_{\rm eng}$. The adopted formalism includes 
the formation of forward and reverse shocks at the jet head, where the former shock heats the stellar material and 
the latter shock heats the jet. This inflates a pressurized cocoon of shock heated material 
surrounding the jet that collimates the jet head. Even though the jet is introduced moving 
relativistically, it slows down to mildly relativistic speeds due to its interaction with the 
stellar interior. After penetrating out of the stellar envelope, the jet material propagates 
relativistically.

Figure \ref{fig:breakup_combined} shows the temporal evolution of the jet head position 
$z_h$ for the 20R9W2 model. We first test for the threshold jet power by injecting the jet with a constant power in 
the range 45$\leq\log L_j ({\rm erg\,s}^{-1}) \leq50$ and note the amount of time it takes for 
the jet to break out of the star, as shown by the solid lines in the figure. 
The goal here is to identify the value of $L_j$ for which {a hydrodynamic jet} 
{is able to break out of the star over a typical breakout time of $t_{\rm bo}\sim10$\,s \citep{Bromberg+15}.}
As expected, more powerful jets break out over shorter times since their higher power drives a faster head, according to 
the relation $v_h \approx c / (1 + \tilde{L}^{-1/2})$. In this case, $\tilde{L} \propto 
L_j / \rho_a$ for a fixed size of the jet head, and since $\rho_a$ decreases 
(Eq.\,\ref{eq:density}), $v_h$ increases as the jet advances, facilitating earlier breakout. 
As shown in the figure,  jets with power $L_j < 10^{48}\,{\rm erg\,s}^{-1}$ take 
longer than the typical engine activity time to breakout for this particular stellar 
model and therefore may be choked. {The argument presented here is not definitive proof that they are indeed choked, but their long breakout times are inconsistent with observations \citep{Bromberg+15}, and therefore, offer an indirect way to disfavor them.}

When we use the time-dependent jet power, as obtained from our accretion model and shown 
in Fig.\,\ref{fig:Jet-Power}, the jet is able to breakout over a much shorter break out 
time of $t_{\rm bo} \approx 3$\,s, consistent with recent GRMHD collapsar simulations \citep{Urrutia+25}. The propagation of the jet head is shown for the different 
values of the parameter $s$ ($0.0, 0.3$, and 0.7) with negligible differences. These profiles 
reflect a more realistic scenario where the activity of the central engine decreases 
over time due to the reduction in material available for accretion.

The threshold jet power alone is not enough to guarantee a successful breakout. For that the jet must be powered 
for at least as long as the jet breakout time of the star, which can vary depending on the jet power, stellar radius, 
and whether the jet is hydrodynamic or Poynting flux dominated \citep{Bromberg-Tchekhovskoy-16}. The typical jet breakout 
time from Wolf-Rayet progenitors that have $R_{\rm WR}\sim10^{11}$\,cm is of the order of $t_{\rm bo}\sim10$\,s. 
\citep{Bromberg+15}. If the jet is powered at $L_j^{\rm thr}$ over this period, then the total energy of bipolar 
jets is $E_j=2L_j^{\rm thr}t_{\rm bo}$. A large fraction of this energy is deposited inside the star and goes into 
powering a quasi-spherical cocoon that also breaks out of the star. In order to produce the GRB the engine activity 
time must be larger than the breakout time, so that the minimum energy that is available for the GRB is around 
$E_{\rm j,min}=2L_j^{\rm th}(t_{\rm eng}-t_{\rm bo})$. The radiated $\gamma$-ray energy is some fraction of this, and 
the minimum isotropic-equivalent energy for a uniform jet with half-opening angle $\theta_j$ is then
\begin{eqnarray}\label{eq:E_gamma_iso}
    E_{\gamma,\rm min}^{\rm iso} &\approx& 
    \frac{\epsilon_\gamma 2E_{\rm j,min}}{\theta_j^2} \approx \frac{4\epsilon_\gamma L_j^{\rm thr} (t_{\rm eng}-t_{\rm bo})}{\theta_j^2} \\
    &\approx& 4\times10^{50}(1+z)^{-1}\epsilon_{\gamma,-1}L_{j,48}^{\rm thr}\,t_{\rm GRB, 1}\theta_{j,-1}^{-2}\,{\rm erg}\,,
\end{eqnarray}
where we have assumed a $\gamma$-ray efficiency of $\epsilon_\gamma = 0.1\epsilon_{\gamma,-1}$, 
$(t_{\rm eng}-t_{\rm bo}) = t_{z,\rm GRB} = t_{\rm GRB}/(1+z) = 10(1+z)^{-1}\,t_{\rm GRB, 1}$\,s, 
and $\theta_j = 0.1\theta_{j,-1}$. The above estimate is in agreement with the low-end of the 
$E_\gamma^{\rm iso}$ distribution of long GRBs \citep[e.g.][]{Poolakkil+21}. If stellar properties 
of Pop III progenitors are indeed similar to that of Pop I/II, then this observed limit should hold.

\subsubsection{Accretion Disc Mass}

Even with a disc in place, a minimum disc mass $M_d$ is required to maintain accretion 
at the rates and duration needed for a GRB-producing jet. As was done above, we relate 
the mass accretion rate with the jet power, so that $\dot M = L_j/\eta_jc^2$. Since some 
of the disc mass can be lost to outflows, only a fraction $\eta_d$ makes it to the BH over the 
duration the engine is active ($t_{\rm eng}$), which yields the requisite disc mass to be 
$M_d = \dot M\,t_{\rm eng}/\eta_d$. The engine must be active for longer than the jet break 
out time, so that $t_{\rm eng} = t_{\rm bo} + t_{z,\rm GRB} = (1 + t_{\rm bo}/t_{z,\rm GRB})t_{z,\rm GRB} = \eta_t t_{z,\rm GRB}$, 
with $1\leq\eta_t\lesssim2$, where the lower limit is valid for GRBs longer than the typically 
measured timescales of $t_{\rm GRB} = (1+z)t_{z,\rm GRB}\sim10$\,s and the upper limit applies 
when the GRB is powered for at least as long as the breakout time, with $t_{z,\rm GRB}\sim t_{\rm bo}$. 
Combining this with Eq.\,(\ref{eq:E_gamma_iso}), we get an estimate of the minimum disc mass, 
\begin{equation}
    \frac{M_d}{M_\odot} = \frac{\eta_t}{\eta_d\eta_j}\frac{\theta_j^2}{4}\frac{E_\gamma^{\rm iso}}{\epsilon_\gamma c^2} 
    \approx 0.2\,\eta_t\left(\frac{\eta_d}{0.7}\right)^{-1}\eta_{j,-3}^{-1}\theta_{j,-1}^2\epsilon_{\gamma,-1}^{-1}E_{\gamma,52}^{\rm iso}\,,
\end{equation}
where $\eta_j=10^{-3}$ when $a_*\sim 0.75$ and we have assumed that at most $30\%$ of the disc 
mass can be lost to outflows. This analytical estimate aligns with general expectations that a disc 
mass on the order of $\sim0.1-1\,M_{\odot}$, depending on the jet power efficiency ($\eta_j\sim10^{-3}$–$10^{-2}$), 
is necessary to sustain accretion and power a successful GRB \citep{Kumar+08, MacFadyen-Woosley-99}. 
Such disc masses can liberate energies in excess of $10^{51}$ erg, sufficient for producing typical 
GRBs. Conversely, significantly smaller disc masses ($M_d\ll0.1\ M_{\odot}$) would be insufficient to 
sustain the necessary jet power long enough for jet breakout, leading to choked jets and a failed GRB \citep{Bromberg+11}. 

\subsubsection{Phase Diagram for a Successful GRB}
The conditions for a successful jet in a collapsar are intricately linked. It requires a sufficiently rapidly rotating 
central engine and an accretion disc with enough mass to sustain an outflow with power greater than the threshold jet power, 
for timescales longer than the jet breakout time. 
To that end, we select the stellar models in our grid that satisfy the following essential requirements: 
i) disc mass $M_d \gtrsim 1\ M_\odot$, ii) engine time $t_{\rm eng}>t_{\rm bo}\sim 10$ s, while maintaining 
iii) the jet power above the threshold, with $L_j>L_{j}^{\mathrm{thr}}$. 

\begin{figure}
\centering
\includegraphics[width=0.49\textwidth]{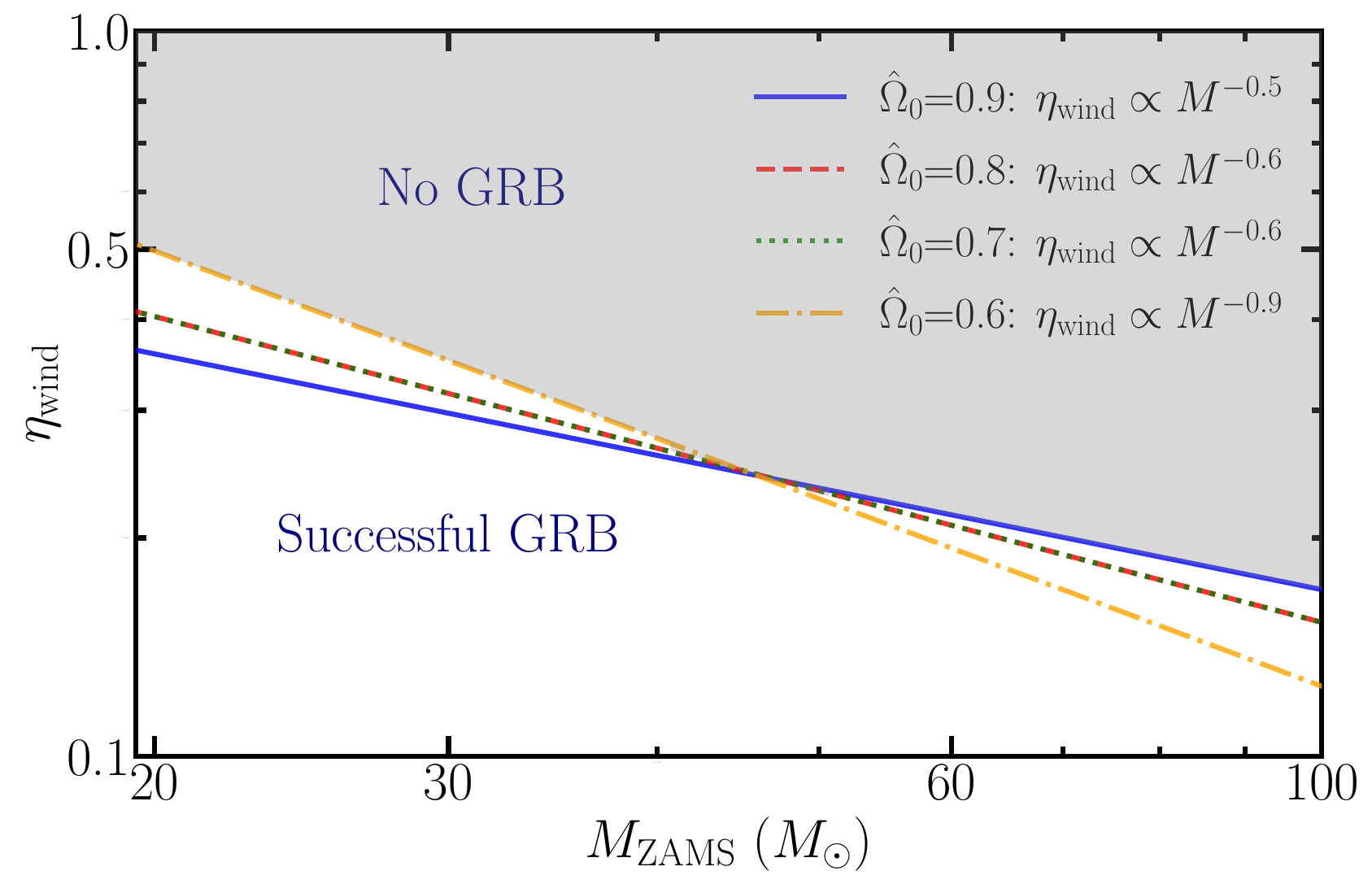}
\caption{
    Phase diagram showing the conditions required for a successful gamma-ray burst (GRB) 
    as a function of the initial mass and the wind mass-loss efficiency ($\eta_{\rm wind}$). Each curve 
    represents the critical boundary for a given initial rotation rate: $\hat{\Omega}_0 = 0.9$ (blue), 
    0.8 (red), 0.7 (green), and 0.6 (orange). The shaded region above each curve corresponds to parameter 
    combinations that fail to produce a GRB: (i) $M_d \gtrsim 1\ M_{\odot}$, (ii) $t_{\rm eng}>t_{\rm bo}\sim 10$ s, and (iii) $L_j>L_j^{\rm thr}$. See sec.~\ref{sec:criteria_GRB} for a detailed discussion. 
    }
\label{fig:PhaseDiagram}
\end{figure}

Figure \ref{fig:PhaseDiagram} combines all of the above requirements into a phase 
diagram over three fundamental qualities of the progenitor star, namely the initial 
mass ($M_{\rm ZAMS}$), initial rotation ($\hat{\Omega}_0$), and wind mass-loss 
efficiency ($\eta_{\rm wind}$). The lines correspond to different initial rotation rates 
$\hat{\Omega}_0$, where models lying in the shaded area above each curve fail to produce 
a GRB. For $M_{\rm ZAMS}\lesssim50\,M_\odot$, 
having relatively lower initial rotation allows for a larger $\eta_{\rm wind}$ that could still produce a successful GRB. 
The trend reverse for $M_{\rm ZAMS}\gtrsim50\,M_\odot$. For lower stellar masses, it is possible to have a higher 
$\eta_{\rm wind}$. As a consequence, if the wind mass loss efficiency were to match that observed in the local Universe 
for massive stars, then that would restrict GRB producing population in Pop III progenitors with $M_{\rm ZAMS}\sim 10\,M_\odot$ 
and $\hat{\Omega}_0\sim0.6$. In contrast, GRB progenitors on the high mass end in our grid require $\eta_{\rm wind}\lesssim0.2$.

\section{The GRB Rate at high redshifts}\label{sec:GRB-rate}
The intrinsic production rate of GRBs at any redshift, as given by the number of events per unit 
observer-frame time $t_{\rm obs}$ and per unit redshift, depends on the birthrate and rate of death of the progenitors. 
Since the lifespan of massive stars is much smaller and unable to cause any significant delay, the rate of GRBs 
is expected follow the formation rate of Pop III stars, such that
\begin{equation}
    \Psi_{\rm GRB}(z) \equiv \frac{d^2N_{\rm GRB}}{dt_{\rm obs}\,dz} 
    = \eta_{\rm GRB}(z)\frac{\dot\rho_\star(z)}{(1+z)} \frac{dV}{dz}\,,
\end{equation}
where $\dot\rho_\star(z)$ is the local star formation rate density (SFRD), given by stellar mass per unit 
comoving volume per unit comoving time $t_{\rm em}$, and $\eta_{\rm GRB}(z)$ is the efficiency of 
turning massive stars into GRBs. In general, $\eta_{\rm GRB}$ is redshift dependent, however, in 
what follows we make the simplifying assumption that properties of the progenitor Pop III stars do 
not evolve significantly over redshift, which allows us to obtain a mean value for the GRB production 
efficiency. The cosmological time-dilation factor of $(1+z)^{-1}$ converts the 
rate from comoving time to observer-frame time $t_{\rm obs}=(1+z)t_{\rm em}$, and $dV/dz$ is the 
comoving volume element per unit redshift, 
\begin{equation}
    \frac{dV}{dz} = 4\pi r^2(z)\left|\frac{dr}{dz}\right|\,,
\end{equation}
where $r(z)$ is the comoving distance to redshift $z$, given by
\begin{equation}
    r(z) = \frac{c}{H_0}\int_0^z\frac{dz'}{E(z')}\,,
\end{equation}
with $c/H_0$ giving the Hubble distance and $E(z) = \sqrt{\Omega_m(1+z)^3+\Omega_\Lambda}$ when 
assuming a flat Universe. In this case, the comoving distance has a simple relation with the luminosity 
distance, where $d_L(z) = (1+z)r(z)$. 

\begin{figure}
    \centering
    \includegraphics[width=0.48\textwidth]{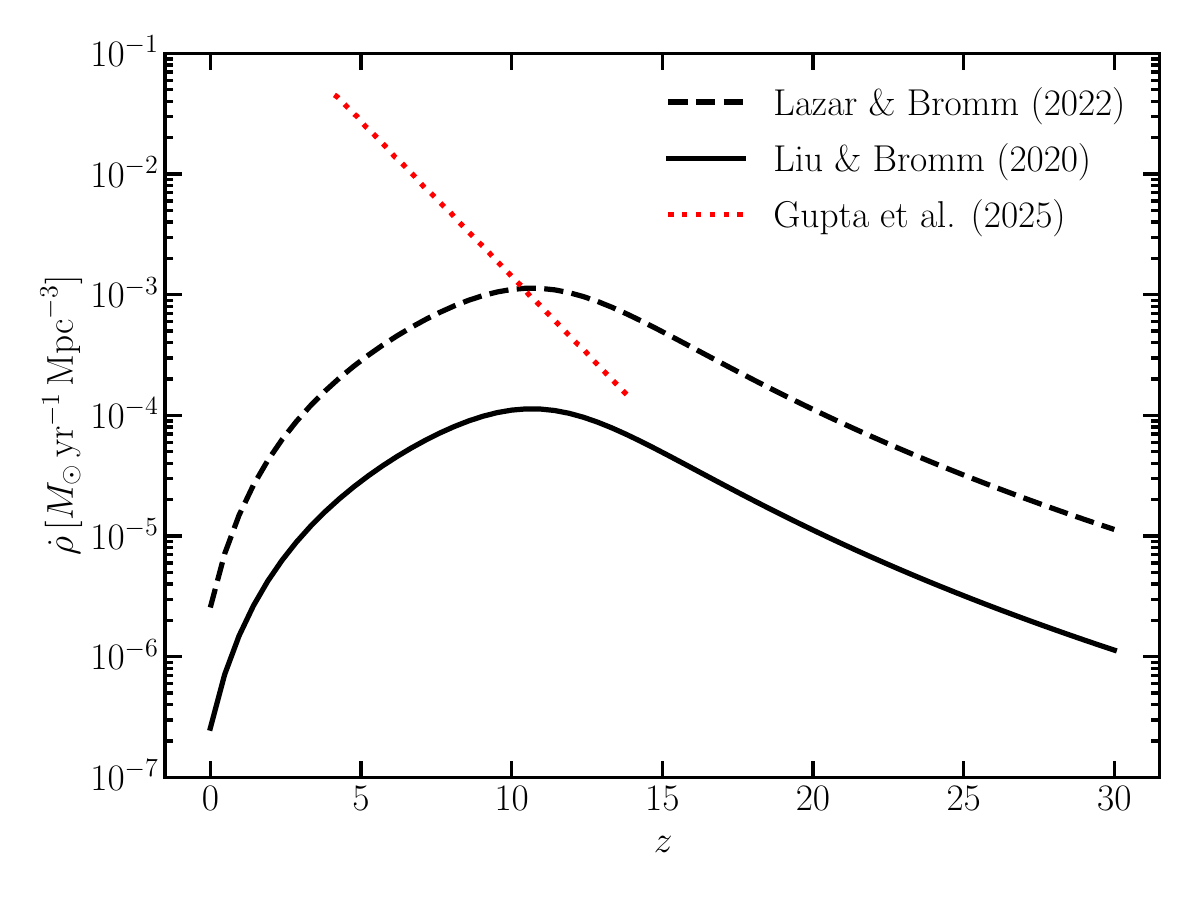}
    \caption{
        Star formation rate density of Pop III stars from \citealt{Lazar-Bromm-22} (dashed), 
        same but with normalization reduced by a factor of 10 (solid) as in \citet{Liu-Bromm-20}, 
        and from \citealt{Gupta+25} (dotted) that uses recent results from JWST observations.
    } 
    \label{fig:SFRD}
\end{figure}

To calculate the SFR of Pop III stars we follow the treatment by 
\citet{Lazar-Bromm-22} where they adopt the generalized SFRD from 
\citet{Madau-Dickinson-14}
\begin{equation}
    \frac{\dot\rho_\star(z)}{M_\odot\,{\rm yr}^{-1}\,{\rm Mpc}^{-3}} = \frac{a(1+z)^b}{1 + [(1+z)/c]^d}\,,
\end{equation}
with $a=7657$, $b=-5.92$, $c=12.83$, and $d=-8.55$, as shown in Fig.\,\ref{fig:SFRD} (dashed line). However, 
when comparing the recent estimates from \citet{Gupta+25}, that uses deep observations made by the JWST, we 
find that the \citet{Lazar-Bromm-22} SFRD overproduces the rate of star formation beyond $z\simeq10$. \citet{Lazar-Bromm-22} 
used the fit to the cosmological numerical simulations of \citet{Liu-Bromm-20}, but raised the normalization 
by a factor of 10 at $z=10$. Here we use the original fit of \citet{Liu-Bromm-20}, with $a=765.7$, that better agrees with 
limits from JWST. While the above is a measure of the rate at which stars are forming, the distribution in stellar mass of the number of 
formed stars is given by the initial mass function (IMF). Since there are no direct observations, 
the true IMF of Pop III stars is unknown. Following \citet{Lazar-Bromm-22}, here we consider a generalized 
IMF
\begin{equation}
    \xi(m)\equiv\frac{dN}{dm}\propto m^{-\alpha}\exp\left[-\left(\frac{m}{m_{\rm char}}\right)^{-\beta}\right]\,,
\end{equation}
that asymptotes to $m^{-\alpha}$ for large $m$ and is suppressed exponentially below a characteristic mass $m_{\rm char}$. 
There is large uncertainty regarding the power-law index of the Pop III IMF. The power-law index of the present day IMF 
is $\alpha=2.35$ \citep{Salpeter-55} and it is bottom-heavy where the average mass,
\begin{equation}
    \langle m \rangle = \frac{\int m\xi(m)\,dm}{\int \xi(m)\,dm}\,,
\end{equation}
is dominated by low-mass stars with $\langle m\rangle \sim 0.5M_\odot$ \citep{Bromm-13}. 
Inferences regarding the IMF of Pop III stars have thus far been obtained from 
cosmological numerical simulations. Most of which generally find the IMF to be top-heavy, with recent simulations 
finding $\alpha\sim1.13$ \citep[e.g.][]{Stacy+16,Wollenberg+20} which would yield larger average masses. 
Another important question 
is the range of masses that are produced in these simulations. Due to these being computationally intensive, 
they are limited to smaller runtimes, that probe the evolution of star-forming mini dark-matter halos to 
$\sim10^3-5\times10^3$\, yr with resolutions of few to several tens of astronomical units. Early works found the 
IMF dominated by very massive Pop III stars with $m\gtrsim100M_\odot$ \citep[e.g.][]{Abel+02,Bromm+02}, 
however, more recent simulations show a greater diversity in stellar mass with $1M_\odot \lesssim m \lesssim 10^3M_\odot$ 
\citep[e.g.][]{Hirano+14,Hosokawa+16,Stacy+16}.

\begin{figure}
    \centering
    \includegraphics[width=0.48\textwidth]{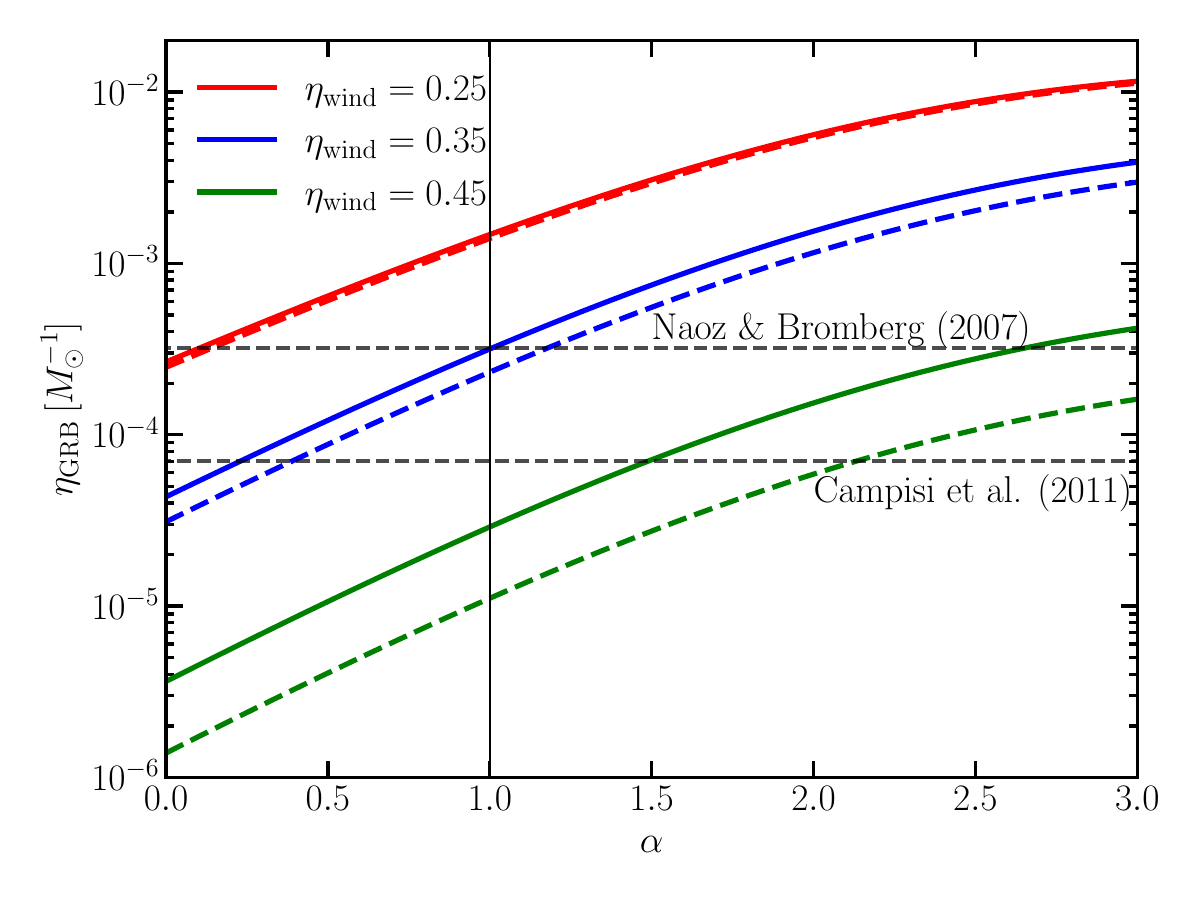}
    \caption{GRB efficiency for different power-law indices $\alpha$ of the generalized Pop III IMF, shown 
    for different values of the wind efficiency $\eta_{\rm wind}$ when assuming $m_{\rm char} = 20M_\odot$. 
    The solid curves assume a flat distribution in initial rotation, i.e. $P_{\rm rot}\propto\hat\Omega_0^\delta$ 
    with $\delta=0$, and the dashed curves assume a steeper distribution with $\delta=5$, so that 50\% of 
    the distribution has $\hat\Omega_0>0.9$. The vertical line shows our fiducial value for the IMF that 
    yields a top-heavy and flat distribution $m\xi(m)=dN/d\ln m$, in stellar mass. The horizontal dashed 
    lines show the upper limits obtained in two separate works.
    }
    \label{fig:eta_GRB}
\end{figure}

\begin{figure}
    \centering
    \includegraphics[width=0.48\textwidth]{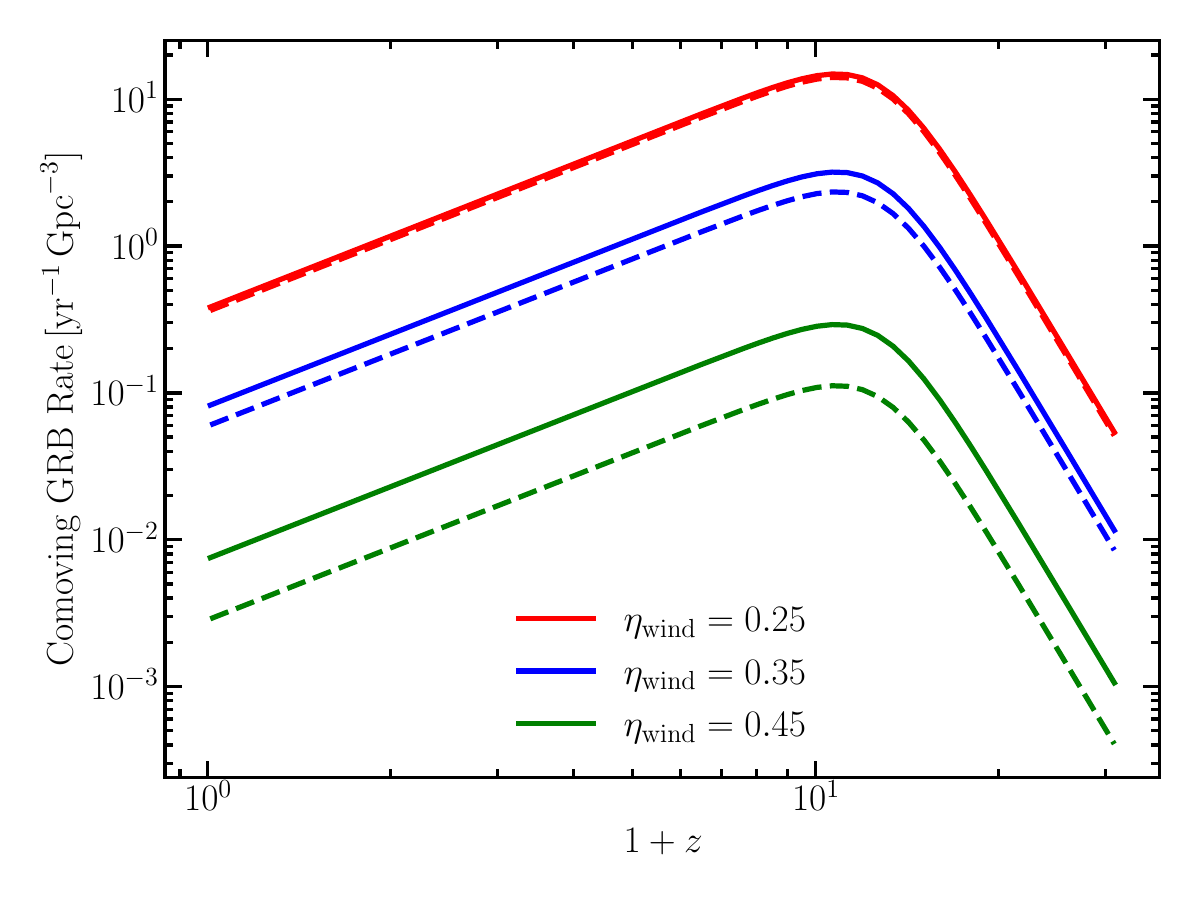}
    \includegraphics[width=0.48\textwidth]{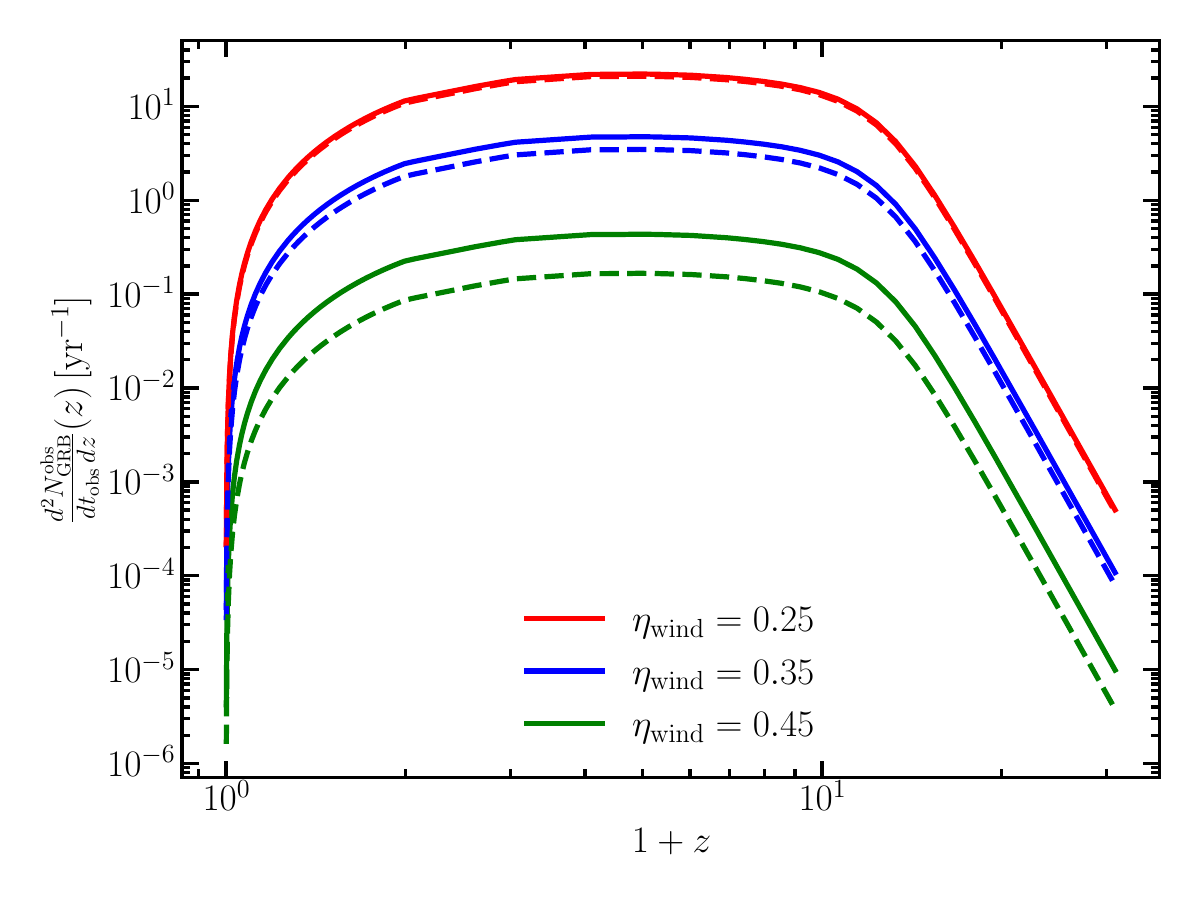}
    \includegraphics[width=0.48\textwidth]{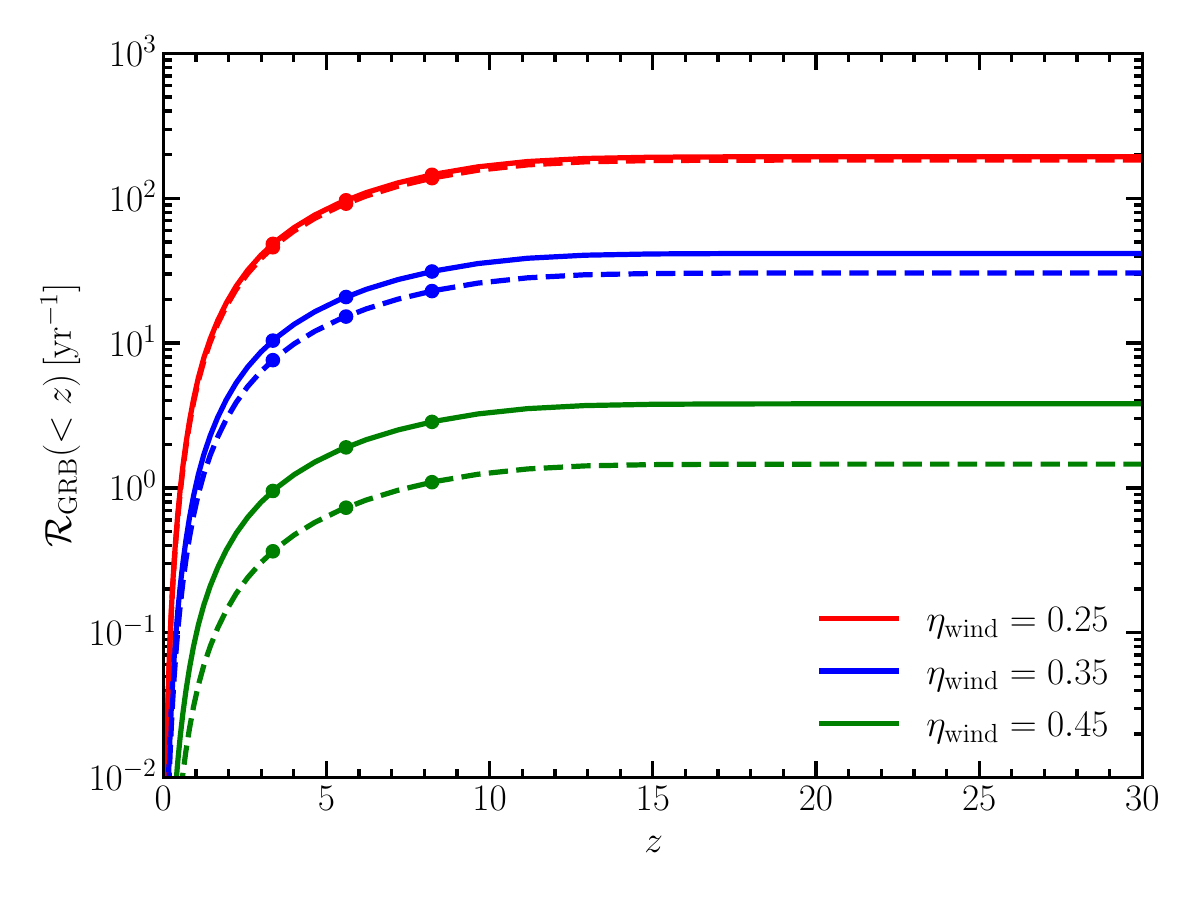}
    \caption{(Top) Intrinsic comoving GRB rate density, shown for a top-heavy IMF power-law index $\alpha=1$, 
    for different wind efficiency parameter $\eta_{\rm wind}$ values and initial rotation 
    distribution, with $\delta=0$ (solid) and $\delta=5$ (dashed). All other parameters 
    are the same as assumed for Fig.\,\ref{fig:eta_GRB}. 
    (Middle) The observable all-sky Pop III GRB rate at any given redshift for \textit{Swift}/BAT 
    sensitivity. 
    (Bottom) The integrated all-sky rate of detection by \textit{Swift}/BAT up to a given $z$. The 
    three dots show the redshifts that contribute 25\%, 50\%, and 75\% of the total rate.
    }
    \label{fig:GRB_rate}
\end{figure}

Given the IMF, the fraction of stars in a given range of stellar mass, $m_{\rm low}\leq m \leq m_{\rm high}$, 
that will collapse to a rapidly rotating BH which is able to launch a jet and successfully produce a GRB can be calculated, 
\begin{equation}
    \eta_{\rm GRB} = \dfrac{\int_{m_{\rm low}}^{m_{\rm high}(\hat\Omega_0)}\xi(m)\,P_{\rm rot}(\hat\Omega_0)\,dm\,d\hat\Omega_0}
    {\int_{m_{\min}}^{m_{\max}} m\,\xi(m)\,dm}\,,
\end{equation}
where $\eta_{\rm GRB}$ is the efficiency of producing a GRB per unit stellar mass. Here $m_{\min} = 1.0M_\odot$ and 
$m_{\max}=300M_\odot$ are the minimum and maximum masses of Pop III stars \citep{Bromm-13,Lazar-Bromm-22}.
We obtain the limits on the GRB progenitor stellar masses from the phase diagram in Fig.\,\ref{fig:PhaseDiagram}, 
that shows the range of initial masses that can yield a successful GRB for a range of wind efficiencies and initial 
rotation rates. 

Since the distribution of initial rotation rates with stellar mass is not uniform, appropriate 
weights must be given to different stellar masses. From numerical simulations there is some evidence that Pop III 
stars may have been rapid rotators with surface angular speeds approaching the critical limit, 
such that $\hat\Omega_0\sim1$ \citep{Stacy+11,Stacy+13,Hirano-Bromm-2018}. 
In general, the distribution of Pop III stars with $\hat\Omega_0$ can be expressed as
\begin{equation}
    P_{\rm rot}(\hat\Omega_0)\propto\hat\Omega_0^{\delta}\,,\quad\quad {\rm for}\quad\quad 0.6 \leq\hat\Omega_0\leq 1\,,
\end{equation} 
where we have taken the limits on $\hat\Omega_0$ that are relevant for powering GRB jets. Due to resolution 
and computational time constraints in cosmological simulations, the exact distribution, and therefore 
the exact value of $\delta$, is not clear. If critical rotation is indeed favored then $\delta$ can 
be quite steep. Here we consider two value of $\delta=\{0,\,5\}$, where $\delta=0$ yields a flat 
distribution with 25\% of all Pop III stars at a given redshift born with $\hat\Omega_0>0.9$, while 
for $\delta=5$ the same fraction rises to 50\%. As demonstrated in Fig.\,\ref{fig:PhaseDiagram}, 
the high-mass limit, $m_{\rm high}(\hat\Omega_0)$, on the GRB progenitors is sensitive to the initial rotation 
rate, whereas the lower limit, $m_{\rm low} = 20M_\odot$, is insensitive and fixed to yield a 
BH remnant after core-collapse.

Figure\,\ref{fig:eta_GRB} shows the GRB production efficiency as a function of the 
IMF power-law index $\alpha$ for different wind efficiencies $\eta_{\rm wind}$ and initial rotation 
distributions, while assuming $m_{\rm char} = 20M_\odot$. 
An increasing value of $\eta_{\rm wind}$ leads to reduced efficiency due to a reduction in 
the range of masses that can produce a successful GRB, as can be seen in Fig.\,\ref{fig:PhaseDiagram}. 
Due to the same effect, giving more weight to very rapidly rotating progenitors, by having a 
steep distribution in $\hat\Omega_0$, further reduces the efficiency. 

The intrinsic comoving rate density of GRBs as a function of redshift is shown in 
the top panel of Fig.\,\ref{fig:GRB_rate}. Using this we obtain the all-sky 
rate while accounting for beaming, flux sensitivity of the detector, 
and the luminosity function of GRBs, such that
\begin{eqnarray}
    \mathcal{R}_{\rm GRB}(<z) &\equiv& \frac{dN_{\rm GRB}^{\rm obs}}{dt_{\rm obs}} \\
    &=& \eta_{\rm beam}\int_0^zdz\,\Psi_{\rm GRB}(z)\int_{L_{\rm lim}(z)}\Phi(L)d\log L \nonumber \,.
\end{eqnarray}
The beaming fraction is given by $\eta_{\rm beam} = \Delta\Omega_{\rm GRB}/4\pi = 1-\cos(\theta_{\max})\approx\theta_{\max}^2/2$, 
where $\Delta\Omega_{\rm GRB}$ is the solid angle into which emission from the bipolar jets is beamed.  
In general, GRB jets have angular structure comprising of a narrow energetic core of angular size $\theta_c$ 
and power-law wings at $\theta>\theta_c$, 
with $\theta$ measured from the jet symmetry axis. Unless the GRB is only several tens of Mpc away, one prime 
example being GW170817/GRB\,170817A ($d_L\simeq40$\,Mpc), most GRBs are found at cosmological distances 
($d_L\sim{\rm Gpc}$) and so they can only be observed if they are viewed from within the core with 
viewing angle $\theta_{\rm obs}\lesssim\theta_c$ \citep[e.g.][]{BN19,OConnor+24}. In addition, the bulk Lorentz 
factor of the outflow $\Gamma\gtrsim100$ as demanded by compactness constraints \citep{Piran-99}, and therefore 
$\theta_{\max} = \theta_c + \Gamma^{-1}\approx\theta_c$. The size of the jet core is most easily 
determined from observing jet breaks in the afterglow lightcurve, that occur when the angular size 
($1/\Gamma$) of the observable region of the jet's surface around the line-of-sight includes the edge 
of the jet core, yielding the condition $\Gamma(\theta_c-\theta_{\rm obs})=1$. Canonical jet breaks are 
considered in the context of top-hat jets that feature sharp edges with vanishingly small energy outside 
of the jet core. This particular geometry is also the one that yields the sharpest jet breaks as compared 
to jets with shallower power-law wings. Not many GRBs show such sharp jet breaks but in a small fraction 
that do, jet core sizes of $\theta_c\sim0.1$\,rad have been inferred \citep{Frail+01}, with a mean 
beaming fraction $\langle\eta_{\rm beam}\rangle\sim1/500$. 

The luminosity function of GRBs is given by \citet{Wanderman-Piran-10}, which yields the number of GRBs 
with isotropic-equivalent luminosities in the interval $\log L$ and $\log L + d\log L$,
\begin{equation}
    \Phi(L) \equiv \frac{dN}{d\log L} =
    \begin{cases}
        \left(\frac{L}{L_\star}\right)^{-0.2}, & L < L_\star \\
        \left(\frac{L}{L_\star}\right)^{-1.4}, & L > L_\star
    \end{cases}\,,
\end{equation}
where $L_\star = 10^{52.5}\,{\rm erg\,s}^{-1}$ is a characteristic isotropic-equivalent luminosity. 
This function has been demonstrated to hold for at least $z<9$. To infer the limiting luminosity for a given detector, 
$L_{\rm lim} = 4\pi d_L^2 F_{\rm lim}$, we use the limiting bolometric flux, e.g. \textit{Swift} has 
$F_{\rm lim}=1.2\times10^{-8}\,{\rm erg\,cm}^{-2}{\,\rm s}^{-1}$ \citep{Li-08}.

The middle and bottom panels of Fig.\,\ref{fig:GRB_rate} respectively show the differential observable 
all-sky equivalent rate for \textit{Swift}/BAT for a given redshift and integrated rate 
up to a given redshift, for different wind efficiencies $\eta_{\rm wind}$ and initial 
stellar rotation distributions. For each curve shown in the bottom panel, the three dots 
show the range of redshifts, $z\simeq3.3 - 8.2$, that contribute the central 50\% of the 
Pop III GRBs. The rate declines with increasing $\eta_{\rm wind}$ due to 
the corresponding decline in the $m_{\rm high}$ mass of the stellar progenitor for any given initial 
rotation rate. From Fig.\,\ref{fig:PhaseDiagram} it is expected that if $\eta_{\rm wind}\gtrsim0.5$, there will 
be no massive Pop III stars capable of producing a GRB.

\section{Discussion \& Conclusions}\label{sec:discussion}

In this work, we obtain the radial makeup of Pop III stars just prior to 
core-collapse using a 1D stellar evolution code. The density and rotational profiles 
are used to calculate the initial mass and spin of the remnant black hole as well 
as the mass accretion rate onto it. The accretion rate is then converted to the jet 
power using a BH spin dependent efficiency, and the jet is propagated through the 
star in a semi-analytical calculation while keeping track of its collimation due to 
the pressurized cocoon. In the end, we constrain a parameter space comprising the 
initial properties of the progenitor, namely its ZAMS mass, rotation, and efficiency 
of mass loss during its CHE, that yields a successful GRB. 
It is this parameter space that ultimately determines the efficiency of GRB production 
from Pop III progenitors given their initial mass function and initial rotation 
distribution at high redshifts.

We find that the phase space to produce a successful GRB is very sensitive to the 
efficiency of mass loss due to line-driven winds. Even though Pop III stars 
begin with having no metals in their envelopes, rotationally-driven CHE mixes nuclear 
products from the core into the outer stellar layers. As the efficiency of mass 
loss is increased, it starts to limit the maximum stellar mass capable of producing a 
successful GRB, until an initial rotation-dependent $\eta_{\rm wind}(\hat\Omega_0)$. 
For any higher efficiency, the stars start to lose more than the requisite angular 
momentum that is needed to launch and breakout jets from the stellar envelope. Low wind 
efficiency in the range of $0.1\lesssim\eta_{\rm wind}\lesssim0.3$ has been employed 
in earlier works dealing with Pop I/II \citep[e.g.][]{Woosley-Heger-06} and Pop III 
\citep{Yoon+12} progenitors for the same reason. While this range of $\eta_{\rm wind}$ 
still yields successful GRBs for a broad range in initial rotation rates, {albeit over a narrow range of $M_{\rm ZAMS}$}, 
only slower rotators with $\hat\Omega_0<0.9$ {may be able to} produce GRBs {in an even narrower range of stellar masses 
near $M_{\rm ZAMS}\sim20M_\odot$} when $\eta_{\rm wind}\gtrsim0.5$. {Therefore, the want for larger wind efficiency restricts the stellar models to slow 
rotators and masses smaller than $20M_\odot$, where the latter condition becomes prohibitive for collapsar models that require a BH remnant.}

The true value of the wind mass-loss efficiency in Pop III stars is still very uncertain. 
It may also vary with the particular mass-loss scheme used in stellar evolution codes. 
In particular, our study used the \textit{Dutch} scheme that has been empirically calibrated 
using observations of local stars (Pop I and II). Within this scheme, the mass-loss rate 
in Pop III stars is initially negligible. When the star evolves off the main sequence, rotational mixing 
transports carbon from the core to the surface that increases mass loss, thereby causing 
loss of angular momentum. However, recent works from optically thick winds 
\citep[e.g.][]{Sander-Vink-20} have shown that the surface abundance of iron (not the total 
metallicity) is what mostly drives line-driven winds in massive stars. This implies that the 
empirical recipes used may overestimate mass loss, since the surface abundance of iron 
remains essentially constant during stellar evolution. Building on this, \citet{Jeena+23} 
provide an updated discussion of modern mass loss recipes and their limitations in very low 
metallicity regimes. In light of this, we acknowledge the limitations of the mass-loss 
scheme used in this work, and note that a more detailed analysis of alternative 
mass-loss schemes remains out of the scope of this work.

Taking $\eta_{\rm wind}=0.45$ to be the fiducial value, we find the GRB production 
efficiency to be $\eta_{\rm GRB}\sim10^{-5}\,M_\odot^{-1}$ when assuming a flat distribution 
in $dN/d\ln m$ of the Pop III IMF. This estimate is in agreement with earlier upper limits 
\citep[e.g.][]{Naoz-Bromberg-07,Campisi+11}, and it would yield an all-sky detectable rate of 
$\sim2$ Pop III GRBs per year for an instrument with \textit{Swift}/BAT sensitivity. 
Since \textit{Swift}/BAT has a field of view of FOV = 1.4\,str, the detection 
rate is reduced by FOV$/4\pi\approx0.11$, which yields a total of $\sim4-5$ Pop III GRBs 
over the past 20 years of its present operational lifetime.

So far, two very high-$z$ long-duration GRBs with $z=8.2$ and $9.4$ have been observed, 
where the redshift of the latter was determined using the non-observations of the host 
galaxy. For that GRB it is not clear if the stellar progenitor is a Pop III star, although 
no notable differences in its emission were seen \citep{Cuchiara+11}. In fact, below $z\sim15$ 
more metal rich progenitors start to dominate the SFR, making it less likely for this 
GRB to have come from a Pop III star \citep{Belczynski+10}. One possible way to infer if 
the progenitor is a Pop III star is through high-resolution and IR spectroscopy of the 
afterglow. If the afterglow spectrum shows no absorption from iron-group elements, which 
would not be present if earlier SNe and stellar winds from the star itself had not polluted 
the surrounding external medium with metals, then a Pop III progenitor becomes a possibility 
\citep{Toma+16}. GRBs from very massive Pop III stars ($M_{\rm ZAMS}\gtrsim300M_\odot$), 
which could have produced longer lasting and more energetic bursts, may be ruled out \citep{Yoon+15}. 
The properties of the bursts produced by stars in our grid are similar to the more routinely 
detected bursts. This similarity makes it very challenging to distinguish 
between Pop I/II  and Pop III GRBs.

At high-$z$ it becomes challenging to detect the 
host galaxy even for well localized bursts, which means no redshift information. At the 
time of writing, the fraction of GRBs that do not have redshifts is 77 per cent\footnote{GRB data obtained from https://swift.gsfc.nasa.gov/archive/grb\_table}. These include a total of 1962 detected GRBs by 
any space-based observatory operating since the year 2004, out of which only 453 bursts 
have a measured redshift. A fraction of the GRBs without redshifts may have Pop III progenitors, 
assuming that such bursts are not discernible from the more commonly detected ones. This fraction 
is not known and the possibility of it being negligible cannot be ruled out based on existing observations.

If the actual all-sky detection rate of Pop III GRBs is significantly smaller than $\sim1\,{\rm yr}^{-1}$ 
for an instrument similar to \textit{Swift}/BAT, then two conclusions can be drawn. 
First, it can be argued that the wind efficiency must remain higher 
than $\eta_{\rm wind}\approx0.5$ for Pop III stars, and due to which they lose significant angular momentum, 
which ultimately renders them unable to launch relativistic jets and power GRBs. This scenario assumes 
that these stars were born with high initial surface angular velocities ($\hat\Omega_0>0.6$). Second, 
it might also mean that the assumption/inference of rapid rotation is incorrect, and that these stars did not 
have the requisite angular momentum to begin with.

An alternative possibility is that of a Pop III binary in which additional angular momentum can be gained 
from a donor companion either by Roche lobe overflow or common envelope evolution 
\citep{Bromm-Loeb-06,Belczynski+07,Kinugawa+19,Lloyd-Ronning-22}. If the GRB progenitor is spun up by 
binary evolution, then the detection rate for \textit{Swift} has been estimated to be rather low with 
$\sim0.01-0.1\,{\rm yr}^{-1}$ \citep{Belczynski+07}, where the latter value is for the most optimistic 
scenario. {Several works (though not necessarily for Pop III stars) have explored this scenario \citep[e.g.][]{Cantiello+07,Sana+12,Mandel+16,Marchant-Bodensteiner-24} and have shown that binary interactions can induce chemically homogeneous evolution or envelope stripping, both of which favor the formation of rapidly rotating, compact cores. Modeling this scenario lies beyond the scope of the present work.}

Hopes are now pinned on future more sensitive missions, e.g. HiZ-GUNDAM, Gamow, THESEUS, that will be 
capable of detecting the most distant GRBs. The challenge with identifying the host galaxy and obtaining 
a reliable redshift estimate, that can be complicated due to dust extinction \citep{Fausey+23}, 
will still need overcoming.

\section*{Acknowledgements}

We acknowledge useful conversations with Chris Fryer and Ore Gottlieb at the Workshop on GRBs 
and Central Engine Powered Transients held in Playa del Carmen, Mexico (2024). We also thank 
Enrique Moreno Mendez for insightful discussions on this topic. G. M. acknowledges support from an UNAM-DGAPA grant.   P.B. was supported by a grant 
(no. 2020747) from the United States-Israel Binational Science Foundation (BSF), Jerusalem, Israel 
(PB) and by a grant (no. 1649/23) from the Israel Science Foundation.

\section*{Data Availability}
The data will be shared on reasonable request to the corresponding authors.


\FloatBarrier    
\balance        

\bibliographystyle{mnras}
\bibliography{refs} 

@ARTICLE{Abel+02,
       author = {{Abel}, Tom and {Bryan}, Greg L. and {Norman}, Michael L.},
        title = "{The Formation of the First Star in the Universe}",
      journal = {Science},
         year = 2002,
        month = jan,
       volume = {295},
       number = {5552},
        pages = {93-98},
          doi = {10.1126/science.295.5552.93},
archivePrefix = {arXiv},
       eprint = {astro-ph/0112088},
 primaryClass = {astro-ph},
       adsurl = {https://ui.adsabs.harvard.edu/abs/2002Sci...295...93A}
}

@ARTICLE{Aloy+18,
       author = {{Aloy}, M.~A. and {Cuesta-Mart{\'\i}nez}, C. and {Obergaulinger}, M.},
        title = "{On the existence of a luminosity threshold of GRB jets in massive stars}",
      journal = {\mnras},
         year = 2018,
        month = aug,
       volume = {478},
       number = {3},
        pages = {3576-3589},
          doi = {10.1093/mnras/sty1212},
archivePrefix = {arXiv},
       eprint = {1801.06186},
 primaryClass = {astro-ph.HE},
       adsurl = {https://ui.adsabs.harvard.edu/abs/2018MNRAS.478.3576A}
}

@ARTICLE{Amati+18,
       author = {{Amati}, L. and {O'Brien}, P. and {G{\"o}tz}, D. and {Bozzo}, E. and {Tenzer}, C. and {Frontera}, F. and {Ghirlanda}, G. and {Labanti}, C. and {Osborne}, J.~P. and {Stratta}, G. and {Tanvir}, N. and {Willingale}, R. and {Attina}, P. and {Campana}, R. and {Castro-Tirado}, A.~J. and {Contini}, C. and {Fuschino}, F. and {Gomboc}, A. and {Hudec}, R. and {Orleanski}, P. and {Renotte}, E. and {Rodic}, T. and {Bagoly}, Z. and {Blain}, A. and {Callanan}, P. and {Covino}, S. and {Ferrara}, A. and {Le Floch}, E. and {Marisaldi}, M. and {Mereghetti}, S. and {Rosati}, P. and {Vacchi}, A. and {D'Avanzo}, P. and {Giommi}, P. and {Piranomonte}, S. and {Piro}, L. and {Reglero}, V. and {Rossi}, A. and {Santangelo}, A. and {Salvaterra}, R. and {Tagliaferri}, G. and {Vergani}, S. and {Vinciguerra}, S. and {Briggs}, M. and {Campolongo}, E. and {Ciolfi}, R. and {Connaughton}, V. and {Cordier}, B. and {Morelli}, B. and {Orlandini}, M. and {Adami}, C. and {Argan}, A. and {Atteia}, J. -L. and {Auricchio}, N. and {Balazs}, L. and {Baldazzi}, G. and {Basa}, S. and {Basak}, R. and {Bellutti}, P. and {Bernardini}, M.~G. and {Bertuccio}, G. and {Braga}, J. and {Branchesi}, M. and {Brandt}, S. and {Brocato}, E. and {Budtz-Jorgensen}, C. and {Bulgarelli}, A. and {Burderi}, L. and {Camp}, J. and {Capozziello}, S. and {Caruana}, J. and {Casella}, P. and {Cenko}, B. and {Chardonnet}, P. and {Ciardi}, B. and {Colafrancesco}, S. and {Dainotti}, M.~G. and {D'Elia}, V. and {De Martino}, D. and {De Pasquale}, M. and {Del Monte}, E. and {Della Valle}, M. and {Drago}, A. and {Evangelista}, Y. and {Feroci}, M. and {Finelli}, F. and {Fiorini}, M. and {Fynbo}, J. and {Gal-Yam}, A. and {Gendre}, B. and {Ghisellini}, G. and {Grado}, A. and {Guidorzi}, C. and {Hafizi}, M. and {Hanlon}, L. and {Hjorth}, J. and {Izzo}, L. and {Kiss}, L. and {Kumar}, P. and {Kuvvetli}, I. and {Lavagna}, M. and {Li}, T. and {Longo}, F. and {Lyutikov}, M. and {Maio}, U. and {Maiorano}, E. and {Malcovati}, P. and {Malesani}, D. and {Margutti}, R. and {Martin-Carrillo}, A. and {Masetti}, N. and {McBreen}, S. and {Mignani}, R. and {Morgante}, G. and {Mundell}, C. and {Nargaard-Nielsen}, H.~U. and {Nicastro}, L. and {Palazzi}, E. and {Paltani}, S. and {Panessa}, F. and {Pareschi}, G. and {Pe'er}, A. and {Penacchioni}, A.~V. and {Pian}, E. and {Piedipalumbo}, E. and {Piran}, T. and {Rauw}, G. and {Razzano}, M. and {Read}, A. and {Rezzolla}, L. and {Romano}, P. and {Ruffini}, R. and {Savaglio}, S. and {Sguera}, V. and {Schady}, P. and {Skidmore}, W. and {Song}, L. and {Stanway}, E. and {Starling}, R. and {Topinka}, M. and {Troja}, E. and {van Putten}, M. and {Vanzella}, E. and {Vercellone}, S. and {Wilson-Hodge}, C. and {Yonetoku}, D. and {Zampa}, G. and {Zampa}, N. and {Zhang}, B. and {Zhang}, B.~B. and {Zhang}, S. and {Zhang}, S. -N. and {Antonelli}, A. and {Bianco}, F. and {Boci}, S. and {Boer}, M. and {Botticella}, M.~T. and {Boulade}, O. and {Butler}, C. and {Campana}, S. and {Capitanio}, F. and {Celotti}, A. and {Chen}, Y. and {Colpi}, M. and {Comastri}, A. and {Cuby}, J. -G. and {Dadina}, M. and {De Luca}, A. and {Dong}, Y. -W. and {Ettori}, S. and {Gandhi}, P. and {Geza}, E. and {Greiner}, J. and {Guiriec}, S. and {Harms}, J. and {Hernanz}, M. and {Hornstrup}, A. and {Hutchinson}, I. and {Israel}, G. and {Jonker}, P. and {Kaneko}, Y. and {Kawai}, N. and {Wiersema}, K. and {Korpela}, S. and {Lebrun}, V. and {Lu}, F. and {MacFadyen}, A. and {Malaguti}, G. and {Maraschi}, L. and {Melandri}, A. and {Modjaz}, M. and {Morris}, D. and {Omodei}, N. and {Paizis}, A. and {P{\'a}ta}, P. and {Petrosian}, V. and {Rachevski}, A. and {Rhoads}, J. and {Ryde}, F. and {Sabau-Graziati}, L.},
        title = "{The THESEUS space mission concept: science case, design and expected performances}",
      journal = {Advances in Space Research},
         year = 2018,
        month = jul,
       volume = {62},
       number = {1},
        pages = {191-244},
          doi = {10.1016/j.asr.2018.03.010},
archivePrefix = {arXiv},
       eprint = {1710.04638},
 primaryClass = {astro-ph.IM},
       adsurl = {https://ui.adsabs.harvard.edu/abs/2018AdSpR..62..191A}
}

@ARTICLE{Bardeen+72,
       author = {{Bardeen}, James M. and {Press}, William H. and {Teukolsky}, Saul A.},
        title = "{Rotating Black Holes: Locally Nonrotating Frames, Energy Extraction, and Scalar Synchrotron Radiation}",
      journal = {\apj},
         year = 1972,
        month = dec,
       volume = {178},
        pages = {347-370},
          doi = {10.1086/151796},
       adsurl = {https://ui.adsabs.harvard.edu/abs/1972ApJ...178..347B}
}

@ARTICLE{Barkana-Loeb-01,
       author = {{Barkana}, R. and {Loeb}, A.},
        title = "{In the beginning: the first sources of light and the reionization of the universe}",
      journal = {\physrep},
         year = 2001,
        month = jul,
       volume = {349},
       number = {2},
        pages = {125-238},
          doi = {10.1016/S0370-1573(01)00019-9},
archivePrefix = {arXiv},
       eprint = {astro-ph/0010468},
 primaryClass = {astro-ph},
       adsurl = {https://ui.adsabs.harvard.edu/abs/2001PhR...349..125B}
}

@ARTICLE{Belczynski+07,
       author = {{Belczynski}, Krzysztof and {Bulik}, Tomasz and {Heger}, Alexander and {Fryer}, Chris},
        title = "{The Lack of Gamma-Ray Bursts from Population III Binaries}",
      journal = {\apj},
         year = 2007,
        month = aug,
       volume = {664},
       number = {2},
        pages = {986-999},
          doi = {10.1086/517500},
archivePrefix = {arXiv},
       eprint = {astro-ph/0610014},
 primaryClass = {astro-ph},
       adsurl = {https://ui.adsabs.harvard.edu/abs/2007ApJ...664..986B}
}

@ARTICLE{Belczynski+10,
       author = {{Belczynski}, Krzysztof and {Holz}, Daniel E. and {Fryer}, Chris L. and {Berger}, Edo and {Hartmann}, Dieter H. and {O'Shea}, Brian},
        title = "{On the Origin of the Highest Redshift Gamma-Ray Bursts}",
      journal = {\apj},
         year = 2010,
        month = jan,
       volume = {708},
       number = {1},
        pages = {117-126},
          doi = {10.1088/0004-637X/708/1/117},
archivePrefix = {arXiv},
       eprint = {0812.2470},
 primaryClass = {astro-ph},
       adsurl = {https://ui.adsabs.harvard.edu/abs/2010ApJ...708..117B}
}

@ARTICLE{Blandford-Znajek-77,
       author = {{Blandford}, R.~D. and {Znajek}, R.~L.},
        title = "{Electromagnetic extraction of energy from Kerr black holes.}",
      journal = {\mnras},
         year = 1977,
        month = may,
       volume = {179},
        pages = {433-456},
          doi = {10.1093/mnras/179.3.433},
       adsurl = {https://ui.adsabs.harvard.edu/abs/1977MNRAS.179..433B}
}

@ARTICLE{Breu-Rezzolla-16,
       author = {{Breu}, Cosima and {Rezzolla}, Luciano},
        title = "{Maximum mass, moment of inertia and compactness of relativistic stars}",
      journal = {\mnras},
         year = 2016,
        month = jun,
       volume = {459},
       number = {1},
        pages = {646-656},
          doi = {10.1093/mnras/stw575},
archivePrefix = {arXiv},
       eprint = {1601.06083},
 primaryClass = {gr-qc},
       adsurl = {https://ui.adsabs.harvard.edu/abs/2016MNRAS.459..646B}
}

@ARTICLE{Bromm+99,
       author = {{Bromm}, Volker and {Coppi}, Paolo S. and {Larson}, Richard B.},
        title = "{Forming the First Stars in the Universe: The Fragmentation of Primordial Gas}",
      journal = {\apjl},
         year = 1999,
        month = dec,
       volume = {527},
       number = {1},
        pages = {L5-L8},
          doi = {10.1086/312385},
archivePrefix = {arXiv},
       eprint = {astro-ph/9910224},
 primaryClass = {astro-ph},
       adsurl = {https://ui.adsabs.harvard.edu/abs/1999ApJ...527L...5B}
}

@ARTICLE{Bromm+02,
       author = {{Bromm}, Volker and {Coppi}, Paolo S. and {Larson}, Richard B.},
        title = "{The Formation of the First Stars. I. The Primordial Star-forming Cloud}",
      journal = {\apj},
         year = 2002,
        month = jan,
       volume = {564},
       number = {1},
        pages = {23-51},
          doi = {10.1086/323947},
archivePrefix = {arXiv},
       eprint = {astro-ph/0102503},
 primaryClass = {astro-ph},
       adsurl = {https://ui.adsabs.harvard.edu/abs/2002ApJ...564...23B}
}

@ARTICLE{Bromm-Larson-04,
       author = {{Bromm}, Volker and {Larson}, Richard B.},
        title = "{The First Stars}",
      journal = {\araa},
         year = 2004,
        month = sep,
       volume = {42},
       number = {1},
        pages = {79-118},
          doi = {10.1146/annurev.astro.42.053102.134034},
archivePrefix = {arXiv},
       eprint = {astro-ph/0311019},
 primaryClass = {astro-ph},
       adsurl = {https://ui.adsabs.harvard.edu/abs/2004ARA&A..42...79B}
}

@ARTICLE{Bromm-Loeb-06,
       author = {{Bromm}, Volker and {Loeb}, Abraham},
        title = "{High-Redshift Gamma-Ray Bursts from Population III Progenitors}",
      journal = {\apj},
         year = 2006,
        month = may,
       volume = {642},
       number = {1},
        pages = {382-388},
          doi = {10.1086/500799},
archivePrefix = {arXiv},
       eprint = {astro-ph/0509303},
 primaryClass = {astro-ph},
       adsurl = {https://ui.adsabs.harvard.edu/abs/2006ApJ...642..382B}
}

@ARTICLE{Bromberg+11,
       author = {{Bromberg}, Omer and {Nakar}, Ehud and {Piran}, Tsvi and {Sari}, Re'em},
        title = "{The Propagation of Relativistic Jets in External Media}",
      journal = {\apj},
         year = 2011,
        month = oct,
       volume = {740},
       number = {2},
          eid = {100},
        pages = {100},
          doi = {10.1088/0004-637X/740/2/100},
archivePrefix = {arXiv},
       eprint = {1107.1326},
 primaryClass = {astro-ph.HE},
       adsurl = {https://ui.adsabs.harvard.edu/abs/2011ApJ...740..100B}
}

@ARTICLE{Bromberg+15,
       author = {{Bromberg}, Omer and {Granot}, Jonathan and {Piran}, Tsvi},
        title = "{On the composition of GRBs' Collapsar jets}",
      journal = {\mnras},
         year = 2015,
        month = jun,
       volume = {450},
       number = {1},
        pages = {1077-1084},
          doi = {10.1093/mnras/stv226},
archivePrefix = {arXiv},
       eprint = {1407.0123},
 primaryClass = {astro-ph.HE},
       adsurl = {https://ui.adsabs.harvard.edu/abs/2015MNRAS.450.1077B}
}

@ARTICLE{Bromberg-Tchekhovskoy-16,
       author = {{Bromberg}, Omer and {Tchekhovskoy}, Alexander},
        title = "{Relativistic MHD simulations of core-collapse GRB jets: 3D instabilities and magnetic dissipation}",
      journal = {\mnras},
         year = 2016,
        month = feb,
       volume = {456},
       number = {2},
        pages = {1739-1760},
          doi = {10.1093/mnras/stv2591},
archivePrefix = {arXiv},
       eprint = {1508.02721},
 primaryClass = {astro-ph.HE},
       adsurl = {https://ui.adsabs.harvard.edu/abs/2016MNRAS.456.1739B}
}

@ARTICLE{Bromm-13,
       author = {{Bromm}, Volker},
        title = "{Formation of the first stars}",
      journal = {Reports on Progress in Physics},
         year = 2013,
        month = nov,
       volume = {76},
       number = {11},
          eid = {112901},
        pages = {112901},
          doi = {10.1088/0034-4885/76/11/112901},
archivePrefix = {arXiv},
       eprint = {1305.5178},
 primaryClass = {astro-ph.CO},
       adsurl = {https://ui.adsabs.harvard.edu/abs/2013RPPh...76k2901B}
}

@ARTICLE{Brott+11,
       author = {{Brott}, I. and {de Mink}, S.~E. and {Cantiello}, M. and {Langer}, N. and {de Koter}, A. and {Evans}, C.~J. and {Hunter}, I. and {Trundle}, C. and {Vink}, J.~S.},
        title = "{Rotating massive main-sequence stars. I. Grids of evolutionary models and isochrones}",
      journal = {\aap},
         year = 2011,
        month = jun,
       volume = {530},
          eid = {A115},
        pages = {A115},
          doi = {10.1051/0004-6361/201016113},
archivePrefix = {arXiv},
       eprint = {1102.0530},
 primaryClass = {astro-ph.SR},
       adsurl = {https://ui.adsabs.harvard.edu/abs/2011A&A...530A.115B}
}

@ARTICLE{Busmann+2025,
       author = {{Busmann}, Malte and {O'Connor}, Brendan and {Sommer}, Julian and {Gruen}, Daniel and {Beniamini}, Paz and {Gill}, Ramandeep and {Moss}, Michael J. and {Palmese}, Antonella and {Riffeser}, Arno and {Yang}, Yu-Han and {Troja}, Eleonora and {Dichiara}, Simone and {Ricci}, Roberto and {Klingler}, Noel and {G{\"o}ssl}, Claus and {Hu}, Lei and {Rau}, Arne and {Ries}, Christoph and {Ryan}, Geoffrey and {Schmidt}, Michael and {Yadav}, Muskan and {Zeimann}, Gregory R.},
        title = "{The curious case of EP241021a: Unraveling the mystery of its exceptional rebrightening}",
      journal = {arXiv e-prints},
         year = 2025,
        month = mar,
          eid = {arXiv:2503.14588},
        pages = {arXiv:2503.14588},
          doi = {10.48550/arXiv.2503.14588},
archivePrefix = {arXiv},
       eprint = {2503.14588},
 primaryClass = {astro-ph.HE},
       adsurl = {https://ui.adsabs.harvard.edu/abs/2025arXiv250314588B}
}

@ARTICLE{BN19,
       author = {{Beniamini}, Paz and {Nakar}, Ehud},
        title = "{Observational constraints on the structure of gamma-ray burst jets}",
      journal = {\mnras},
         year = 2019,
        month = feb,
       volume = {482},
       number = {4},
        pages = {5430-5440},
          doi = {10.1093/mnras/sty3110},
archivePrefix = {arXiv},
       eprint = {1808.07493},
 primaryClass = {astro-ph.HE},
       adsurl = {https://ui.adsabs.harvard.edu/abs/2019MNRAS.482.5430B}
}

@ARTICLE{Campisi+11,
       author = {{Campisi}, M.~A. and {Maio}, U. and {Salvaterra}, R. and {Ciardi}, B.},
        title = "{Population III stars and the long gamma-ray burst rate}",
      journal = {\mnras},
         year = 2011,
        month = oct,
       volume = {416},
       number = {4},
        pages = {2760-2767},
          doi = {10.1111/j.1365-2966.2011.19238.x},
archivePrefix = {arXiv},
       eprint = {1106.1439},
 primaryClass = {astro-ph.CO},
       adsurl = {https://ui.adsabs.harvard.edu/abs/2011MNRAS.416.2760C}
}

@ARTICLE{Cano+17,
       author = {{Cano}, Zach and {Wang}, Shan-Qin and {Dai}, Zi-Gao and {Wu}, Xue-Feng},
        title = "{The Observer's Guide to the Gamma-Ray Burst Supernova Connection}",
      journal = {Advances in Astronomy},
         year = 2017,
        month = jan,
       volume = {2017},
          eid = {8929054},
        pages = {8929054},
          doi = {10.1155/2017/8929054},
archivePrefix = {arXiv},
       eprint = {1604.03549},
 primaryClass = {astro-ph.HE},
       adsurl = {https://ui.adsabs.harvard.edu/abs/2017AdAst2017E...5C}
}

@ARTICLE{Cantiello+07,
       author = {{Cantiello}, M. and {Yoon}, S.-C. and {Langer}, N. and {Livio}, M.},
        title = "{Binary star progenitors of long gamma-ray bursts}",
      journal = {\aap},
         year = 2007,
        month = apr,
       volume = {465},
       number = {2},
        pages = {L29-L33},
          doi = {10.1051/0004-6361:20077115},
archivePrefix = {arXiv},
       eprint = {astro-ph/0702540},
 primaryClass = {astro-ph},
       adsurl = {https://ui.adsabs.harvard.edu/abs/2007A&A...465L..29C}
}

@ARTICLE{Chabrier-03,
       author = {{Chabrier}, Gilles},
        title = "{Galactic Stellar and Substellar Initial Mass Function}",
      journal = {\pasp},
         year = 2003,
        month = jul,
       volume = {115},
       number = {809},
        pages = {763-795},
          doi = {10.1086/376392},
archivePrefix = {arXiv},
       eprint = {astro-ph/0304382},
 primaryClass = {astro-ph},
       adsurl = {https://ui.adsabs.harvard.edu/abs/2003PASP..115..763C}
}

@ARTICLE{Chatzopoulos+12,
       author = {{Chatzopoulos}, E. and {Wheeler}, J. Craig},
        title = "{Effects of Rotation on the Minimum Mass of Primordial Progenitors of Pair-instability Supernovae}",
      journal = {\apj},
         year = 2012,
        month = mar,
       volume = {748},
       number = {1},
          eid = {42},
        pages = {42},
          doi = {10.1088/0004-637X/748/1/42},
archivePrefix = {arXiv},
       eprint = {1201.1328},
 primaryClass = {astro-ph.HE},
       adsurl = {https://ui.adsabs.harvard.edu/abs/2012ApJ...748...42C}
}

@ARTICLE{Ciardi-Loeb-00,
       author = {{Ciardi}, Benedetta and {Loeb}, Abraham},
        title = "{Expected Number and Flux Distribution of Gamma-Ray Burst Afterglows with High Redshifts}",
      journal = {\apj},
         year = 2000,
        month = sep,
       volume = {540},
       number = {2},
        pages = {687-696},
          doi = {10.1086/309384},
archivePrefix = {arXiv},
       eprint = {astro-ph/0002412},
 primaryClass = {astro-ph},
       adsurl = {https://ui.adsabs.harvard.edu/abs/2000ApJ...540..687C}
}

@ARTICLE{Coleman-Fernandez-24,
       author = {{Dean}, Coleman and {Fern{\'a}ndez}, Rodrigo},
        title = "{Collapsar disk outflows: Viscous hydrodynamic evolution in axisymmetry}",
      journal = {\prd},
         year = 2024,
        month = apr,
       volume = {109},
       number = {8},
          eid = {083010},
        pages = {083010},
          doi = {10.1103/PhysRevD.109.083010},
archivePrefix = {arXiv},
       eprint = {2403.08877},
 primaryClass = {astro-ph.HE},
       adsurl = {https://ui.adsabs.harvard.edu/abs/2024PhRvD.109h3010D}
}

@ARTICLE{Cuchiara+11,
       author = {{Cucchiara}, A. and {Levan}, A.~J. and {Fox}, D.~B. and {Tanvir}, N.~R. and {Ukwatta}, T.~N. and {Berger}, E. and {Kr{\"u}hler}, T. and {K{\"u}pc{\"u} Yolda{\c{s}}}, A. and {Wu}, X.~F. and {Toma}, K. and {Greiner}, J. and {Olivares}, F.~E. and {Rowlinson}, A. and {Amati}, L. and {Sakamoto}, T. and {Roth}, K. and {Stephens}, A. and {Fritz}, Alexander and {Fynbo}, J.~P.~U. and {Hjorth}, J. and {Malesani}, D. and {Jakobsson}, P. and {Wiersema}, K. and {O'Brien}, P.~T. and {Soderberg}, A.~M. and {Foley}, R.~J. and {Fruchter}, A.~S. and {Rhoads}, J. and {Rutledge}, R.~E. and {Schmidt}, B.~P. and {Dopita}, M.~A. and {Podsiadlowski}, P. and {Willingale}, R. and {Wolf}, C. and {Kulkarni}, S.~R. and {D'Avanzo}, P.},
        title = "{A Photometric Redshift of z \raisebox{-0.5ex}\textasciitilde 9.4 for GRB 090429B}",
      journal = {\apj},
         year = 2011,
        month = jul,
       volume = {736},
       number = {1},
          eid = {7},
        pages = {7},
          doi = {10.1088/0004-637X/736/1/7},
archivePrefix = {arXiv},
       eprint = {1105.4915},
 primaryClass = {astro-ph.CO},
       adsurl = {https://ui.adsabs.harvard.edu/abs/2011ApJ...736....7C}
}

@ARTICLE{Dejager+18,
       author = {{de Jager}, C. and {Nieuwenhuijzen}, H. and {van der Hucht}, K.~A.},
        title = "{Mass loss rates in the Hertzsprung-Russell diagram.}",
      journal = {\aaps},
         year = 1988,
        month = feb,
       volume = {72},
        pages = {259-289},
       adsurl = {https://ui.adsabs.harvard.edu/abs/1988A&AS...72..259D}
}

@ARTICLE{DeSouza+11,
       author = {{de Souza}, R.~S. and {Yoshida}, N. and {Ioka}, K.},
        title = "{Populations III.1 and III.2 gamma-ray bursts: constraints on the event rate for future radio and X-ray surveys}",
      journal = {\aap},
         year = 2011,
        month = sep,
       volume = {533},
          eid = {A32},
        pages = {A32},
          doi = {10.1051/0004-6361/201117242},
archivePrefix = {arXiv},
       eprint = {1105.2395},
 primaryClass = {astro-ph.CO},
       adsurl = {https://ui.adsabs.harvard.edu/abs/2011A&A...533A..32D}
}

@ARTICLE{Ekstrom+08,
       author = {{Ekstr{\"o}m}, S. and {Meynet}, G. and {Chiappini}, C. and {Hirschi}, R. and {Maeder}, A.},
        title = "{Effects of rotation on the evolution of primordial stars}",
      journal = {\aap},
         year = 2008,
        month = oct,
       volume = {489},
       number = {2},
        pages = {685-698},
          doi = {10.1051/0004-6361:200809633},
archivePrefix = {arXiv},
       eprint = {0807.0573},
 primaryClass = {astro-ph},
       adsurl = {https://ui.adsabs.harvard.edu/abs/2008A&A...489..685E}
}

@ARTICLE{Fausey+23,
       author = {{Fausey}, H.~M. and {van der Horst}, A.~J. and {White}, N.~E. and {Seiffert}, M. and {Willems}, P. and {Young}, E.~T. and {Kann}, D.~A. and {Ghirlanda}, G. and {Salvaterra}, R. and {Tanvir}, N.~R. and {Levan}, A. and {Moss}, M. and {Chang}, T. -C. and {Fruchter}, A. and {Guiriec}, S. and {Hartmann}, D.~H. and {Kouveliotou}, C. and {Granot}, J. and {Lidz}, A.},
        title = "{Photometric redshift estimation for gamma-ray bursts from the early Universe}",
      journal = {\mnras},
         year = 2023,
        month = dec,
       volume = {526},
       number = {3},
        pages = {4599-4612},
          doi = {10.1093/mnras/stad2996},
archivePrefix = {arXiv},
       eprint = {2310.03093},
 primaryClass = {astro-ph.IM},
       adsurl = {https://ui.adsabs.harvard.edu/abs/2023MNRAS.526.4599F}
}

@ARTICLE{Frail+01,
       author = {{Frail}, D.~A. and {Kulkarni}, S.~R. and {Sari}, R. and {Djorgovski}, S.~G. and {Bloom}, J.~S. and {Galama}, T.~J. and {Reichart}, D.~E. and {Berger}, E. and {Harrison}, F.~A. and {Price}, P.~A. and {Yost}, S.~A. and {Diercks}, A. and {Goodrich}, R.~W. and {Chaffee}, F.},
        title = "{Beaming in Gamma-Ray Bursts: Evidence for a Standard Energy Reservoir}",
      journal = {\apjl},
         year = 2001,
        month = nov,
       volume = {562},
       number = {1},
        pages = {L55-L58},
          doi = {10.1086/338119},
archivePrefix = {arXiv},
       eprint = {astro-ph/0102282},
 primaryClass = {astro-ph},
       adsurl = {https://ui.adsabs.harvard.edu/abs/2001ApJ...562L..55F}
}

@ARTICLE{Fryer-99,
       author = {{Fryer}, Chris L.},
        title = "{Mass Limits For Black Hole Formation}",
      journal = {\apj},
         year = 1999,
        month = sep,
       volume = {522},
       number = {1},
        pages = {413-418},
          doi = {10.1086/307647},
archivePrefix = {arXiv},
       eprint = {astro-ph/9902315},
 primaryClass = {astro-ph},
       adsurl = {https://ui.adsabs.harvard.edu/abs/1999ApJ...522..413F}
}

@ARTICLE{Fryer+01,
       author = {{Fryer}, C.~L. and {Woosley}, S.~E. and {Heger}, A.},
        title = "{Pair-Instability Supernovae, Gravity Waves, and Gamma-Ray Transients}",
      journal = {\apj},
         year = 2001,
        month = mar,
       volume = {550},
       number = {1},
        pages = {372-382},
          doi = {10.1086/319719},
archivePrefix = {arXiv},
       eprint = {astro-ph/0007176},
 primaryClass = {astro-ph},
       adsurl = {https://ui.adsabs.harvard.edu/abs/2001ApJ...550..372F}
}

@ARTICLE{Fryer+22,
       author = {{Fryer}, Chris L. and {Lien}, Amy Y. and {Fruchter}, Andrew and {Ghirlanda}, Giancarlo and {Hartmann}, Dieter and {Salvaterra}, Ruben and {Upton Sanderbeck}, Phoebe R. and {Johnson}, Jarrett L.},
        title = "{Properties of High-redshift Gamma-Ray Bursts}",
      journal = {\apj},
         year = 2022,
        month = apr,
       volume = {929},
       number = {2},
          eid = {111},
        pages = {111},
          doi = {10.3847/1538-4357/ac5d5c},
archivePrefix = {arXiv},
       eprint = {2112.00643},
 primaryClass = {astro-ph.HE},
       adsurl = {https://ui.adsabs.harvard.edu/abs/2022ApJ...929..111F}
}

@ARTICLE{Ghirlanda-Salvaterra-22,
       author = {{Ghirlanda}, Giancarlo and {Salvaterra}, Ruben},
        title = "{The Cosmic History of Long Gamma-Ray Bursts}",
      journal = {\apj},
         year = 2022,
        month = jun,
       volume = {932},
       number = {1},
          eid = {10},
        pages = {10},
          doi = {10.3847/1538-4357/ac6e43},
       adsurl = {https://ui.adsabs.harvard.edu/abs/2022ApJ...932...10G}
}

@ARTICLE{Ghodla-Eldridge-24,
       author = {{Ghodla}, Sohan and {Eldridge}, J.~J.},
        title = "{The effect of stellar rotation on black hole mass and spin}",
      journal = {\mnras},
         year = 2024,
        month = nov,
       volume = {534},
       number = {3},
        pages = {1868-1888},
          doi = {10.1093/mnras/stae2198},
archivePrefix = {arXiv},
       eprint = {2312.10400},
 primaryClass = {astro-ph.HE},
       adsurl = {https://ui.adsabs.harvard.edu/abs/2024MNRAS.534.1868G}
}

@ARTICLE{Glebbeek+09,
  author = {Glebbeek, E. and Gaburov, E. and de Mink, S.E. and Pols, O.R. and Portegies Zwart, S.F.},
  title = {The evolution of runaway stellar collision products},
  journal = {Astronomy and Astrophysics},
  volume = {497},
  pages = {255--270},
  year = {2009},
  doi = {10.1051/0004-6361/200810001}
}

@ARTICLE{Gnedin-Ostriker-97,
       author = {{Gnedin}, Nickolay Y. and {Ostriker}, Jeremiah P.},
        title = "{Reionization of the Universe and the Early Production of Metals}",
      journal = {\apj},
         year = 1997,
        month = sep,
       volume = {486},
       number = {2},
        pages = {581-598},
          doi = {10.1086/304548},
archivePrefix = {arXiv},
       eprint = {astro-ph/9612127},
 primaryClass = {astro-ph},
       adsurl = {https://ui.adsabs.harvard.edu/abs/1997ApJ...486..581G}
}

@ARTICLE{Gou+04,
       author = {{Gou}, L.~J. and {M{\'e}sz{\'a}ros}, P. and {Abel}, T. and {Zhang}, B.},
        title = "{Detectability of Long Gamma-Ray Burst Afterglows from Very High Redshifts}",
      journal = {\apj},
         year = 2004,
        month = apr,
       volume = {604},
       number = {2},
        pages = {508-520},
          doi = {10.1086/382061},
archivePrefix = {arXiv},
       eprint = {astro-ph/0307489},
 primaryClass = {astro-ph},
       adsurl = {https://ui.adsabs.harvard.edu/abs/2004ApJ...604..508G}
}

@ARTICLE{Gupta+25,
       author = {{Gupta}, Om and {Beniamini}, Paz and {Kumar}, Pawan and {Finkelstein}, Steven L.},
        title = "{The Cosmic Evolution of Fast Radio Bursts Inferred from the CHIME/FRB Baseband Catalog 1}",
      journal = {\apj},
         year = 2025,
        month = jun,
       volume = {986},
       number = {1},
          eid = {100},
        pages = {100},
          doi = {10.3847/1538-4357/add14c},
archivePrefix = {arXiv},
       eprint = {2501.09810},
 primaryClass = {astro-ph.HE},
       adsurl = {https://ui.adsabs.harvard.edu/abs/2025ApJ...986..100G}
}

@ARTICLE{Halevi+23,
       author = {{Halevi}, Goni and {Wu}, Belinda and {M{\"o}sta}, Philipp and {Gottlieb}, Ore and {Tchekhovskoy}, Alexander and {Aguilera-Dena}, David R.},
        title = "{Density Profiles of Collapsed Rotating Massive Stars Favor Long Gamma-Ray Bursts}",
      journal = {\apjl},
         year = 2023,
        month = feb,
       volume = {944},
       number = {2},
          eid = {L38},
        pages = {L38},
          doi = {10.3847/2041-8213/acb702},
archivePrefix = {arXiv},
       eprint = {2211.11781},
 primaryClass = {astro-ph.HE},
       adsurl = {https://ui.adsabs.harvard.edu/abs/2023ApJ...944L..38H}
}

@ARTICLE{Hartwig+22,
       author = {{Hartwig}, Tilman and {Magg}, Mattis and {Chen}, Li-Hsin and {Tarumi}, Yuta and {Bromm}, Volker and {Glover}, Simon C.~O. and {Ji}, Alexander P. and {Klessen}, Ralf S. and {Latif}, Muhammad A. and {Volonteri}, Marta and {Yoshida}, Naoki},
        title = "{Public Release of A-SLOTH: Ancient Stars and Local Observables by Tracing Halos}",
      journal = {\apj},
         year = 2022,
        month = sep,
       volume = {936},
       number = {1},
          eid = {45},
        pages = {45},
          doi = {10.3847/1538-4357/ac7150},
archivePrefix = {arXiv},
       eprint = {2206.00223},
 primaryClass = {astro-ph.GA},
       adsurl = {https://ui.adsabs.harvard.edu/abs/2022ApJ...936...45H}
}

@ARTICLE{Heger+00,
       author = {{Heger}, A. and {Langer}, N. and {Woosley}, S.~E.},
        title = "{Presupernova Evolution of Rotating Massive Stars. I. Numerical Method and Evolution of the Internal Stellar Structure}",
      journal = {\apj},
         year = 2000,
        month = jan,
       volume = {528},
       number = {1},
        pages = {368-396},
          doi = {10.1086/308158},
archivePrefix = {arXiv},
       eprint = {astro-ph/9904132},
 primaryClass = {astro-ph},
       adsurl = {https://ui.adsabs.harvard.edu/abs/2000ApJ...528..368H}
}

@ARTICLE{Heger-Woosley-02,
       author = {{Heger}, A. and {Woosley}, S.~E.},
        title = "{The Nucleosynthetic Signature of Population III}",
      journal = {\apj},
         year = 2002,
        month = mar,
       volume = {567},
       number = {1},
        pages = {532-543},
          doi = {10.1086/338487},
archivePrefix = {arXiv},
       eprint = {astro-ph/0107037},
 primaryClass = {astro-ph},
       adsurl = {https://ui.adsabs.harvard.edu/abs/2002ApJ...567..532H}
}

@ARTICLE{Heger+03,
       author = {{Heger}, A. and {Fryer}, C.~L. and {Woosley}, S.~E. and {Langer}, N. and {Hartmann}, D.~H.},
        title = "{How Massive Single Stars End Their Life}",
      journal = {\apj},
         year = 2003,
        month = jul,
       volume = {591},
       number = {1},
        pages = {288-300},
          doi = {10.1086/375341},
archivePrefix = {arXiv},
       eprint = {astro-ph/0212469},
 primaryClass = {astro-ph},
       adsurl = {https://ui.adsabs.harvard.edu/abs/2003ApJ...591..288H}
}

@ARTICLE{Hirano+14,
       author = {{Hirano}, Shingo and {Hosokawa}, Takashi and {Yoshida}, Naoki and {Umeda}, Hideyuki and {Omukai}, Kazuyuki and {Chiaki}, Gen and {Yorke}, Harold W.},
        title = "{One Hundred First Stars: Protostellar Evolution and the Final Masses}",
      journal = {\apj},
         year = 2014,
        month = feb,
       volume = {781},
       number = {2},
          eid = {60},
        pages = {60},
          doi = {10.1088/0004-637X/781/2/60},
archivePrefix = {arXiv},
       eprint = {1308.4456},
 primaryClass = {astro-ph.CO},
       adsurl = {https://ui.adsabs.harvard.edu/abs/2014ApJ...781...60H}
}

@ARTICLE{Hirano-Bromm-2018,
       author = {{Hirano}, Shingo and {Bromm}, Volker},
        title = "{Angular momentum transfer in primordial discs and the rotation of the first stars}",
      journal = {\mnras},
         year = 2018,
        month = may,
       volume = {476},
       number = {3},
        pages = {3964-3973},
          doi = {10.1093/mnras/sty487},
archivePrefix = {arXiv},
       eprint = {1802.07279},
 primaryClass = {astro-ph.GA},
       adsurl = {https://ui.adsabs.harvard.edu/abs/2018MNRAS.476.3964H}
}

@ARTICLE{Hosokawa+16,
       author = {{Hosokawa}, Takashi and {Hirano}, Shingo and {Kuiper}, Rolf and {Yorke}, Harold W. and {Omukai}, Kazuyuki and {Yoshida}, Naoki},
        title = "{Formation of Massive Primordial Stars: Intermittent UV Feedback with Episodic Mass Accretion}",
      journal = {\apj},
         year = 2016,
        month = jun,
       volume = {824},
       number = {2},
          eid = {119},
        pages = {119},
          doi = {10.3847/0004-637X/824/2/119},
archivePrefix = {arXiv},
       eprint = {1510.01407},
 primaryClass = {astro-ph.GA},
       adsurl = {https://ui.adsabs.harvard.edu/abs/2016ApJ...824..119H}
}

@ARTICLE{Inoue-04,
       author = {{Inoue}, Susumu},
        title = "{Probing the cosmic reionization history and local environment of gamma-ray bursts through radio dispersion}",
      journal = {\mnras},
         year = 2004,
        month = mar,
       volume = {348},
       number = {3},
        pages = {999-1008},
          doi = {10.1111/j.1365-2966.2004.07359.x},
archivePrefix = {arXiv},
       eprint = {astro-ph/0309364},
 primaryClass = {astro-ph},
       adsurl = {https://ui.adsabs.harvard.edu/abs/2004MNRAS.348..999I}
}

@ARTICLE{Ioka-03,
       author = {{Ioka}, Kunihito},
        title = "{The Cosmic Dispersion Measure from Gamma-Ray Burst Afterglows: Probing the Reionization History and the Burst Environment}",
      journal = {\apjl},
         year = 2003,
        month = dec,
       volume = {598},
       number = {2},
        pages = {L79-L82},
          doi = {10.1086/380598},
archivePrefix = {arXiv},
       eprint = {astro-ph/0309200},
 primaryClass = {astro-ph},
       adsurl = {https://ui.adsabs.harvard.edu/abs/2003ApJ...598L..79I}
}

@ARTICLE{Ioka-Meszaros-05,
       author = {{Ioka}, Kunihito and {M{\'e}sz{\'a}ros}, Peter},
        title = "{Radio Afterglows of Gamma-Ray Bursts and Hypernovae at High Redshift and Their Potential for 21 Centimeter Absorption Studies}",
      journal = {\apj},
         year = 2005,
        month = feb,
       volume = {619},
       number = {2},
        pages = {684-696},
          doi = {10.1086/426785},
archivePrefix = {arXiv},
       eprint = {astro-ph/0408487},
 primaryClass = {astro-ph},
       adsurl = {https://ui.adsabs.harvard.edu/abs/2005ApJ...619..684I}
}

@ARTICLE{Iwamoto+98,
       author = {{Iwamoto}, K. and {Mazzali}, P.~A. and {Nomoto}, K. and {Umeda}, H. and {Nakamura}, T. and {Patat}, F. and {Danziger}, I.~J. and {Young}, T.~R. and {Suzuki}, T. and {Shigeyama}, T. and {Augusteijn}, T. and {Doublier}, V. and {Gonzalez}, J. -F. and {Boehnhardt}, H. and {Brewer}, J. and {Hainaut}, O.~R. and {Lidman}, C. and {Leibundgut}, B. and {Cappellaro}, E. and {Turatto}, M. and {Galama}, T.~J. and {Vreeswijk}, P.~M. and {Kouveliotou}, C. and {van Paradijs}, J. and {Pian}, E. and {Palazzi}, E. and {Frontera}, F.},
        title = "{A hypernova model for the supernova associated with the {\ensuremath{\gamma}}-ray burst of 25 April 1998}",
      journal = {\nat},
         year = 1998,
        month = oct,
       volume = {395},
       number = {6703},
        pages = {672-674},
          doi = {10.1038/27155},
archivePrefix = {arXiv},
       eprint = {astro-ph/9806382},
 primaryClass = {astro-ph},
       adsurl = {https://ui.adsabs.harvard.edu/abs/1998Natur.395..672I}
}

@ARTICLE{Jacquemin-Ide+24,
       author = {{Jacquemin-Ide}, Jonatan and {Gottlieb}, Ore and {Lowell}, Beverly and {Tchekhovskoy}, Alexander},
        title = "{Collapsar Gamma-Ray Bursts Grind Their Black Hole Spins to a Halt}",
      journal = {\apj},
         year = 2024,
        month = feb,
       volume = {961},
       number = {2},
          eid = {212},
        pages = {212},
          doi = {10.3847/1538-4357/ad02f0},
archivePrefix = {arXiv},
       eprint = {2302.07281},
 primaryClass = {astro-ph.HE},
       adsurl = {https://ui.adsabs.harvard.edu/abs/2024ApJ...961..212J}
}

@ARTICLE{Jeena+23,
       author = {{Jeena}, S.~K. and {Banerjee}, Projjwal and {Chiaki}, Gen and {Heger}, Alexander},
        title = "{Rapidly rotating massive Population III stars: a solution for high carbon enrichment in CEMP-no stars}",
      journal = {\mnras},
         year = 2023,
        month = dec,
       volume = {526},
       number = {3},
        pages = {4467-4483},
          doi = {10.1093/mnras/stad3028},
archivePrefix = {arXiv},
       eprint = {2306.06433},
 primaryClass = {astro-ph.SR},
       adsurl = {https://ui.adsabs.harvard.edu/abs/2023MNRAS.526.4467J}
}

@ARTICLE{Jeon+21,
       author = {{Jeon}, Myoungwon and {Bromm}, Volker and {Besla}, Gurtina and {Yoon}, Jinmi and {Choi}, Yumi},
        title = "{The role of faint population III supernovae in forming CEMP stars in ultra-faint dwarf galaxies}",
      journal = {\mnras},
         year = 2021,
        month = mar,
       volume = {502},
       number = {1},
        pages = {1-14},
          doi = {10.1093/mnras/staa4017},
archivePrefix = {arXiv},
       eprint = {2012.10012},
 primaryClass = {astro-ph.GA},
       adsurl = {https://ui.adsabs.harvard.edu/abs/2021MNRAS.502....1J}
}

@ARTICLE{Just+22,
       author = {{Just}, O. and {Aloy}, M.~A. and {Obergaulinger}, M. and {Nagataki}, S.},
        title = "{r-process Viable Outflows are Suppressed in Global Alpha-viscosity Models of Collapsar Disks}",
      journal = {\apjl},
         year = 2022,
        month = aug,
       volume = {934},
       number = {2},
          eid = {L30},
        pages = {L30},
          doi = {10.3847/2041-8213/ac83a1},
archivePrefix = {arXiv},
       eprint = {2205.14158},
 primaryClass = {astro-ph.HE},
       adsurl = {https://ui.adsabs.harvard.edu/abs/2022ApJ...934L..30J}
}

@ARTICLE{Kinugawa+19,
       author = {{Kinugawa}, Tomoya and {Harikane}, Yuichi and {Asano}, Katsuaki},
        title = "{Long Gamma-Ray Burst Rate at Very High Redshift}",
      journal = {\apj},
         year = 2019,
        month = jun,
       volume = {878},
       number = {2},
          eid = {128},
        pages = {128},
          doi = {10.3847/1538-4357/ab2188},
archivePrefix = {arXiv},
       eprint = {1901.03516},
 primaryClass = {astro-ph.HE},
       adsurl = {https://ui.adsabs.harvard.edu/abs/2019ApJ...878..128K}
}

@ARTICLE{Klessen-Glover-23,
       author = {{Klessen}, Ralf S. and {Glover}, Simon C.~O.},
        title = "{The First Stars: Formation, Properties, and Impact}",
      journal = {\araa},
         year = 2023,
        month = aug,
       volume = {61},
        pages = {65-130},
          doi = {10.1146/annurev-astro-071221-053453},
archivePrefix = {arXiv},
       eprint = {2303.12500},
 primaryClass = {astro-ph.CO},
       adsurl = {https://ui.adsabs.harvard.edu/abs/2023ARA&A..61...65K}
}

@ARTICLE{Kumar-Zhang-15,
       author = {{Kumar}, Pawan and {Zhang}, Bing},
        title = "{The physics of gamma-ray bursts \& relativistic jets}",
      journal = {\physrep},
         year = 2015,
        month = feb,
       volume = {561},
        pages = {1-109},
          doi = {10.1016/j.physrep.2014.09.008},
archivePrefix = {arXiv},
       eprint = {1410.0679},
 primaryClass = {astro-ph.HE},
       adsurl = {https://ui.adsabs.harvard.edu/abs/2015PhR...561....1K}
}

@ARTICLE{Khokhlov+99,
       author = {{Khokhlov}, A.~M. and {H{\"o}flich}, P.~A. and {Oran}, E.~S. and {Wheeler}, J.~C. and {Wang}, L. and {Chtchelkanova}, A. Yu.},
        title = "{Jet-induced Explosions of Core Collapse Supernovae}",
      journal = {\apjl},
         year = 1999,
        month = oct,
       volume = {524},
       number = {2},
        pages = {L107-L110},
          doi = {10.1086/312305},
archivePrefix = {arXiv},
       eprint = {astro-ph/9904419},
 primaryClass = {astro-ph},
       adsurl = {https://ui.adsabs.harvard.edu/abs/1999ApJ...524L.107K}
}

@ARTICLE{Lamb-Reichart-00,
       author = {{Lamb}, Donald Q. and {Reichart}, Daniel E.},
        title = "{Gamma-Ray Bursts as a Probe of the Very High Redshift Universe}",
      journal = {\apj},
         year = 2000,
        month = jun,
       volume = {536},
       number = {1},
        pages = {1-18},
          doi = {10.1086/308918},
archivePrefix = {arXiv},
       eprint = {astro-ph/9909002},
 primaryClass = {astro-ph},
       adsurl = {https://ui.adsabs.harvard.edu/abs/2000ApJ...536....1L}
}

@ARTICLE{Langer+85,
       author = {{Langer}, N. and {El Eid}, M.~F. and {Fricke}, K.~J.},
        title = "{Evolution of massive stars with semiconvective diffusion}",
      journal = {\aap},
         year = 1985,
        month = apr,
       volume = {145},
       number = {1},
        pages = {179-191},
       adsurl = {https://ui.adsabs.harvard.edu/abs/1985A&A...145..179L}
}

@ARTICLE{Langer-92,
       author = {{Langer}, N.},
        title = "{Helium enrichment in massive early type stars.}",
      journal = {\aap},
         year = 1992,
        month = nov,
       volume = {265},
        pages = {L17-L20},
       adsurl = {https://ui.adsabs.harvard.edu/abs/1992A&A...265L..17L}
}

@ARTICLE{Langer-98,
       author = {{Langer}, N.},
        title = "{Coupled mass and angular momentum loss of massive main sequence stars}",
      journal = {\aap},
         year = 1998,
        month = jan,
       volume = {329},
        pages = {551-558},
       adsurl = {https://ui.adsabs.harvard.edu/abs/1998A&A...329..551L}
}

@ARTICLE{Langer-12,
       author = {{Langer}, N.},
        title = "{Presupernova Evolution of Massive Single and Binary Stars}",
      journal = {\araa},
         year = 2012,
        month = sep,
       volume = {50},
        pages = {107-164},
          doi = {10.1146/annurev-astro-081811-125534},
archivePrefix = {arXiv},
       eprint = {1206.5443},
 primaryClass = {astro-ph.SR},
       adsurl = {https://ui.adsabs.harvard.edu/abs/2012ARA&A..50..107L}
}

@ARTICLE{Lazar-Bromm-22,
       author = {{Lazar}, Alexandres and {Bromm}, Volker},
        title = "{Probing the initial mass function of the first stars with transients}",
      journal = {\mnras},
         year = 2022,
        month = apr,
       volume = {511},
       number = {2},
        pages = {2505-2514},
          doi = {10.1093/mnras/stac176},
archivePrefix = {arXiv},
       eprint = {2110.11956},
 primaryClass = {astro-ph.HE},
       adsurl = {https://ui.adsabs.harvard.edu/abs/2022MNRAS.511.2505L}
}

@ARTICLE{Lei+17,
       author = {{Lei}, Wei-Hua and {Zhang}, Bing and {Wu}, Xue-Feng and {Liang}, En-Wei},
        title = "{Hyperaccreting Black Hole as Gamma-Ray Burst Central Engine. II. Temporal Evolution of the Central Engine Parameters during the Prompt and Afterglow Phases}",
      journal = {\apj},
         year = 2017,
        month = nov,
       volume = {849},
       number = {1},
          eid = {47},
        pages = {47},
          doi = {10.3847/1538-4357/aa9074},
archivePrefix = {arXiv},
       eprint = {1708.05043},
 primaryClass = {astro-ph.HE},
       adsurl = {https://ui.adsabs.harvard.edu/abs/2017ApJ...849...47L}
}

@ARTICLE{Leng-Giannios-14,
       author = {{Leng}, M. and {Giannios}, D.},
        title = "{Testing the neutrino annihilation model for launching GRB jets.}",
      journal = {\mnras},
         year = 2014,
        month = nov,
       volume = {445},
        pages = {L1-L5},
          doi = {10.1093/mnrasl/slu122},
archivePrefix = {arXiv},
       eprint = {1408.4509},
 primaryClass = {astro-ph.HE},
       adsurl = {https://ui.adsabs.harvard.edu/abs/2014MNRAS.445L...1L}
}

@ARTICLE{Li-08,
       author = {{Li}, Li-Xin},
        title = "{Star formation history up to z = 7.4: implications for gamma-ray bursts and cosmic metallicity evolution}",
      journal = {\mnras},
         year = 2008,
        month = aug,
       volume = {388},
       number = {4},
        pages = {1487-1500},
          doi = {10.1111/j.1365-2966.2008.13488.x},
archivePrefix = {arXiv},
       eprint = {0710.3587},
 primaryClass = {astro-ph},
       adsurl = {https://ui.adsabs.harvard.edu/abs/2008MNRAS.388.1487L}
}

@ARTICLE{Liu+15,
       author = {{Liu}, Tong and {Hou}, Shu-Jin and {Xue}, Li and {Gu}, Wei-Min},
        title = "{Jet Luminosity of Gamma-ray Bursts: the Blandford-Znajek Mechanism versus the Neutrino Annihilation Process}",
      journal = {\apjs},
         year = 2015,
        month = may,
       volume = {218},
       number = {1},
          eid = {12},
        pages = {12},
          doi = {10.1088/0067-0049/218/1/12},
archivePrefix = {arXiv},
       eprint = {1504.04067},
 primaryClass = {astro-ph.HE},
       adsurl = {https://ui.adsabs.harvard.edu/abs/2015ApJS..218...12L}
}

@ARTICLE{Liu-Bromm-20,
       author = {{Liu}, Boyuan and {Bromm}, Volker},
        title = "{When did Population III star formation end?}",
      journal = {\mnras},
         year = 2020,
        month = sep,
       volume = {497},
       number = {3},
        pages = {2839-2854},
          doi = {10.1093/mnras/staa2143},
archivePrefix = {arXiv},
       eprint = {2006.15260},
 primaryClass = {astro-ph.GA},
       adsurl = {https://ui.adsabs.harvard.edu/abs/2020MNRAS.497.2839L}
}

@ARTICLE{Lloyd-Ronning+02,
       author = {{Lloyd-Ronning}, Nicole M. and {Fryer}, Chris L. and {Ramirez-Ruiz}, Enrico},
        title = "{Cosmological Aspects of Gamma-Ray Bursts: Luminosity Evolution and an Estimate of the Star Formation Rate at High Redshifts}",
      journal = {\apj},
         year = 2002,
        month = aug,
       volume = {574},
       number = {2},
        pages = {554-565},
          doi = {10.1086/341059},
archivePrefix = {arXiv},
       eprint = {astro-ph/0108200},
 primaryClass = {astro-ph},
       adsurl = {https://ui.adsabs.harvard.edu/abs/2002ApJ...574..554L}
}

@ARTICLE{Lloyd-Ronning+20,
       author = {{Lloyd-Ronning}, Nicole M. and {Johnson}, Jarrett L. and {Aykutalp}, Aycin},
        title = "{The consequences of gamma-ray burst jet opening angle evolution on the inferred star formation rate}",
      journal = {\mnras},
         year = 2020,
        month = nov,
       volume = {498},
       number = {4},
        pages = {5041-5047},
          doi = {10.1093/mnras/staa2787},
archivePrefix = {arXiv},
       eprint = {2006.00022},
 primaryClass = {astro-ph.HE},
       adsurl = {https://ui.adsabs.harvard.edu/abs/2020MNRAS.498.5041L}
}

@ARTICLE{Lloyd-Ronning-22,
       author = {{Lloyd-Ronning}, Nicole},
        title = "{Radio-loud versus Radio-quiet Gamma-Ray Bursts: The Role of Binary Progenitors}",
      journal = {\apj},
         year = 2022,
        month = apr,
       volume = {928},
       number = {2},
          eid = {104},
        pages = {104},
          doi = {10.3847/1538-4357/ac54b3},
archivePrefix = {arXiv},
       eprint = {2109.14122},
 primaryClass = {astro-ph.HE},
       adsurl = {https://ui.adsabs.harvard.edu/abs/2022ApJ...928..104L}
}

@ARTICLE{Lowell+24,
       author = {{Lowell}, Beverly and {Jacquemin-Ide}, Jonatan and {Tchekhovskoy}, Alexander and {Duncan}, Alex},
        title = "{Rapid Black Hole Spin-down by Thick Magnetically Arrested Disks}",
      journal = {\apj},
         year = 2024,
        month = jan,
       volume = {960},
       number = {1},
          eid = {82},
        pages = {82},
          doi = {10.3847/1538-4357/ad09af},
archivePrefix = {arXiv},
       eprint = {2302.01351},
 primaryClass = {astro-ph.HE},
       adsurl = {https://ui.adsabs.harvard.edu/abs/2024ApJ...960...82L}
}

@ARTICLE{MacFadyen-Woosley-99,
       author = {{MacFadyen}, A.~I. and {Woosley}, S.~E.},
        title = "{Collapsars: Gamma-Ray Bursts and Explosions in ``Failed Supernovae''}",
      journal = {\apj},
         year = 1999,
        month = oct,
       volume = {524},
       number = {1},
        pages = {262-289},
          doi = {10.1086/307790},
archivePrefix = {arXiv},
       eprint = {astro-ph/9810274},
 primaryClass = {astro-ph},
       adsurl = {https://ui.adsabs.harvard.edu/abs/1999ApJ...524..262M}
}

@ARTICLE{Madau-Dickinson-14,
       author = {{Madau}, Piero and {Dickinson}, Mark},
        title = "{Cosmic Star-Formation History}",
      journal = {\araa},
         year = 2014,
        month = aug,
       volume = {52},
        pages = {415-486},
          doi = {10.1146/annurev-astro-081811-125615},
archivePrefix = {arXiv},
       eprint = {1403.0007},
 primaryClass = {astro-ph.CO},
       adsurl = {https://ui.adsabs.harvard.edu/abs/2014ARA&A..52..415M}
}

@ARTICLE{Maeder-87,
       author = {{Maeder}, A.},
        title = "{Evidences for a bifurcation in massive star evolution. The ON-blue stragglers.}",
      journal = {\aap},
         year = 1987,
        month = may,
       volume = {178},
        pages = {159-169},
       adsurl = {https://ui.adsabs.harvard.edu/abs/1987A&A...178..159M}
}

@ARTICLE{Maeder-99,
       author = {{Maeder}, Andr{\'e}},
        title = "{Stellar evolution with rotation IV: von Zeipel's theorem and anisotropic losses of mass and angular momentum}",
      journal = {\aap},
         year = 1999,
        month = jul,
       volume = {347},
        pages = {185-193},
       adsurl = {https://ui.adsabs.harvard.edu/abs/1999A&A...347..185M}
}

@ARTICLE{Maeder-Meynet-00,
       author = {{Maeder}, Andr{\'e} and {Meynet}, Georges},
        title = "{The Evolution of Rotating Stars}",
      journal = {\araa},
         year = 2000,
        month = jan,
       volume = {38},
        pages = {143-190},
          doi = {10.1146/annurev.astro.38.1.143},
archivePrefix = {arXiv},
       eprint = {astro-ph/0004204},
 primaryClass = {astro-ph},
       adsurl = {https://ui.adsabs.harvard.edu/abs/2000ARA&A..38..143M}
}

@ARTICLE{Mandel+16,
       author = {{Mandel}, Ilya and {de Mink}, Selma E.},
        title = "{Merging binary black holes formed through chemically homogeneous evolution in short-period stellar binaries}",
      journal = {\mnras},
         year = 2016,
        month = may,
       volume = {458},
       number = {3},
        pages = {2634-2647},
          doi = {10.1093/mnras/stw379},
archivePrefix = {arXiv},
       eprint = {1601.00007},
 primaryClass = {astro-ph.HE},
       adsurl = {https://ui.adsabs.harvard.edu/abs/2016MNRAS.458.2634M}
}

@ARTICLE{Marchant-Bodensteiner-24,
       author = {{Marchant}, Pablo and {Bodensteiner}, Julia},
        title = "{The Evolution of Massive Binary Stars}",
      journal = {\araa},
         year = 2024,
        month = sep,
       volume = {62},
       number = {1},
        pages = {21-61},
          doi = {10.1146/annurev-astro-052722-105936},
archivePrefix = {arXiv},
       eprint = {2311.01865},
 primaryClass = {astro-ph.SR},
       adsurl = {https://ui.adsabs.harvard.edu/abs/2024ARA&A..62...21M}
}

@ARTICLE{Margalit-Metzger17,
       author = {{Margalit}, Ben and {Metzger}, Brian D.},
        title = "{Constraining the Maximum Mass of Neutron Stars from Multi-messenger Observations of GW170817}",
      journal = {\apjl},
         year = 2017,
        month = dec,
       volume = {850},
       number = {2},
          eid = {L19},
        pages = {L19},
          doi = {10.3847/2041-8213/aa991c},
archivePrefix = {arXiv},
       eprint = {1710.05938},
 primaryClass = {astro-ph.HE},
       adsurl = {https://ui.adsabs.harvard.edu/abs/2017ApJ...850L..19M}
}

@ARTICLE{Marigo+01,
       author = {{Marigo}, P. and {Girardi}, L. and {Chiosi}, C. and {Wood}, P.~R.},
        title = "{Zero-metallicity stars. I. Evolution at constant mass}",
      journal = {\aap},
         year = 2001,
        month = may,
       volume = {371},
        pages = {152-173},
          doi = {10.1051/0004-6361:20010309},
archivePrefix = {arXiv},
       eprint = {astro-ph/0102253},
 primaryClass = {astro-ph},
       adsurl = {https://ui.adsabs.harvard.edu/abs/2001A&A...371..152M}
}

@ARTICLE{Matzner-03,
       author = {{Matzner}, Christopher D.},
        title = "{Supernova hosts for gamma-ray burst jets: dynamical constraints}",
      journal = {\mnras},
         year = 2003,
        month = oct,
       volume = {345},
       number = {2},
        pages = {575-589},
          doi = {10.1046/j.1365-8711.2003.06969.x},
archivePrefix = {arXiv},
       eprint = {astro-ph/0203085},
 primaryClass = {astro-ph},
       adsurl = {https://ui.adsabs.harvard.edu/abs/2003MNRAS.345..575M}
}

@ARTICLE{McKinney+12,
       author = {{McKinney}, Jonathan C. and {Tchekhovskoy}, Alexander and {Blandford}, Roger D.},
        title = "{General relativistic magnetohydrodynamic simulations of magnetically choked accretion flows around black holes}",
      journal = {\mnras},
         year = 2012,
        month = jul,
       volume = {423},
       number = {4},
        pages = {3083-3117},
          doi = {10.1111/j.1365-2966.2012.21074.x},
archivePrefix = {arXiv},
       eprint = {1201.4163},
 primaryClass = {astro-ph.HE},
       adsurl = {https://ui.adsabs.harvard.edu/abs/2012MNRAS.423.3083M}
}

@ARTICLE{Meynet-Maeder-97,
       author = {{Meynet}, G. and {Maeder}, A.},
        title = "{Stellar evolution with rotation. I. The computational method and the inhibiting effect of the {\ensuremath{\mu}}-gradient.}",
      journal = {\aap},
         year = 1997,
        month = may,
       volume = {321},
        pages = {465-476},
       adsurl = {https://ui.adsabs.harvard.edu/abs/1997A&A...321..465M}
}

@ARTICLE{McKinney-05,
       author = {{McKinney}, Jonathan C.},
        title = "{Total and Jet Blandford-Znajek Power in the Presence of an Accretion Disk}",
      journal = {\apjl},
         year = 2005,
        month = sep,
       volume = {630},
       number = {1},
        pages = {L5-L8},
          doi = {10.1086/468184},
archivePrefix = {arXiv},
       eprint = {astro-ph/0506367},
 primaryClass = {astro-ph},
       adsurl = {https://ui.adsabs.harvard.edu/abs/2005ApJ...630L...5M}
}

@ARTICLE{Metzger+18,
       author = {{Metzger}, Brian D. and {Beniamini}, Paz and {Giannios}, Dimitrios},
        title = "{Effects of Fallback Accretion on Protomagnetar Outflows in Gamma-Ray Bursts and Superluminous Supernovae}",
      journal = {\apj},
         year = 2018,
        month = apr,
       volume = {857},
       number = {2},
          eid = {95},
        pages = {95},
          doi = {10.3847/1538-4357/aab70c},
archivePrefix = {arXiv},
       eprint = {1802.07750},
 primaryClass = {astro-ph.HE},
       adsurl = {https://ui.adsabs.harvard.edu/abs/2018ApJ...857...95M}
}

@ARTICLE{Meynet-Maeder-02,
       author = {{Meynet}, G. and {Maeder}, A.},
        title = "{Stellar evolution with rotation. VIII. Models at Z = 10$^{-5}$ and CNO yields for early galactic evolution}",
      journal = {\aap},
         year = 2002,
        month = aug,
       volume = {390},
        pages = {561-583},
          doi = {10.1051/0004-6361:20020755},
archivePrefix = {arXiv},
       eprint = {astro-ph/0205370},
 primaryClass = {astro-ph},
       adsurl = {https://ui.adsabs.harvard.edu/abs/2002A&A...390..561M}
}

@ARTICLE{Murphy+21,
       author = {{Murphy}, Laura J. and {Groh}, Jose H. and {Ekstr{\"o}m}, Sylvia and {Meynet}, Georges and {Pezzotti}, Camila and {Georgy}, Cyril and {Choplin}, Arthur and {Eggenberger}, Patrick and {Farrell}, Eoin and {Haemmerl{\'e}}, Lionel and {Hirschi}, Raphael and {Maeder}, Andr{\'e} and {Martinet}, S{\'e}bastien},
        title = "{Grids of stellar models with rotation - V. Models from 1.7 to 120 M$_{{\ensuremath{\odot}}}$ at zero metallicity}",
      journal = {\mnras},
         year = 2021,
        month = feb,
       volume = {501},
       number = {2},
        pages = {2745-2763},
          doi = {10.1093/mnras/staa3803},
archivePrefix = {arXiv},
       eprint = {2012.07420},
 primaryClass = {astro-ph.SR},
       adsurl = {https://ui.adsabs.harvard.edu/abs/2021MNRAS.501.2745M}
}

@ARTICLE{Narayan-Yi-94,
       author = {{Narayan}, Ramesh and {Yi}, Insu},
        title = "{Advection-dominated Accretion: A Self-similar Solution}",
      journal = {\apjl},
         year = 1994,
        month = jun,
       volume = {428},
        pages = {L13},
          doi = {10.1086/187381},
archivePrefix = {arXiv},
       eprint = {astro-ph/9403052},
 primaryClass = {astro-ph},
       adsurl = {https://ui.adsabs.harvard.edu/abs/1994ApJ...428L..13N}
}

@ARTICLE{Narayan+01,
       author = {{Narayan}, Ramesh and {Piran}, Tsvi and {Kumar}, Pawan},
        title = "{Accretion Models of Gamma-Ray Bursts}",
      journal = {\apj},
         year = 2001,
        month = aug,
       volume = {557},
       number = {2},
        pages = {949-957},
          doi = {10.1086/322267},
archivePrefix = {arXiv},
       eprint = {astro-ph/0103360},
 primaryClass = {astro-ph},
       adsurl = {https://ui.adsabs.harvard.edu/abs/2001ApJ...557..949N}
}

@ARTICLE{Narayan+03,
       author = {{Narayan}, Ramesh and {Igumenshchev}, Igor V. and {Abramowicz}, Marek A.},
        title = "{Magnetically Arrested Disk: an Energetically Efficient Accretion Flow}",
      journal = {\pasj},
         year = 2003,
        month = dec,
       volume = {55},
        pages = {L69-L72},
          doi = {10.1093/pasj/55.6.L69},
archivePrefix = {arXiv},
       eprint = {astro-ph/0305029},
 primaryClass = {astro-ph},
       adsurl = {https://ui.adsabs.harvard.edu/abs/2003PASJ...55L..69N}
}

@ARTICLE{Narayan+12,
       author = {{Narayan}, Ramesh and {S{\k{a}}dowski}, Aleksander and {Penna}, Robert F. and {Kulkarni}, Akshay K.},
        title = "{GRMHD simulations of magnetized advection-dominated accretion on a non-spinning black hole: role of outflows}",
      journal = {\mnras},
         year = 2012,
        month = nov,
       volume = {426},
       number = {4},
        pages = {3241-3259},
          doi = {10.1111/j.1365-2966.2012.22002.x},
archivePrefix = {arXiv},
       eprint = {1206.1213},
 primaryClass = {astro-ph.HE},
       adsurl = {https://ui.adsabs.harvard.edu/abs/2012MNRAS.426.3241N}
}

@ARTICLE{Naoz-Bromberg-07,
       author = {{Naoz}, S. and {Bromberg}, O.},
        title = "{An observational limit on the earliest gamma-ray bursts}",
      journal = {\mnras},
         year = 2007,
        month = sep,
       volume = {380},
       number = {2},
        pages = {757-762},
          doi = {10.1111/j.1365-2966.2007.12110.x},
archivePrefix = {arXiv},
       eprint = {astro-ph/0702357},
 primaryClass = {astro-ph},
       adsurl = {https://ui.adsabs.harvard.edu/abs/2007MNRAS.380..757N}
}

@INPROCEEDINGS{Nomoto+03,
       author = {{Nomoto}, Ken'ichi and {Maeda}, Keiichi and {Umeda}, Hideyuki and {Ohkubo}, Takuya and {Deng}, Jingsong and {Mazzali}, Paolo},
        title = "{Hypernovae and their nucleosynthesis}",
    booktitle = {A Massive Star Odyssey: From Main Sequence to Supernova},
         year = 2003,
       editor = {{van der Hucht}, Karel and {Herrero}, Artemio and {Esteban}, C{\'e}sar},
       series = {IAU Symposium},
       volume = {212},
        month = jan,
        pages = {395},
          doi = {10.48550/arXiv.astro-ph/0209064},
archivePrefix = {arXiv},
       eprint = {astro-ph/0209064},
 primaryClass = {astro-ph},
       adsurl = {https://ui.adsabs.harvard.edu/abs/2003IAUS..212..395N}
}

@ARTICLE{Nugis-Lamers-00,
       author = {{Nugis}, T. and {Lamers}, H.~J.~G.~L.~M.},
        title = "{Mass-loss rates of Wolf-Rayet stars as a function of stellar parameters}",
      journal = {\aap},
         year = 2000,
        month = aug,
       volume = {360},
        pages = {227-244},
       adsurl = {https://ui.adsabs.harvard.edu/abs/2000A&A...360..227N}
}

@ARTICLE{OConnor+11,
       author = {{O'Connor}, Evan and {Ott}, Christian D.},
        title = "{Black Hole Formation in Failing Core-Collapse Supernovae}",
      journal = {\apj},
         year = 2011,
        month = apr,
       volume = {730},
       number = {2},
          eid = {70},
        pages = {70},
          doi = {10.1088/0004-637X/730/2/70},
archivePrefix = {arXiv},
       eprint = {1010.5550},
 primaryClass = {astro-ph.HE},
       adsurl = {https://ui.adsabs.harvard.edu/abs/2011ApJ...730...70O}
}

@ARTICLE{OConnor+24,
       author = {{O'Connor}, Brendan and {Beniamini}, Paz and {Gill}, Ramandeep},
        title = "{X-ray afterglow limits on the viewing angles of short gamma-ray bursts}",
      journal = {\mnras},
         year = 2024,
        month = sep,
       volume = {533},
       number = {2},
        pages = {1629-1648},
          doi = {10.1093/mnras/stae1941},
archivePrefix = {arXiv},
       eprint = {2406.05297},
 primaryClass = {astro-ph.HE},
       adsurl = {https://ui.adsabs.harvard.edu/abs/2024MNRAS.533.1629O}
}

@ARTICLE{Ostriker-Gnedin-96,
       author = {{Ostriker}, Jeremiah P. and {Gnedin}, Nickolay Y.},
        title = "{Reheating of the Universe and Population III}",
      journal = {\apjl},
         year = 1996,
        month = dec,
       volume = {472},
        pages = {L63},
          doi = {10.1086/310375},
archivePrefix = {arXiv},
       eprint = {astro-ph/9608047},
 primaryClass = {astro-ph},
       adsurl = {https://ui.adsabs.harvard.edu/abs/1996ApJ...472L..63O}
}

@ARTICLE{Paxton+11,
       author = {{Paxton}, Bill and {Bildsten}, Lars and {Dotter}, Aaron and {Herwig}, Falk and {Lesaffre}, Pierre and {Timmes}, Frank},
        title = "{Modules for Experiments in Stellar Astrophysics (MESA)}",
      journal = {\apjs},
         year = 2011,
        month = jan,
       volume = {192},
       number = {1},
          eid = {3},
        pages = {3},
          doi = {10.1088/0067-0049/192/1/3},
archivePrefix = {arXiv},
       eprint = {1009.1622},
 primaryClass = {astro-ph.SR},
       adsurl = {https://ui.adsabs.harvard.edu/abs/2011ApJS..192....3P}
}

@ARTICLE{Paxton+13,
       author = {{Paxton}, Bill and {Cantiello}, Matteo and {Arras}, Phil and {Bildsten}, Lars and {Brown}, Edward F. and {Dotter}, Aaron and {Mankovich}, Christopher and {Montgomery}, M.~H. and {Stello}, Dennis and {Timmes}, F.~X. and {Townsend}, Richard},
        title = "{Modules for Experiments in Stellar Astrophysics (MESA): Planets, Oscillations, Rotation, and Massive Stars}",
      journal = {\apjs},
         year = 2013,
        month = sep,
       volume = {208},
       number = {1},
          eid = {4},
        pages = {4},
          doi = {10.1088/0067-0049/208/1/4},
archivePrefix = {arXiv},
       eprint = {1301.0319},
 primaryClass = {astro-ph.SR},
       adsurl = {https://ui.adsabs.harvard.edu/abs/2013ApJS..208....4P}
}

@ARTICLE{Paxton+15,
       author = {{Paxton}, Bill and {Marchant}, Pablo and {Schwab}, Josiah and {Bauer}, Evan B. and {Bildsten}, Lars and {Cantiello}, Matteo and {Dessart}, Luc and {Farmer}, R. and {Hu}, H. and {Langer}, N. and {Townsend}, R.~H.~D. and {Townsley}, Dean M. and {Timmes}, F.~X.},
        title = "{Modules for Experiments in Stellar Astrophysics (MESA): Binaries, Pulsations, and Explosions}",
      journal = {\apjs},
         year = 2015,
        month = sep,
       volume = {220},
       number = {1},
          eid = {15},
        pages = {15},
          doi = {10.1088/0067-0049/220/1/15},
archivePrefix = {arXiv},
       eprint = {1506.03146},
 primaryClass = {astro-ph.SR},
       adsurl = {https://ui.adsabs.harvard.edu/abs/2015ApJS..220...15P}
}

@ARTICLE{Paxton+18,
       author = {{Paxton}, Bill and {Schwab}, Josiah and {Bauer}, Evan B. and {Bildsten}, Lars and {Blinnikov}, Sergei and {Duffell}, Paul and {Farmer}, R. and {Goldberg}, Jared A. and {Marchant}, Pablo and {Sorokina}, Elena and {Thoul}, Anne and {Townsend}, Richard H.~D. and {Timmes}, F.~X.},
        title = "{Modules for Experiments in Stellar Astrophysics (MESA): Convective Boundaries, Element Diffusion, and Massive Star Explosions}",
      journal = {\apjs},
         year = 2018,
        month = feb,
       volume = {234},
       number = {2},
          eid = {34},
        pages = {34},
          doi = {10.3847/1538-4365/aaa5a8},
archivePrefix = {arXiv},
       eprint = {1710.08424},
 primaryClass = {astro-ph.SR},
       adsurl = {https://ui.adsabs.harvard.edu/abs/2018ApJS..234...34P}
}

@ARTICLE{Paxton+19,
       author = {{Paxton}, Bill and {Smolec}, R. and {Schwab}, Josiah and {Gautschy}, A. and {Bildsten}, Lars and {Cantiello}, Matteo and {Dotter}, Aaron and {Farmer}, R. and {Goldberg}, Jared A. and {Jermyn}, Adam S. and {Kanbur}, S.~M. and {Marchant}, Pablo and {Thoul}, Anne and {Townsend}, Richard H.~D. and {Wolf}, William M. and {Zhang}, Michael and {Timmes}, F.~X.},
        title = "{Modules for Experiments in Stellar Astrophysics (MESA): Pulsating Variable Stars, Rotation, Convective Boundaries, and Energy Conservation}",
      journal = {\apjs},
         year = 2019,
        month = jul,
       volume = {243},
       number = {1},
          eid = {10},
        pages = {10},
          doi = {10.3847/1538-4365/ab2241},
archivePrefix = {arXiv},
       eprint = {1903.01426},
 primaryClass = {astro-ph.SR},
       adsurl = {https://ui.adsabs.harvard.edu/abs/2019ApJS..243...10P}
}

@ARTICLE{Piran-99,
       author = {{Piran}, T.},
        title = "{Gamma-ray bursts and the fireball model}",
      journal = {\physrep},
         year = 1999,
        month = jun,
       volume = {314},
       number = {6},
        pages = {575-667},
          doi = {10.1016/S0370-1573(98)00127-6},
archivePrefix = {arXiv},
       eprint = {astro-ph/9810256},
 primaryClass = {astro-ph},
       adsurl = {https://ui.adsabs.harvard.edu/abs/1999PhR...314..575P}
}

@ARTICLE{Poolakkil+21,
       author = {{Poolakkil}, S. and {Preece}, R. and {Fletcher}, C. and {Goldstein}, A. and {Bhat}, P.~N. and {Bissaldi}, E. and {Briggs}, M.~S. and {Burns}, E. and {Cleveland}, W.~H. and {Giles}, M.~M. and {Hui}, C.~M. and {Kocevski}, D. and {Lesage}, S. and {Mailyan}, B. and {Malacaria}, C. and {Paciesas}, W.~S. and {Roberts}, O.~J. and {Veres}, P. and {von Kienlin}, A. and {Wilson-Hodge}, C.~A.},
        title = "{The Fermi-GBM Gamma-Ray Burst Spectral Catalog: 10 yr of Data}",
      journal = {\apj},
         year = 2021,
        month = may,
       volume = {913},
       number = {1},
          eid = {60},
        pages = {60},
          doi = {10.3847/1538-4357/abf24d},
archivePrefix = {arXiv},
       eprint = {2103.13528},
 primaryClass = {astro-ph.HE},
       adsurl = {https://ui.adsabs.harvard.edu/abs/2021ApJ...913...60P}
}

@ARTICLE{Popham+99,
       author = {{Popham}, Robert and {Woosley}, S.~E. and {Fryer}, Chris},
        title = "{Hyperaccreting Black Holes and Gamma-Ray Bursts}",
      journal = {\apj},
         year = 1999,
        month = jun,
       volume = {518},
       number = {1},
        pages = {356-374},
          doi = {10.1086/307259},
archivePrefix = {arXiv},
       eprint = {astro-ph/9807028},
 primaryClass = {astro-ph},
       adsurl = {https://ui.adsabs.harvard.edu/abs/1999ApJ...518..356P}
}

@ARTICLE{Rezzolla+18,
       author = {{Rezzolla}, Luciano and {Most}, Elias R. and {Weih}, Lukas R.},
        title = "{Using Gravitational-wave Observations and Quasi-universal Relations to Constrain the Maximum Mass of Neutron Stars}",
      journal = {\apjl},
         year = 2018,
        month = jan,
       volume = {852},
       number = {2},
          eid = {L25},
        pages = {L25},
          doi = {10.3847/2041-8213/aaa401},
archivePrefix = {arXiv},
       eprint = {1711.00314},
 primaryClass = {astro-ph.HE},
       adsurl = {https://ui.adsabs.harvard.edu/abs/2018ApJ...852L..25R}
}

@ARTICLE{Robitaille-Whitnew-10,
       author = {{Robitaille}, Thomas P. and {Whitney}, Barbara A.},
        title = "{The Present-Day Star Formation Rate of the Milky Way Determined from Spitzer-Detected Young Stellar Objects}",
      journal = {\apjl},
         year = 2010,
        month = feb,
       volume = {710},
       number = {1},
        pages = {L11-L15},
          doi = {10.1088/2041-8205/710/1/L11},
archivePrefix = {arXiv},
       eprint = {1001.3672},
 primaryClass = {astro-ph.GA},
       adsurl = {https://ui.adsabs.harvard.edu/abs/2010ApJ...710L..11R}
}

@ARTICLE{Salpeter-55,
       author = {{Salpeter}, Edwin E.},
        title = "{The Luminosity Function and Stellar Evolution.}",
      journal = {\apj},
         year = 1955,
        month = jan,
       volume = {121},
        pages = {161},
          doi = {10.1086/145971},
       adsurl = {https://ui.adsabs.harvard.edu/abs/1955ApJ...121..161S}
}

@ARTICLE{Sana+12,
       author = {{Sana}, H. and {de Mink}, S.~E. and {de Koter}, A. and {Langer}, N. and {Evans}, C.~J. and {Gieles}, M. and {Gosset}, E. and {Izzard}, R.~G. and {Le Bouquin}, J.-B. and {Schneider}, F.~R.~N.},
        title = "{Binary Interaction Dominates the Evolution of Massive Stars}",
      journal = {Science},
         year = 2012,
        month = jul,
       volume = {337},
       number = {6093},
        pages = {444},
          doi = {10.1126/science.1223344},
archivePrefix = {arXiv},
       eprint = {1207.6397},
 primaryClass = {astro-ph.SR},
       adsurl = {https://ui.adsabs.harvard.edu/abs/2012Sci...337..444S}
}

@ARTICLE{Sander-Vink-20,
       author = {{Sander}, Andreas A.~C. and {Vink}, Jorick S.},
        title = "{On the nature of massive helium star winds and Wolf-Rayet-type mass-loss}",
      journal = {\mnras},
         year = 2020,
        month = nov,
       volume = {499},
       number = {1},
        pages = {873-892},
          doi = {10.1093/mnras/staa2712},
archivePrefix = {arXiv},
       eprint = {2009.01849},
 primaryClass = {astro-ph.SR},
       adsurl = {https://ui.adsabs.harvard.edu/abs/2020MNRAS.499..873S}
}

@ARTICLE{Schauer+20,
       author = {{Schauer}, Anna T.~P. and {Drory}, Niv and {Bromm}, Volker},
        title = "{The Ultimately Large Telescope: What Kind of Facility Do We Need to Detect Population III Stars?}",
      journal = {\apj},
         year = 2020,
        month = dec,
       volume = {904},
       number = {2},
          eid = {145},
        pages = {145},
          doi = {10.3847/1538-4357/abbc0b},
archivePrefix = {arXiv},
       eprint = {2007.02946},
 primaryClass = {astro-ph.GA},
       adsurl = {https://ui.adsabs.harvard.edu/abs/2020ApJ...904..145S}
}

@ARTICLE{Shibata+17,
       author = {{Shibata}, Masaru and {Fujibayashi}, Sho and {Hotokezaka}, Kenta and {Kiuchi}, Kenta and {Kyutoku}, Koutarou and {Sekiguchi}, Yuichiro and {Tanaka}, Masaomi},
        title = "{Modeling GW170817 based on numerical relativity and its implications}",
      journal = {\prd},
         year = 2017,
        month = dec,
       volume = {96},
       number = {12},
          eid = {123012},
        pages = {123012},
          doi = {10.1103/PhysRevD.96.123012},
archivePrefix = {arXiv},
       eprint = {1710.07579},
 primaryClass = {astro-ph.HE},
       adsurl = {https://ui.adsabs.harvard.edu/abs/2017PhRvD..96l3012S}
}

@ARTICLE{Spergel+15,
       author = {{Spergel}, D. and {Gehrels}, N. and {Baltay}, C. and {Bennett}, D. and {Breckinridge}, J. and {Donahue}, M. and {Dressler}, A. and {Gaudi}, B.~S. and {Greene}, T. and {Guyon}, O. and {Hirata}, C. and {Kalirai}, J. and {Kasdin}, N.~J. and {Macintosh}, B. and {Moos}, W. and {Perlmutter}, S. and {Postman}, M. and {Rauscher}, B. and {Rhodes}, J. and {Wang}, Y. and {Weinberg}, D. and {Benford}, D. and {Hudson}, M. and {Jeong}, W. -S. and {Mellier}, Y. and {Traub}, W. and {Yamada}, T. and {Capak}, P. and {Colbert}, J. and {Masters}, D. and {Penny}, M. and {Savransky}, D. and {Stern}, D. and {Zimmerman}, N. and {Barry}, R. and {Bartusek}, L. and {Carpenter}, K. and {Cheng}, E. and {Content}, D. and {Dekens}, F. and {Demers}, R. and {Grady}, K. and {Jackson}, C. and {Kuan}, G. and {Kruk}, J. and {Melton}, M. and {Nemati}, B. and {Parvin}, B. and {Poberezhskiy}, I. and {Peddie}, C. and {Ruffa}, J. and {Wallace}, J.~K. and {Whipple}, A. and {Wollack}, E. and {Zhao}, F.},
        title = "{Wide-Field InfrarRed Survey Telescope-Astrophysics Focused Telescope Assets WFIRST-AFTA 2015 Report}",
      journal = {arXiv e-prints},
         year = 2015,
        month = mar,
          eid = {arXiv:1503.03757},
        pages = {arXiv:1503.03757},
          doi = {10.48550/arXiv.1503.03757},
archivePrefix = {arXiv},
       eprint = {1503.03757},
 primaryClass = {astro-ph.IM},
       adsurl = {https://ui.adsabs.harvard.edu/abs/2015arXiv150303757S}
}

@ARTICLE{Spruit-02,
  author = {Spruit, H. C.},
  title = "{Dynamo action by differential rotation in a stably stratified stellar interior}",
  journal = {Astronomy \& Astrophysics},
  volume = {381},
  pages = {923--932},
  year = {2002},
  doi = {10.1051/0004-6361:20011400}
}

@ARTICLE{Stacy+11,
       author = {{Stacy}, Athena and {Bromm}, Volker and {Loeb}, Abraham},
        title = "{Rotation speed of the first stars}",
      journal = {\mnras},
         year = 2011,
        month = may,
       volume = {413},
       number = {1},
        pages = {543-553},
          doi = {10.1111/j.1365-2966.2010.18152.x},
archivePrefix = {arXiv},
       eprint = {1010.0997},
 primaryClass = {astro-ph.CO},
       adsurl = {https://ui.adsabs.harvard.edu/abs/2011MNRAS.413..543S}
}

@ARTICLE{Stacy+13,
       author = {{Stacy}, Athena and {Greif}, Thomas H. and {Klessen}, Ralf S. and {Bromm}, Volker and {Loeb}, Abraham},
        title = "{Rotation and internal structure of Population III protostars}",
      journal = {\mnras},
         year = 2013,
        month = may,
       volume = {431},
       number = {2},
        pages = {1470-1486},
          doi = {10.1093/mnras/stt264},
archivePrefix = {arXiv},
       eprint = {1209.1439},
 primaryClass = {astro-ph.CO},
       adsurl = {https://ui.adsabs.harvard.edu/abs/2013MNRAS.431.1470S}
}

@ARTICLE{Stacy+16,
       author = {{Stacy}, Athena and {Bromm}, Volker and {Lee}, Aaron T.},
        title = "{Building up the Population III initial mass function from cosmological initial conditions}",
      journal = {\mnras},
         year = 2016,
        month = oct,
       volume = {462},
       number = {2},
        pages = {1307-1328},
          doi = {10.1093/mnras/stw1728},
archivePrefix = {arXiv},
       eprint = {1603.09475},
 primaryClass = {astro-ph.GA},
       adsurl = {https://ui.adsabs.harvard.edu/abs/2016MNRAS.462.1307S}
}

@ARTICLE{Tanvir+09,
       author = {{Tanvir}, N.~R. and {Fox}, D.~B. and {Levan}, A.~J. and {Berger}, E. and {Wiersema}, K. and {Fynbo}, J.~P.~U. and {Cucchiara}, A. and {Kr{\"u}hler}, T. and {Gehrels}, N. and {Bloom}, J.~S. and {Greiner}, J. and {Evans}, P.~A. and {Rol}, E. and {Olivares}, F. and {Hjorth}, J. and {Jakobsson}, P. and {Farihi}, J. and {Willingale}, R. and {Starling}, R.~L.~C. and {Cenko}, S.~B. and {Perley}, D. and {Maund}, J.~R. and {Duke}, J. and {Wijers}, R.~A.~M.~J. and {Adamson}, A.~J. and {Allan}, A. and {Bremer}, M.~N. and {Burrows}, D.~N. and {Castro-Tirado}, A.~J. and {Cavanagh}, B. and {de Ugarte Postigo}, A. and {Dopita}, M.~A. and {Fatkhullin}, T.~A. and {Fruchter}, A.~S. and {Foley}, R.~J. and {Gorosabel}, J. and {Kennea}, J. and {Kerr}, T. and {Klose}, S. and {Krimm}, H.~A. and {Komarova}, V.~N. and {Kulkarni}, S.~R. and {Moskvitin}, A.~S. and {Mundell}, C.~G. and {Naylor}, T. and {Page}, K. and {Penprase}, B.~E. and {Perri}, M. and {Podsiadlowski}, P. and {Roth}, K. and {Rutledge}, R.~E. and {Sakamoto}, T. and {Schady}, P. and {Schmidt}, B.~P. and {Soderberg}, A.~M. and {Sollerman}, J. and {Stephens}, A.~W. and {Stratta}, G. and {Ukwatta}, T.~N. and {Watson}, D. and {Westra}, E. and {Wold}, T. and {Wolf}, C.},
        title = "{A {\ensuremath{\gamma}}-ray burst at a redshift of z\raisebox{-0.5ex}\textasciitilde8.2}",
      journal = {\nat},
         year = 2009,
        month = oct,
       volume = {461},
       number = {7268},
        pages = {1254-1257},
          doi = {10.1038/nature08459},
archivePrefix = {arXiv},
       eprint = {0906.1577},
 primaryClass = {astro-ph.CO},
       adsurl = {https://ui.adsabs.harvard.edu/abs/2009Natur.461.1254T}
}

@ARTICLE{Tchekhovskoy+11,
       author = {{Tchekhovskoy}, Alexander and {Narayan}, Ramesh and {McKinney}, Jonathan C.},
        title = "{Efficient generation of jets from magnetically arrested accretion on a rapidly spinning black hole}",
      journal = {\mnras},
         year = 2011,
        month = nov,
       volume = {418},
       number = {1},
        pages = {L79-L83},
          doi = {10.1111/j.1745-3933.2011.01147.x},
archivePrefix = {arXiv},
       eprint = {1108.0412},
 primaryClass = {astro-ph.HE},
       adsurl = {https://ui.adsabs.harvard.edu/abs/2011MNRAS.418L..79T}
}

@ARTICLE{Toma+16,
       author = {{Toma}, Kenji and {Yoon}, Sung-Chul and {Bromm}, Volker},
        title = "{Gamma-Ray Bursts and Population III Stars}",
      journal = {\ssr},
         year = 2016,
        month = dec,
       volume = {202},
       number = {1-4},
        pages = {159-180},
          doi = {10.1007/s11214-016-0250-7},
archivePrefix = {arXiv},
       eprint = {1603.04640},
 primaryClass = {astro-ph.HE},
       adsurl = {https://ui.adsabs.harvard.edu/abs/2016SSRv..202..159T}
}

@ARTICLE{Tumlinson-06,
       author = {{Tumlinson}, Jason},
        title = "{Chemical Evolution in Hierarchical Models of Cosmic Structure. I. Constraints on the Early Stellar Initial Mass Function}",
      journal = {\apj},
         year = 2006,
        month = apr,
       volume = {641},
       number = {1},
        pages = {1-20},
          doi = {10.1086/500383},
archivePrefix = {arXiv},
       eprint = {astro-ph/0507442},
 primaryClass = {astro-ph},
       adsurl = {https://ui.adsabs.harvard.edu/abs/2006ApJ...641....1T}
}

@ARTICLE{Tumlinson-Shull-00,
       author = {{Tumlinson}, Jason and {Shull}, J. Michael},
        title = "{Zero-Metallicity Stars and the Effects of the First Stars on Reionization}",
      journal = {\apjl},
         year = 2000,
        month = jan,
       volume = {528},
       number = {2},
        pages = {L65-L68},
          doi = {10.1086/312432},
archivePrefix = {arXiv},
       eprint = {astro-ph/9911339},
 primaryClass = {astro-ph},
       adsurl = {https://ui.adsabs.harvard.edu/abs/2000ApJ...528L..65T}
}

@misc{Urrutia+25,
      title={Numerical simulations of jet launching and breakout from collapsars}, 
      author={Gerardo Urrutia and Agnieszka Janiuk and Hector Olivares},
      year={2025},
      eprint={2507.10231},
      archivePrefix={arXiv},
      primaryClass={astro-ph.HE},
      url={https://arxiv.org/abs/2507.10231}, 
}

@ARTICLE{vonZeipel-24,
       author = {{von Zeipel}, H.},
        title = "{The radiative equilibrium of a rotating system of gaseous masses}",
      journal = {\mnras},
         year = 1924,
        month = jun,
       volume = {84},
        pages = {665-683},
          doi = {10.1093/mnras/84.9.665},
       adsurl = {https://ui.adsabs.harvard.edu/abs/1924MNRAS..84..665V}
}

@ARTICLE{Wanderman-Piran-10,
       author = {{Wanderman}, David and {Piran}, Tsvi},
        title = "{The luminosity function and the rate of Swift's gamma-ray bursts}",
      journal = {\mnras},
         year = 2010,
        month = aug,
       volume = {406},
       number = {3},
        pages = {1944-1958},
          doi = {10.1111/j.1365-2966.2010.16787.x},
archivePrefix = {arXiv},
       eprint = {0912.0709},
 primaryClass = {astro-ph.HE},
       adsurl = {https://ui.adsabs.harvard.edu/abs/2010MNRAS.406.1944W}
}

@ARTICLE{Whalen+13,
       author = {{Whalen}, Daniel J. and {Fryer}, Chris L. and {Holz}, Daniel E. and {Heger}, Alexander and {Woosley}, S.~E. and {Stiavelli}, Massimo and {Even}, Wesley and {Frey}, Lucille H.},
        title = "{Seeing the First Supernovae at the Edge of the Universe with JWST}",
      journal = {\apjl},
         year = 2013,
        month = jan,
       volume = {762},
       number = {1},
          eid = {L6},
        pages = {L6},
          doi = {10.1088/2041-8205/762/1/L6},
archivePrefix = {arXiv},
       eprint = {1209.3457},
 primaryClass = {astro-ph.CO},
       adsurl = {https://ui.adsabs.harvard.edu/abs/2013ApJ...762L...6W}
}

@INPROCEEDINGS{White+21,
       author = {{White}, N.~E. and {Bauer}, F.~E. and {Baumgartner}, W. and {Bautz}, M. and {Berger}, E. and {Cenko}, B. and {Chang}, T. -C. and {Falcone}, A. and {Fausey}, H. and {Feldman}, C. and {Fox}, D. and {Fox}, O. and {Fruchter}, A. and {Fryer}, C. and {Ghirlanda}, G. and {Gorski}, K. and {Grant}, C. and {Guiriec}, S. and {Hart}, M. and {Hartmann}, D. and {Hennawi}, J. and {Kann}, D.~A. and {Kaplan}, D. and {Kennea}, J. and {Kocevski}, D. and {Kouveliotou}, C. and {Lawrence}, C. and {Levan}, A.~J. and {Lidz}, A. and {Lien}, A. and {Littenberg}, T.~B. and {Mas-Ribas}, L. and {Moss}, M. and {O'Brien}, P. and {O'Meara}, J. and {Palmer}, D.~M. and {Pasham}, D. and {Racusin}, J. and {Remillard}, R. and {Roberts}, O.~J. and {Roming}, P. and {Rud}, M. and {Salvaterra}, R. and {Sambruna}, R. and {Seiffert}, M. . . and {Sun}, G. and {Tanvir}, N.~R. and {Terrile}, R. and {Thomas}, N. and {van der Horst}, A. and {Verstrand}, W.~T. and {Willems}, P. and {Wilson-Hodge}, C. and {Young}, E.~T. and {Amati}, L. and {Bozzo}, E. and {Karczewski}, O. {\L}. and {Hernandez-Monteagudo}, C. and {Rebolo Lopez}, R. and {Genova-Santos}, R. and {Martin}, A. and {Granot}, J. and {Bemiamini}, P. and {Gil}, R. and {Burns}, E.},
        title = "{The Gamow Explorer: a Gamma-Ray Burst Observatory to study the high redshift universe and enable multi-messenger astrophysics}",
    booktitle = {UV, X-Ray, and Gamma-Ray Space Instrumentation for Astronomy XXII},
         year = 2021,
       editor = {{Siegmund}, Oswald H.},
       series = {Society of Photo-Optical Instrumentation Engineers (SPIE) Conference Series},
       volume = {11821},
        month = aug,
          eid = {1182109},
        pages = {1182109},
          doi = {10.1117/12.2599293},
archivePrefix = {arXiv},
       eprint = {2111.06497},
 primaryClass = {astro-ph.HE},
       adsurl = {https://ui.adsabs.harvard.edu/abs/2021SPIE11821E..09W}
}

@ARTICLE{Wollenberg+20,
       author = {{Wollenberg}, Katharina M.~J. and {Glover}, Simon C.~O. and {Clark}, Paul C. and {Klessen}, Ralf S.},
        title = "{Formation sites of Population III star formation: The effects of different levels of rotation and turbulence on the fragmentation behaviour of primordial gas}",
      journal = {\mnras},
         year = 2020,
        month = may,
       volume = {494},
       number = {2},
        pages = {1871-1893},
          doi = {10.1093/mnras/staa289},
archivePrefix = {arXiv},
       eprint = {1912.06377},
 primaryClass = {astro-ph.GA},
       adsurl = {https://ui.adsabs.harvard.edu/abs/2020MNRAS.494.1871W}
}

@ARTICLE{Wu+25,
       author = {{Wu}, Zhao-Feng and {Damoulakis}, Michail and {Beniamini}, Paz and {Giannios}, Dimitrios},
        title = "{Maximal Jet Energy of Gamma-Ray Bursts through the Blandford{\textendash}Znajek Mechanism}",
      journal = {\apjl},
         year = 2025,
        month = feb,
       volume = {980},
       number = {2},
          eid = {L28},
        pages = {L28},
          doi = {10.3847/2041-8213/adaeb8},
archivePrefix = {arXiv},
       eprint = {2411.12850},
 primaryClass = {astro-ph.HE},
       adsurl = {https://ui.adsabs.harvard.edu/abs/2025ApJ...980L..28W}
}

@INPROCEEDINGS{Yonetoku+14,
       author = {{Yonetoku}, Daisuke and {Mihara}, Tatehiro and {Sawano}, Tatsuya and {Ikeda}, Hirokazu and {Harayama}, Atsushi and {Takata}, Shunsuke and {Yoshida}, Kazuki and {Seta}, Hiroki and {Toyanago}, Asuka and {Kagawa}, Yasuaki and {Kawai}, Kentaro and {Kawai}, Nobuyuki and {Sakamoto}, Takanori and {Serino}, Motoko and {Kurosawa}, Shunsuke and {Gunji}, Shuichi and {Tanimori}, Toru and {Murakami}, Toshio and {Yatsu}, Yoichi and {Yamaoka}, Kazutaka and {Yoshida}, Atsumasa and {Kawabata}, Koji and {Matsumoto}, Toshio and {Tsumura}, Koji and {Matsuura}, Shuji and {Shirahata}, Mai and {Okita}, Hirofumi and {Yanagisawa}, Kensi and {Yoshida}, Michitoshi and {Motohara}, Kentaro},
        title = "{High-z gamma-ray bursts for unraveling the dark ages mission HiZ-GUNDAM}",
    booktitle = {Space Telescopes and Instrumentation 2014: Ultraviolet to Gamma Ray},
         year = 2014,
       editor = {{Takahashi}, Tadayuki and {den Herder}, Jan-Willem A. and {Bautz}, Mark},
       series = {Society of Photo-Optical Instrumentation Engineers (SPIE) Conference Series},
       volume = {9144},
        month = jul,
          eid = {91442S},
        pages = {91442S},
          doi = {10.1117/12.2055041},
archivePrefix = {arXiv},
       eprint = {1406.4202},
 primaryClass = {astro-ph.IM},
       adsurl = {https://ui.adsabs.harvard.edu/abs/2014SPIE.9144E..2SY}
}

@ARTICLE{Yoon+06,
       author = {{Yoon}, S. -C. and {Langer}, N. and {Norman}, C.},
        title = "{Single star progenitors of long gamma-ray bursts. I. Model grids and redshift dependent GRB rate}",
      journal = {\aap},
         year = 2006,
        month = dec,
       volume = {460},
       number = {1},
        pages = {199-208},
          doi = {10.1051/0004-6361:20065912},
archivePrefix = {arXiv},
       eprint = {astro-ph/0606637},
 primaryClass = {astro-ph},
       adsurl = {https://ui.adsabs.harvard.edu/abs/2006A&A...460..199Y}
}

@ARTICLE{Yoon+15,
       author = {{Yoon}, Sung-Chul and {Kang}, Jisu and {Kozyreva}, Alexandra},
        title = "{Can Very Massive Population III Stars Produce a Super-Collapsar?}",
      journal = {\apj},
         year = 2015,
        month = mar,
       volume = {802},
       number = {1},
          eid = {16},
        pages = {16},
          doi = {10.1088/0004-637X/802/1/16},
archivePrefix = {arXiv},
       eprint = {1504.01202},
 primaryClass = {astro-ph.HE},
       adsurl = {https://ui.adsabs.harvard.edu/abs/2015ApJ...802...16Y}
}

@ARTICLE{Yoshida+04,
       author = {{Yoshida}, Naoki and {Bromm}, Volker and {Hernquist}, Lars},
        title = "{The Era of Massive Population III Stars: Cosmological Implications and Self-Termination}",
      journal = {\apj},
         year = 2004,
        month = apr,
       volume = {605},
       number = {2},
        pages = {579-590},
          doi = {10.1086/382499},
archivePrefix = {arXiv},
       eprint = {astro-ph/0310443},
 primaryClass = {astro-ph},
       adsurl = {https://ui.adsabs.harvard.edu/abs/2004ApJ...605..579Y}
}

@INPROCEEDINGS{Yoshida+16,
       author = {{Yoshida}, Kazuki and {Yonetoku}, Daisuke and {Sawano}, Tatsuya and {Ikeda}, Hirokazu and {Harayama}, Atsushi and {Arimoto}, Makoto and {Kagawa}, Yasuaki and {Ina}, Masao and {Hatori}, Satoshi and {Kume}, Kyo and {Mizushima}, Satoshi and {Hasegawa}, Takashi},
        title = "{Development of wide-field low-energy x-ray imaging detectors for HiZ-GUNDAM}",
    booktitle = {Space Telescopes and Instrumentation 2016: Ultraviolet to Gamma Ray},
         year = 2016,
       editor = {{den Herder}, Jan-Willem A. and {Takahashi}, Tadayuki and {Bautz}, Marshall},
       series = {Society of Photo-Optical Instrumentation Engineers (SPIE) Conference Series},
       volume = {9905},
        month = jul,
          eid = {99050M},
        pages = {99050M},
          doi = {10.1117/12.2231370},
archivePrefix = {arXiv},
       eprint = {1607.06191},
 primaryClass = {astro-ph.IM},
       adsurl = {https://ui.adsabs.harvard.edu/abs/2016SPIE.9905E..0MY}
}

@ARTICLE{Zalamea-Beloborodov-11,
       author = {{Zalamea}, Ivan and {Beloborodov}, Andrei M.},
        title = "{Neutrino heating near hyper-accreting black holes}",
      journal = {\mnras},
         year = 2011,
        month = feb,
       volume = {410},
       number = {4},
        pages = {2302-2308},
          doi = {10.1111/j.1365-2966.2010.17600.x},
archivePrefix = {arXiv},
       eprint = {1003.0710},
 primaryClass = {astro-ph.HE},
       adsurl = {https://ui.adsabs.harvard.edu/abs/2011MNRAS.410.2302Z}
}

@ARTICLE{Yoon+12,
  author  = {Yoon, S.-C. and Dierks, A. and Langer, N.},
  title   = {Evolution of Massive Stars: Consequences of Rotationally Induced Mixing},
  journal = {Astronomy and Astrophysics},
  volume  = {542},
  pages   = {A113},
  year    = {2012},
  doi     = {10.1051/0004-6361/201218724}
}

@ARTICLE{Kumar+08,
  author  = {Kumar, P. and Narayan, R. and Johnson, J.L.},
  title   = {Fallback in Collapsar Models: A Possible Cause for Rapid Decline and Late-time Flaring in GRB Light Curves},
  journal = {Monthly Notices of the Royal Astronomical Society},
  volume  = {388},
  pages   = {1729--1742},
  year    = {2008},
  doi     = {10.1111/j.1365-2966.2008.13520.x}
}

@ARTICLE{Woosley-Heger-06,
  author  = {Woosley, S. E. and Heger, A.},
  title   = {The Progenitor Stars of Gamma-Ray Bursts},
  journal = {The Astrophysical Journal},
  volume  = {637},
  pages   = {914--921},
  year    = {2006},
  doi     = {10.1086/498893}
}

@ARTICLE{Woosley+06b,
       author = {{Woosley}, S.~E. and {Bloom}, J.~S.},
        title = "{The Supernova Gamma-Ray Burst Connection}",
      journal = {\araa},
         year = 2006,
        month = sep,
       volume = {44},
       number = {1},
        pages = {507-556},
          doi = {10.1146/annurev.astro.43.072103.150558},
archivePrefix = {arXiv},
       eprint = {astro-ph/0609142},
 primaryClass = {astro-ph},
       adsurl = {https://ui.adsabs.harvard.edu/abs/2006ARA&A..44..507W}
}

@ARTICLE{Woosley-17,
       author = {{Woosley}, S.~E.},
        title = "{Pulsational Pair-instability Supernovae}",
      journal = {\apj},
         year = 2017,
        month = feb,
       volume = {836},
       number = {2},
          eid = {244},
        pages = {244},
          doi = {10.3847/1538-4357/836/2/244},
archivePrefix = {arXiv},
       eprint = {1608.08939},
 primaryClass = {astro-ph.HE},
       adsurl = {https://ui.adsabs.harvard.edu/abs/2017ApJ...836..244W}
}


\appendix

\section{Some extra material}

\begin{table*}
\centering
\caption{Final stellar parameters for models with $\eta_{\rm wind} = 0.2$ and different initial rotation rates $\hat{\Omega}_0$. Each row corresponds to a different initial mass $M_{\rm ZAMS}\ {\rm (a)}$. The values reported are taken at the pre-core-collapse stage. The columns show (b) the final stellar mass $M_{\rm pre-cc}$, (c) the specific angular momentum at a mass coordinate of 3 $M_\odot$, (d) the compactness parameter $\xi_{2.5}$, (e) the carbon-oxygen core mass $M_{\rm CO}$, (f) the initial black hole mass $M_{\rm BH,0}$, (g) the dimensionless spin parameter $a_0$, and (h) the disc mass $M_{\rm disc}$.}
\resizebox{\textwidth}{!}{
\begin{tabular}{|c|ccccccc|ccccccc|ccccccc|ccccccc|}
\hline
$M_{\rm ZAMS}\ ^{\mathrm{a}}$ & \multicolumn{7}{c|}{$\hat{\Omega}_0 = 0.6$} & \multicolumn{7}{c|}{$\hat{\Omega}_0 = 0.7$} & \multicolumn{7}{c|}{$\hat{\Omega}_0 = 0.8$} & \multicolumn{7}{c|}{$\hat{\Omega}_0 = 0.9$} \\
\cline{2-29}
 & $M_{\mathrm{pre-cc}}\ ^{\mathrm{b}}$ & $\log j_{\mathrm{3M_\odot}}\ ^{\mathrm{c}}$ & $\xi_{2.5}\ ^{\mathrm{d}}$ & $M_{\rm CO}\ ^{\mathrm{e}}$ & $M_{\rm BH,0}\ ^{\mathrm{f}}$ & $a_0\ ^{\mathrm{g}}$ & $M_{\rm disc}\ ^{\mathrm{h}}$ & $M_{\mathrm{pre-cc}}\ ^{\mathrm{b}}$ & $\log j_{\mathrm{3M_\odot}}\ ^{\mathrm{c}}$ & $\xi_{2.5}\ ^{\mathrm{d}}$ & $M_{\rm CO}\ ^{\mathrm{e}}$ & $M_{\rm BH,0}\ ^{\mathrm{f}}$ & $a_0\ ^{\mathrm{g}}$ & $M_{\rm disc}\ ^{\mathrm{h}}$ & $M_{\mathrm{pre-cc}}\ ^{\mathrm{b}}$ & $\log j_{\mathrm{3M_\odot}}\ ^{\mathrm{c}}$ & $\xi_{2.5}\ ^{\mathrm{d}}$ & $M_{\rm CO}\ ^{\mathrm{e}}$ & $M_{\rm BH,0}\ ^{\mathrm{f}}$ & $a_0\ ^{\mathrm{g}}$ & $M_{\rm disc}\ ^{\mathrm{h}}$ & $M_{\mathrm{pre-cc}}\ ^{\mathrm{b}}$ & $\log j_{\mathrm{3M_\odot}}\ ^{\mathrm{c}}$ & $\xi_{2.5}\ ^{\mathrm{d}}$ & $M_{\rm CO}\ ^{\mathrm{e}}$ & $M_{\rm BH,0}\ ^{\mathrm{f}}$ & $a_0\ ^{\mathrm{g}}$ & $M_{\rm disc}\ ^{\mathrm{h}}$ \\
 & $(M_{\odot})$ & (cm$^2$\,s$^{-1}$) & & $(M_{\odot})$ & $(M_{\odot})$ & & $(M_{\odot})$ & $(M_{\odot})$ & (cm$^2$\,s$^{-1}$) & & $(M_{\odot})$ & $(M_{\odot})$ & & $(M_{\odot})$ & $(M_{\odot})$ & (cm$^2$\,s$^{-1}$) & & $(M_{\odot})$ & $(M_{\odot})$ & & $(M_{\odot})$ & $(M_{\odot})$ & (cm$^2$\,s$^{-1}$) & & $(M_{\odot})$ & $(M_{\odot})$ & & $(M_{\odot})$ \\
\hline
20 & 19.94 & 16.15 & 0.27 & 6.96 & 10.60 & 0.59 & 9.34 & 19.21 & 16.18 & 0.32 & 9.97 & 6.34 & 0.79 & 12.87 & 14.50 & 16.24 & 0.37 & 14.40 & 7.46 & 0.69 & 7.04 & 14.57 & 16.24 & 0.37 & 14.42 & 7.49 & 0.72 & 7.08 \\
25 & 18.08 & 16.23 & 0.38 & 17.88 & 10.93 & 0.67 & 7.15 & 17.74 & 16.24 & 0.34 & 17.56 & 9.94 & 0.69 & 7.80 & 17.80 & 16.25 & 0.34 & 17.58 & 10.06 & 0.70 & 7.74 & 20.65 & 16.19 & 0.34 & 20.45 & 12.94 & 0.66 & 7.71 \\
30 & 26.15 & 16.05 & 0.30 & 16.34 & 24.66 & 0.47 & 1.49 & 20.68 & 16.10 & 0.34 & 20.48 & 13.79 & 0.63 & 6.89 & 20.25 & 16.14 & 0.43 & 20.27 & 13.07 & 0.64 & 7.18 & 20.89 & 16.14 & 0.39 & 20.65 & 13.52 & 0.63 & 7.37 \\
35 & 25.18 & 15.99 & 0.41 & 24.97 & 23.55 & 0.58 & 1.63 & 23.67 & 16.06 & 0.34 & 23.34 & 16.91 & 0.61 & 6.76 & 23.69 & 16.04 & 0.35 & 23.23 & 17.18 & 0.60 & 6.51 & 24.24 & 16.06 & 0.34 & 23.98 & 16.74 & 0.64 & 7.50 \\
40 & 26.75 & 16.00 & 0.42 & 26.51 & 24.99 & 0.56 & 1.76 & 26.63 & 15.97 & 0.35 & 26.37 & 24.79 & 0.60 & 1.84 & 26.42 & 15.98 & 0.34 & 26.17 & 24.46 & 0.59 & 1.96 & 26.25 & 15.99 & 0.41 & 26.02 & 24.41 & 0.59 & 1.84 \\
45 & 29.69 & 15.95 & 0.37 & 29.42 & 28.02 & 0.56 & 1.67 & 29.25 & 15.94 & 0.35 & 28.98 & 27.34 & 0.54 & 1.91 & 29.52 & 15.92 & 0.35 & 29.23 & 27.69 & 0.56 & 1.83 & 29.23 & 15.95 & 0.36 & 28.95 & 27.47 & 0.56 & 1.76 \\
50 & 32.75 & 15.89 & 0.32 & 32.44 & 31.42 & 0.50 & 1.33 & 32.19 & 15.90 & 0.35 & 31.84 & 30.85 & 0.53 & 1.34 & 32.01 & 15.91 & 0.33 & 31.71 & 30.73 & 0.53 & 1.28 & 31.94 & 15.89 & 0.34 & 31.66 & 30.61 & 0.53 & 1.33 \\
55 & 35.62 & 15.94 & 0.26 & 35.22 & 34.74 & 0.47 & 0.88 & 35.18 & 15.97 & 0.28 & 34.85 & 33.91 & 0.50 & 1.27 & 34.80 & 16.00 & 0.27 & 34.48 & 33.72 & 0.51 & 1.08 & 34.84 & 15.98 & 0.28 & 34.49 & 33.59 & 0.51 & 1.25 \\
60 & 38.06 & 15.94 & 0.20 & 37.71 & 37.07 & 0.48 & 0.99 & 37.82 & 15.96 & 0.21 & 37.22 & 37.08 & 0.50 & 0.74 & 37.55 & 15.96 & 0.19 & 37.22 & 36.83 & 0.49 & 0.72 & 37.52 & 15.96 & 0.22 & 37.10 & 36.48 & 0.48 & 1.04 \\
65 & 42.48 & 15.90 & 0.23 & 40.73 & 42.48 & 0.44 & 0.00 & 40.61 & 15.93 & 0.23 & 38.86 & 40.57 & 0.49 & 0.04 & 40.24 & 15.94 & 0.20 & 39.40 & 39.61 & 0.48 & 0.63 & 40.15 & 15.95 & 0.22 & 39.34 & 39.44 & 0.49 & 0.71 \\
70 & 43.34 & 15.91 & 0.24 & 41.68 & 43.33 & 0.46 & 0.01 & 43.32 & 15.92 & 0.21 & 41.21 & 43.32 & 0.47 & 0.00 & 43.19 & 15.92 & 0.23 & 40.61 & 43.19 & 0.47 & 0.00 & 42.67 & 15.92 & 0.23 & 40.93 & 42.67 & 0.47 & 0.00 \\
75 & 74.04 & 17.04 & 0.01 & 0.00 & - & - & - & 45.77 & 15.90 & 0.23 & 42.82 & 45.77 & 0.44 & 0.00 & 45.66 & 15.90 & 0.20 & 42.44 & 45.66 & 0.44 & 0.00 & 45.45 & 15.90 & 0.23 & 42.90 & 45.44 & 0.44 & 0.01 \\
80 & 55.59 & 15.90 & 0.28 & 53.22 & 54.72 & 0.41 & 0.87 & 48.57 & 15.88 & 0.21 & 45.31 & 48.57 & 0.42 & 0.00 & 48.17 & 15.90 & 0.24 & 45.07 & 48.17 & 0.43 & 0.00 & 48.32 & 15.89 & 0.24 & 44.95 & 48.32 & 0.43 & 0.00 \\
85 & 57.55 & 15.87 & 0.04 & 52.82 & 57.55 & 0.39 & 0.00 & 51.24 & 15.87 & 0.26 & 48.54 & 51.24 & 0.41 & 0.00 & 50.83 & 15.88 & 0.04 & 47.03 & 50.82 & 0.41 & 0.01 & 50.69 & 15.89 & 0.05 & 46.87 & 50.69 & 0.41 & 0.00 \\
90 & 54.15 & 15.86 & 0.04 & 49.72 & 54.15 & 0.38 & 0.00 & 53.95 & 15.86 & 0.24 & 50.47 & 53.95 & 0.38 & 0.00 & 53.66 & 15.87 & 0.04 & 49.20 & 53.66 & 0.38 & 0.00 & 53.95 & 15.86 & 0.26 & 50.79 & 53.95 & 0.39 & 0.00 \\
95 & 57.71 & 15.84 & 0.27 & 54.82 & 57.71 & 0.35 & 0.00 & 56.54 & 15.86 & 0.04 & 52.01 & 56.54 & 0.38 & 0.00 & 56.07 & 15.86 & 0.28 & 53.57 & 56.07 & 0.38 & 0.00 & 56.03 & 15.85 & 0.05 & 51.41 & 56.03 & 0.37 & 0.00 \\
100 & 77.22 & 15.94 & 0.34 & 75.79 & 77.22 & 0.50 & 0.00 & 60.45 & 15.82 & 0.26 & 57.21 & 60.45 & 0.34 & 0.00 & 58.67 & 15.83 & 0.28 & 56.05 & 58.67 & 0.35 & 0.00 & 58.76 & 15.84 & 0.04 & 54.00 & 58.76 & 0.36 & 0.00 \\
\hline
\end{tabular}
}
\label{tab:eta_d0.2}
\end{table*}

\begin{table*}
\centering
\caption{Stellar Parameters for $\eta_{\rm wind} = 0.5$, Different $\hat{\Omega}_0$ Values}
\resizebox{\textwidth}{!}{
\begin{tabular}{|c|ccccccc|ccccccc|ccccccc|ccccccc|}
\hline
$M_{\rm ZAMS}\ ^{\mathrm{a}}$ & \multicolumn{7}{c|}{$\hat{\Omega}_0 = 0.6$} & \multicolumn{7}{c|}{$\hat{\Omega}_0 = 0.7$} & \multicolumn{7}{c|}{$\hat{\Omega}_0 = 0.8$} & \multicolumn{7}{c|}{$\hat{\Omega}_0 = 0.9$} \\
\cline{2-29}
 & $M_{\mathrm{pre-cc}}\ ^{\mathrm{b}}$ & $\log j_{\mathrm{3M_\odot}}\ ^{\mathrm{c}}$ & $\xi_{2.5}\ ^{\mathrm{d}}$ & $M_{\rm CO}\ ^{\mathrm{e}}$ & $M_{\rm BH,0}\ ^{\mathrm{f}}$ & $a_0\ ^{\mathrm{g}}$ & $M_{\rm disc}\ ^{\mathrm{h}}$ & $M_{\mathrm{pre-cc}}\ ^{\mathrm{b}}$ & $\log j_{\mathrm{3M_\odot}}\ ^{\mathrm{c}}$ & $\xi_{2.5}\ ^{\mathrm{d}}$ & $M_{\rm CO}\ ^{\mathrm{e}}$ & $M_{\rm BH,0}\ ^{\mathrm{f}}$ & $a_0\ ^{\mathrm{g}}$ & $M_{\rm disc}\ ^{\mathrm{h}}$ & $M_{\mathrm{pre-cc}}\ ^{\mathrm{b}}$ & $\log j_{\mathrm{3M_\odot}}\ ^{\mathrm{c}}$ & $\xi_{2.5}\ ^{\mathrm{d}}$ & $M_{\rm CO}\ ^{\mathrm{e}}$ & $M_{\rm BH,0}\ ^{\mathrm{f}}$ & $a_0\ ^{\mathrm{g}}$ & $M_{\rm disc}\ ^{\mathrm{h}}$ & $M_{\mathrm{pre-cc}}\ ^{\mathrm{b}}$ & $\log j_{\mathrm{3M_\odot}}\ ^{\mathrm{c}}$ & $\xi_{2.5}\ ^{\mathrm{d}}$ & $M_{\rm CO}\ ^{\mathrm{e}}$ & $M_{\rm BH,0}\ ^{\mathrm{f}}$ & $a_0\ ^{\mathrm{g}}$ & $M_{\rm disc}\ ^{\mathrm{h}}$ \\
 & $(M_{\odot})$ & (cm$^2$\,s$^{-1}$) & & $(M_{\odot})$ & $(M_{\odot})$ & & $(M_{\odot})$ & $(M_{\odot})$ & (cm$^2$\,s$^{-1}$) & & $(M_{\odot})$ & $(M_{\odot})$ & & $(M_{\odot})$ & $(M_{\odot})$ & (cm$^2$\,s$^{-1}$) & & $(M_{\odot})$ & $(M_{\odot})$ & & $(M_{\odot})$ & $(M_{\odot})$ & (cm$^2$\,s$^{-1}$) & & $(M_{\odot})$ & $(M_{\odot})$ & & $(M_{\odot})$ \\
\hline
20 & 19.75 & 16.17 & 0.29 & 6.67 & 10.34 & 0.60 & 9.41 & 14.46 & 15.93 & 0.32 & 10.93 & 13.67 & 0.54 & 0.79 & 10.32 & 15.84 & 0.33 & 7.90 & 10.04 & 0.45 & 0.28 & 10.29 & 15.88 & 0.36 & 7.91 & 9.87 & 0.45 & 0.42 \\
25 & 12.23 & 15.98 & 0.34 & 9.63 & 11.98 & 0.43 & 0.25 & 12.17 & 15.94 & 0.36 & 9.51 & 11.92 & 0.43 & 0.25 & 12.01 & 15.97 & 0.32 & 9.50 & 11.78 & 0.43 & 0.23 & 12.45 & 15.90 & 0.36 & 9.74 & 12.04 & 0.42 & 0.41 \\
30 & 20.70 & 15.86 & 0.34 & 20.50 & 20.04 & 0.48 & 0.66 & 14.09 & 15.89 & 0.35 & 13.95 & 13.64 & 0.39 & 0.45 & 13.85 & 15.91 & 0.33 & 13.72 & 13.63 & 0.41 & 0.22 & 13.97 & 15.88 & 0.35 & 13.82 & 13.73 & 0.40 & 0.24 \\
35 & 16.89 & 15.83 & 0.33 & 16.71 & 16.72 & 0.38 & 0.17 & 15.78 & 15.83 & 0.38 & 15.64 & 15.59 & 0.38 & 0.19 & 15.61 & 15.86 & 0.36 & 15.48 & 15.41 & 0.39 & 0.20 & 15.74 & 15.83 & 0.37 & 15.61 & 15.51 & 0.38 & 0.23 \\
40 & 17.59 & 15.77 & 0.39 & 17.43 & 17.27 & 0.36 & 0.32 & 17.36 & 15.79 & 0.37 & 17.28 & 17.25 & 0.37 & 0.11 & 17.33 & 15.77 & 0.37 & 17.16 & 17.16 & 0.37 & 0.17 & 17.40 & 15.77 & 0.39 & 17.21 & 17.22 & 0.37 & 0.18 \\
45 & 19.24 & 15.76 & 0.40 & 19.06 & 19.12 & 0.36 & 0.12 & 19.07 & 15.74 & 0.40 & 18.90 & 18.97 & 0.36 & 0.10 & 18.75 & 15.76 & 0.39 & 18.68 & 18.72 & 0.36 & 0.03 & 18.88 & 15.77 & 0.36 & 18.70 & 18.79 & 0.36 & 0.09 \\
50 & 21.40 & 15.73 & 0.42 & 21.23 & 21.32 & 0.34 & 0.08 & 20.82 & 15.71 & 0.41 & 20.65 & 20.64 & 0.33 & 0.18 & 20.44 & 15.72 & 0.42 & 20.47 & 20.44 & 0.35 & 0.00 & 20.65 & 15.74 & 0.36 & 20.43 & 20.55 & 0.35 & 0.10 \\
55 & 23.71 & 15.66 & 0.42 & 23.50 & 23.58 & 0.31 & 0.13 & 22.27 & 15.71 & 0.40 & 22.06 & 22.04 & 0.32 & 0.23 & 22.32 & 15.72 & 0.41 & 22.11 & 22.31 & 0.33 & 0.01 & 22.23 & 15.70 & 0.35 & 22.00 & 22.17 & 0.34 & 0.06 \\
60 & 24.24 & 15.71 & 0.37 & 24.01 & 24.16 & 0.32 & 0.08 & 23.97 & 15.69 & 0.41 & 23.77 & 23.97 & 0.32 & 0.00 & 23.87 & 15.66 & 0.36 & 23.60 & 23.81 & 0.32 & 0.06 & 23.69 & 15.70 & 0.36 & 23.44 & 23.65 & 0.32 & 0.04 \\
65 & 28.17 & 15.73 & 0.27 & 27.82 & 28.17 & 0.31 & 0.00 & 25.53 & 15.68 & 0.37 & 25.30 & 25.53 & 0.31 & 0.00 & 25.26 & 15.69 & 0.35 & 25.02 & 25.26 & 0.31 & 0.00 & 25.34 & 15.67 & 0.35 & 25.09 & 25.29 & 0.31 & 0.05 \\
70 & 27.16 & 15.59 & 0.34 & 26.90 & 27.04 & 0.28 & 0.12 & 27.03 & 15.68 & 0.36 & 26.75 & 27.03 & 0.30 & 0.00 & 26.81 & 15.64 & 0.35 & 26.54 & 26.67 & 0.29 & 0.14 & 26.78 & 15.65 & 0.35 & 26.50 & 26.69 & 0.29 & 0.09 \\
75 & 32.01 & 15.67 & 0.29 & 31.67 & 32.01 & 0.27 & 0.00 & 28.43 & 15.65 & 0.35 & 28.16 & 28.43 & 0.29 & 0.00 & 28.38 & 15.61 & 0.36 & 28.05 & 28.32 & 0.29 & 0.06 & 28.15 & 15.63 & 0.35 & 27.86 & 28.15 & 0.28 & 0.00 \\
80 & 36.20 & 15.67 & 0.18 & 35.06 & 36.20 & 0.26 & 0.00 & 29.93 & 15.60 & 0.31 & 29.62 & 29.82 & 0.27 & 0.11 & 29.98 & 15.65 & 0.29 & 29.54 & 29.75 & 0.28 & 0.23 & 29.66 & 15.59 & 0.33 & 29.40 & 29.66 & 0.28 & 0.00 \\
85 & 39.16 & 15.66 & 0.23 & 36.94 & 39.16 & 0.25 & 0.00 & 31.55 & 15.66 & 0.26 & 31.24 & 31.55 & 0.27 & 0.00 & 31.24 & 15.62 & 0.25 & 30.87 & 31.24 & 0.26 & 0.00 & 30.98 & 15.62 & 0.25 & 30.71 & 30.98 & 0.27 & 0.00 \\
90 & 33.21 & 15.65 & 0.24 & 32.84 & 33.21 & 0.25 & 0.00 & 32.69 & 15.64 & 0.24 & 32.36 & 32.69 & 0.25 & 0.00 & 32.73 & 15.65 & 0.27 & 32.45 & 32.73 & 0.26 & 0.00 & 32.51 & 15.66 & 0.19 & 31.20 & 32.51 & 0.26 & 0.00 \\
95 & 34.60 & 15.63 & 0.23 & 34.28 & 34.60 & 0.24 & 0.00 & 34.08 & 15.64 & 0.22 & 33.54 & 34.08 & 0.24 & 0.00 & 34.26 & 15.64 & 0.20 & 33.79 & 34.26 & 0.25 & 0.00 & 33.98 & 15.66 & 0.29 & 33.64 & 33.98 & 0.24 & 0.00 \\
100 & 56.56 & 15.72 & 0.28 & 54.74 & 56.56 & 0.28 & 0.00 & 37.61 & 15.60 & 0.23 & 36.22 & 37.61 & 0.22 & 0.00 & 35.67 & 15.63 & 0.16 & 35.21 & 35.66 & 0.24 & 0.01 & 35.24 & 15.64 & 0.17 & 34.81 & 35.24 & 0.24 & 0.00 \\
\hline
\end{tabular}
}
\label{tab:eta_d0.5}
\end{table*}

\begin{table*}
\centering
\caption{Stellar Parameters for $\eta_{\rm wind} = 1.0$, Different $\hat{\Omega}_0$ Values}
\resizebox{\textwidth}{!}{
\begin{tabular}{|c|ccccccc|ccccccc|ccccccc|ccccccc|}
\hline
$M_{\rm ZAMS}\ ^{\mathrm{a}}$ & \multicolumn{7}{c|}{$\hat{\Omega}_0 = 0.6$} & \multicolumn{7}{c|}{$\hat{\Omega}_0 = 0.7$} & \multicolumn{7}{c|}{$\hat{\Omega}_0 = 0.8$} & \multicolumn{7}{c|}{$\hat{\Omega}_0 = 0.9$} \\
\cline{2-29}
 & $M_{\mathrm{pre-cc}}\ ^{\mathrm{b}}$ & $\log j_{\mathrm{3M_\odot}}\ ^{\mathrm{c}}$ & $\xi_{2.5}\ ^{\mathrm{d}}$ & $M_{\rm CO}\ ^{\mathrm{e}}$ & $M_{\rm BH,0}\ ^{\mathrm{f}}$ & $a_0\ ^{\mathrm{g}}$ & $M_{\rm disc}\ ^{\mathrm{h}}$ & $M_{\mathrm{pre-cc}}\ ^{\mathrm{b}}$ & $\log j_{\mathrm{3M_\odot}}\ ^{\mathrm{c}}$ & $\xi_{2.5}\ ^{\mathrm{d}}$ & $M_{\rm CO}\ ^{\mathrm{e}}$ & $M_{\rm BH,0}\ ^{\mathrm{f}}$ & $a_0\ ^{\mathrm{g}}$ & $M_{\rm disc}\ ^{\mathrm{h}}$ & $M_{\mathrm{pre-cc}}\ ^{\mathrm{b}}$ & $\log j_{\mathrm{3M_\odot}}\ ^{\mathrm{c}}$ & $\xi_{2.5}\ ^{\mathrm{d}}$ & $M_{\rm CO}\ ^{\mathrm{e}}$ & $M_{\rm BH,0}\ ^{\mathrm{f}}$ & $a_0\ ^{\mathrm{g}}$ & $M_{\rm disc}\ ^{\mathrm{h}}$ & $M_{\mathrm{pre-cc}}\ ^{\mathrm{b}}$ & $\log j_{\mathrm{3M_\odot}}\ ^{\mathrm{c}}$ & $\xi_{2.5}\ ^{\mathrm{d}}$ & $M_{\rm CO}\ ^{\mathrm{e}}$ & $M_{\rm BH,0}\ ^{\mathrm{f}}$ & $a_0\ ^{\mathrm{g}}$ & $M_{\rm disc}\ ^{\mathrm{h}}$ \\
 & $(M_{\odot})$ & (cm$^2$\,s$^{-1}$) & & $(M_{\odot})$ & $(M_{\odot})$ & & $(M_{\odot})$ & $(M_{\odot})$ & (cm$^2$\,s$^{-1}$) & & $(M_{\odot})$ & $(M_{\odot})$ & & $(M_{\odot})$ & $(M_{\odot})$ & (cm$^2$\,s$^{-1}$) & & $(M_{\odot})$ & $(M_{\odot})$ & & $(M_{\odot})$ & $(M_{\odot})$ & (cm$^2$\,s$^{-1}$) & & $(M_{\odot})$ & $(M_{\odot})$ & & $(M_{\odot})$ \\
\hline
20 & 20.22 & 16.16 & 0.26 & 6.94 & 10.93 & 0.58 & 9.29 & 19.52 & 16.17 & 0.30 & 7.91 & 12.59 & 0.58 & 6.93 & 7.12 & 15.83 & 0.28 & 5.47 & 7.04 & 0.41 & 0.08 & 6.95 & 15.77 & 0.35 & 5.29 & 6.82 & 0.41 & 0.13 \\
25 & 8.10 & 15.78 & 0.35 & 6.29 & 8.01 & 0.39 & 0.09 & 8.01 & 15.65 & 0.35 & 6.18 & 7.84 & 0.39 & 0.17 & 8.07 & 15.64 & 0.31 & 6.19 & 7.95 & 0.39 & 0.12 & 10.29 & 15.64 & 0.35 & 10.20 & 10.16 & 0.38 & 0.13 \\
30 & 17.35 & 15.78 & 0.36 & 17.27 & 17.26 & 0.38 & 0.09 & 9.07 & 15.78 & 0.30 & 8.95 & 8.92 & 0.36 & 0.15 & 9.02 & 15.78 & 0.37 & 8.93 & 8.94 & 0.38 & 0.08 & 9.04 & 15.70 & 0.36 & 8.93 & 8.87 & 0.37 & 0.17 \\
35 & 10.87 & 15.83 & 0.44 & 10.78 & 10.81 & 0.35 & 0.06 & 9.92 & 15.90 & 0.37 & 9.81 & 9.83 & 0.36 & 0.09 & 9.91 & 15.42 & 0.37 & 9.81 & 9.82 & 0.36 & 0.09 & 9.94 & 15.68 & 0.35 & 9.85 & 9.79 & 0.35 & 0.15 \\
40 & 10.99 & 15.82 & 0.38 & 10.93 & 10.99 & 0.34 & 0.00 & 10.76 & 15.49 & 0.32 & 10.71 & 10.75 & 0.34 & 0.01 & 10.77 & 15.89 & 0.36 & 10.66 & 10.71 & 0.34 & 0.06 & 10.69 & 15.71 & 0.32 & 10.67 & 10.69 & 0.34 & 0.00 \\
45 & 11.73 & 15.60 & 0.37 & 11.60 & 11.15 & 0.30 & 0.58 & 11.54 & 15.67 & 0.34 & 11.51 & 11.51 & 0.32 & 0.03 & 11.62 & 15.57 & 0.38 & 11.44 & 11.47 & 0.32 & 0.15 & 11.51 & 15.82 & 0.36 & 11.45 & 11.49 & 0.33 & 0.02 \\
50 & 13.04 & 15.74 & 0.32 & 12.95 & 13.04 & 0.30 & 0.00 & 12.48 & 15.62 & 0.35 & 12.38 & 12.41 & 0.30 & 0.07 & 12.42 & 15.72 & 0.32 & 12.30 & 12.35 & 0.30 & 0.07 & 12.42 & 15.74 & 0.37 & 12.30 & 12.35 & 0.31 & 0.07 \\
55 & 13.74 & 15.49 & 0.33 & 13.60 & 13.62 & 0.27 & 0.12 & 13.23 & 15.68 & 0.36 & 13.08 & 13.16 & 0.29 & 0.07 & 13.12 & 15.70 & 0.36 & 12.99 & 13.06 & 0.29 & 0.06 & 13.10 & 15.70 & 0.36 & 13.02 & 13.10 & 0.31 & 0.00 \\
60 & 14.00 & 15.56 & 0.37 & 13.88 & 13.93 & 0.27 & 0.07 & 13.95 & 15.70 & 0.33 & 13.82 & 13.87 & 0.28 & 0.08 & 13.89 & 15.62 & 0.36 & 13.77 & 13.86 & 0.28 & 0.03 & 13.85 & 15.61 & 0.38 & 13.71 & 13.83 & 0.28 & 0.02 \\
65 & 15.39 & 15.60 & 0.35 & 15.24 & 15.37 & 0.26 & 0.02 & 14.69 & 15.63 & 0.37 & 14.55 & 14.69 & 0.27 & 0.00 & 14.49 & 15.57 & 0.37 & 14.35 & 14.42 & 0.26 & 0.07 & 14.50 & 15.63 & 0.37 & 14.37 & 14.47 & 0.26 & 0.03 \\
70 & 15.30 & 15.59 & 0.36 & 15.16 & 15.28 & 0.25 & 0.02 & 15.26 & 15.53 & 0.35 & 15.12 & 15.22 & 0.25 & 0.04 & 15.22 & 15.61 & 0.37 & 15.07 & 15.15 & 0.25 & 0.07 & 15.26 & 15.56 & 0.37 & 15.02 & 15.17 & 0.25 & 0.09 \\
75 & 21.21 & 15.51 & 0.33 & 21.03 & 21.21 & 0.22 & 0.00 & 15.90 & 15.49 & 0.37 & 15.75 & 15.84 & 0.23 & 0.06 & 15.84 & 15.49 & 0.38 & 15.70 & 15.83 & 0.24 & 0.01 & 15.85 & 15.56 & 0.37 & 15.69 & 15.83 & 0.25 & 0.02 \\
80 & 25.24 & 15.49 & 0.35 & 24.99 & 25.24 & 0.23 & 0.00 & 16.54 & 15.44 & 0.34 & 16.39 & 16.53 & 0.22 & 0.01 & 16.44 & 15.52 & 0.38 & 16.28 & 16.44 & 0.23 & 0.00 & 16.41 & 15.55 & 0.36 & 16.26 & 16.40 & 0.23 & 0.01 \\
85 & 26.96 & 15.53 & 0.33 & 26.69 & 26.96 & 0.24 & 0.00 & 17.13 & 15.50 & 0.39 & 16.97 & 17.13 & 0.91 & 0.00 & 17.03 & 15.51 & 0.36 & 16.86 & 17.03 & 0.22 & 0.00 & 16.93 & 15.36 & 0.37 & 16.78 & 16.93 & 0.21 & 0.00 \\
90 & 17.84 & 15.49 & 0.34 & 17.69 & 17.84 & 0.20 & 0.00 & 17.66 & 15.49 & 0.39 & 17.48 & 17.62 & 0.21 & 0.04 & 17.62 & 15.49 & 0.39 & 17.44 & 17.62 & 0.20 & 0.00 & 17.53 & 15.55 & 0.40 & 17.37 & 17.53 & 0.22 & 0.00 \\
95 & 20.02 & 15.46 & 0.31 & 19.82 & 20.02 & 0.18 & 0.00 & 18.21 & 15.43 & 0.45 & 18.07 & 18.21 & 0.20 & 0.00 & 18.24 & 15.44 & 0.35 & 18.04 & 18.24 & 0.19 & 0.00 & 18.04 & 15.40 & 0.40 & 17.93 & 18.04 & 0.20 & 0.00 \\
100 & 29.51 & 15.53 & 0.28 & 29.19 & 29.51 & 0.19 & 0.00 & 20.42 & 15.41 & 0.41 & 20.20 & 20.42 & 0.18 & 0.00 & 18.83 & 15.42 & 0.33 & 18.64 & 18.83 & 0.19 & 0.00 & 18.70 & 15.45 & 0.39 & 18.53 & 18.70 & 0.19 & 0.00 \\
\hline
\end{tabular}
}
\label{tab:eta_d1.0}
\end{table*}

\bsp	
\label{lastpage}
\end{document}